\documentclass[prd,aps,amsfonts,showpacs,nofootinbib,longbibliography,notitlepage,twocolumn,superscriptaddress]{revtex4-1}

\usepackage{amsmath,bm,amssymb,amsfonts,mathtools,dsfont,relsize}
\usepackage{graphicx}
\usepackage{stmaryrd}

\allowdisplaybreaks

\usepackage[normalem]{ulem} 
\usepackage[dvipsnames]{xcolor}
\usepackage{comment}

\usepackage[colorlinks=true, urlcolor=violet, linkcolor=blue, citecolor=red, hyperindex=true, linktocpage=true]{hyperref}

\usepackage{amsthm}

\usepackage{soul} 

\usepackage{tikz}
\usetikzlibrary{calc}
\usetikzlibrary{positioning}
\usetikzlibrary{arrows,shapes}

\usepackage{enumerate}
\usepackage{suffix} 

\renewcommand{\thesection}{\arabic{section}}
\renewcommand{\thesubsection}{\thesection.\arabic{subsection}}

\makeatletter
\renewcommand{\p@subsection}{}
\def\l@subsubsection#1#2{}
\makeatother

\makeatletter
\g@addto@macro\bfseries{\boldmath}
\makeatother

\newcommand{\vps}[0]{\vphantom{*}}

\DeclarePairedDelimiter{\abs}{\lvert}{\rvert}

\newcommand{\ident}[0]{\mathds{1}}

\newcommand{\transpose}[0]{\mathsf{T}}
\begin{document}
\title{Long-range-enhanced surface codes}

\author{Yifan Hong}
\email{yifan.hong@colorado.edu}
\affiliation{Department of Physics, University of Colorado, Boulder, CO 80309, USA}
\affiliation{Center for Theory of Quantum Matter, University of Colorado, Boulder, CO 80309, USA}

\author{Matteo Marinelli}
\affiliation{Department of Physics, University of Colorado, Boulder, CO 80309, USA}
\affiliation{JILA and National Institute of Standards and Technology, Boulder, CO 80309, USA}

\author{Adam M. Kaufman}
\affiliation{Department of Physics, University of Colorado, Boulder, CO 80309, USA}
\affiliation{JILA and National Institute of Standards and Technology, Boulder, CO 80309, USA}

\author{Andrew Lucas}
\email{andrew.j.lucas@colorado.edu}
\affiliation{Department of Physics, University of Colorado, Boulder, CO 80309, USA}
\affiliation{Center for Theory of Quantum Matter, University of Colorado, Boulder, CO 80309, USA}

\begin{abstract}
    The surface code is a quantum error-correcting code for one logical qubit, protected by spatially localized parity checks in two dimensions.    Due to fundamental constraints from spatial locality, storing more logical qubits requires either sacrificing the robustness of the surface code against errors or increasing the number of physical qubits.  We bound the minimal number of spatially nonlocal parity checks necessary to add logical qubits to a surface code while maintaining, or improving, robustness to errors. We saturate the lower limit of this bound, when the number of added logical qubits is a constant, using a family of hypergraph product codes, interpolating between the surface code and constant-rate low-density parity-check codes. Fault-tolerant protocols for logical gates in the quantum code can be inherited from its classical parent codes. We provide near-term practical implementations of this code for hardware based on trapped ions or  neutral atoms in mobile optical tweezers.  Long-range-enhanced surface codes outperform conventional surface codes using hundreds of physical qubits, and represent a practical strategy to enhance the robustness of logical qubits to errors in near-term devices.  
\end{abstract}

\maketitle

\section{Introduction} 

Noise is inherent in quantum computers, and if ignored, will always destroy any quantum computational advantage. With advances in quantum hardware enabling controllable systems of hundreds of qubits, the use of quantum error correction to prolong the lifetime of quantum information is becoming feasible. At the hardware-theory interface, a key goal is to design optimal codes that leverage specific hardware-level advantages, such as gate nonlocality, to mitigate the effects of key challenges, like the fidelity of few-qubit gates, or the resources required to increase the number of qubits in the system.

Quantum error correction is done by starting with a Hilbert space of $n$ \emph{physical qubits}, and identifying a subset $2^k<2^n$ of the possible states within Hilbert space as encoding the wave function of $k$ \emph{logical qubits}. The smallest number of physical qubits on which a nontrivial logical operation can act determines the code distance $d$, and such a code is often abbreviated as $\llbracket n,k,d \rrbracket$.  A practical set of codes are \emph{stabilizer codes} \cite{gottesman1997stabilizer} in which the logical codewords are the simultaneous $+1$ eigenstates of a commuting set of Pauli operators called the stabilizer group. A Calderbank-Shor-Steane (CSS) code \cite{Calderbank_1996, Steane_1996} is a stabilizer code for which the generators of this set are strictly products of Pauli $X$s or $Z$s. 

\begin{figure}[t]
\centering
\includegraphics[width=0.35\textwidth]{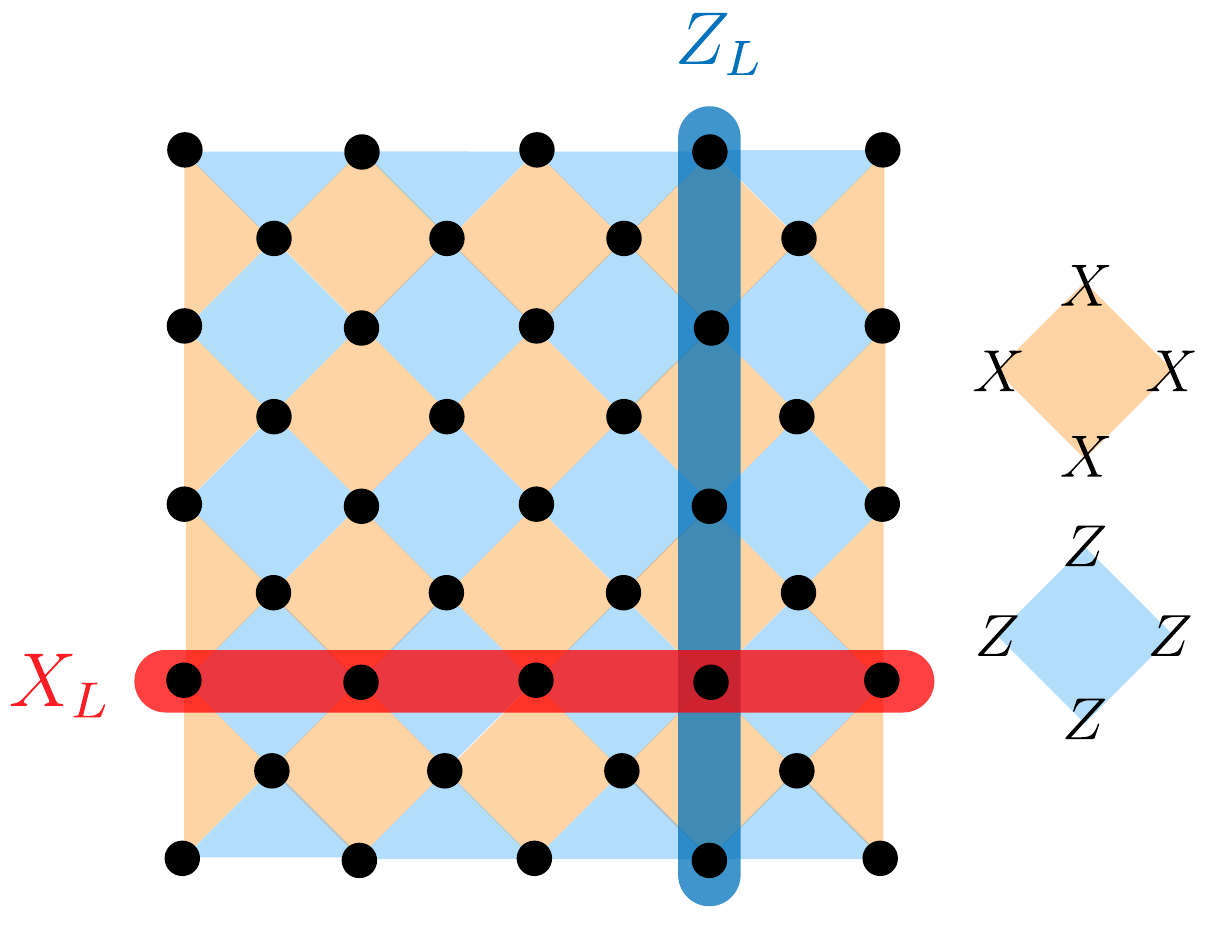}
\caption{The 2D layout of a $\llbracket 41,1,5 \rrbracket$ surface code. Black dots and colored tiles represent physical qubits and stabilizer generators, respectively. The string-like logical operators are also depicted. }
\label{fig:surface-41}
\end{figure}

An important example of a CSS code is the toric code \cite{Kitaev_2003}. Together with its planar cousin, the surface code \cite{bravyi1998, Fowler_2012}, they are leading candidates for near-term implementations of fault-tolerant quantum computation. It has local stabilizer generators supported on a checkerboard-style lattice: see Fig. \ref{fig:surface-41}. The toric code has been realized with neutral atoms \cite{bluvstein2022quantum}, with fault tolerance later achieved in the surface code with superconducting qubits \cite{Google_SC}, although the ``break-even" point after which the logical qubit is more robust than an isolated physical qubit  remains to be reached. Even more recently, progress towards intermediate-scale fault tolerance has been demonstrated with 40 logical qubits encoded in color codes, a close relative of the surface code, using 280 neutral atoms \cite{Bluvstein_2023}. For hardware with highly biased noise (e.g. Pauli $Z$ error much more likely than Pauli $X$ error), there exists elegant modifications to the surface code \cite{Bonilla_Ataides_2021}.

\begin{figure*}[t]
\centering
\includegraphics[width=0.99\textwidth]{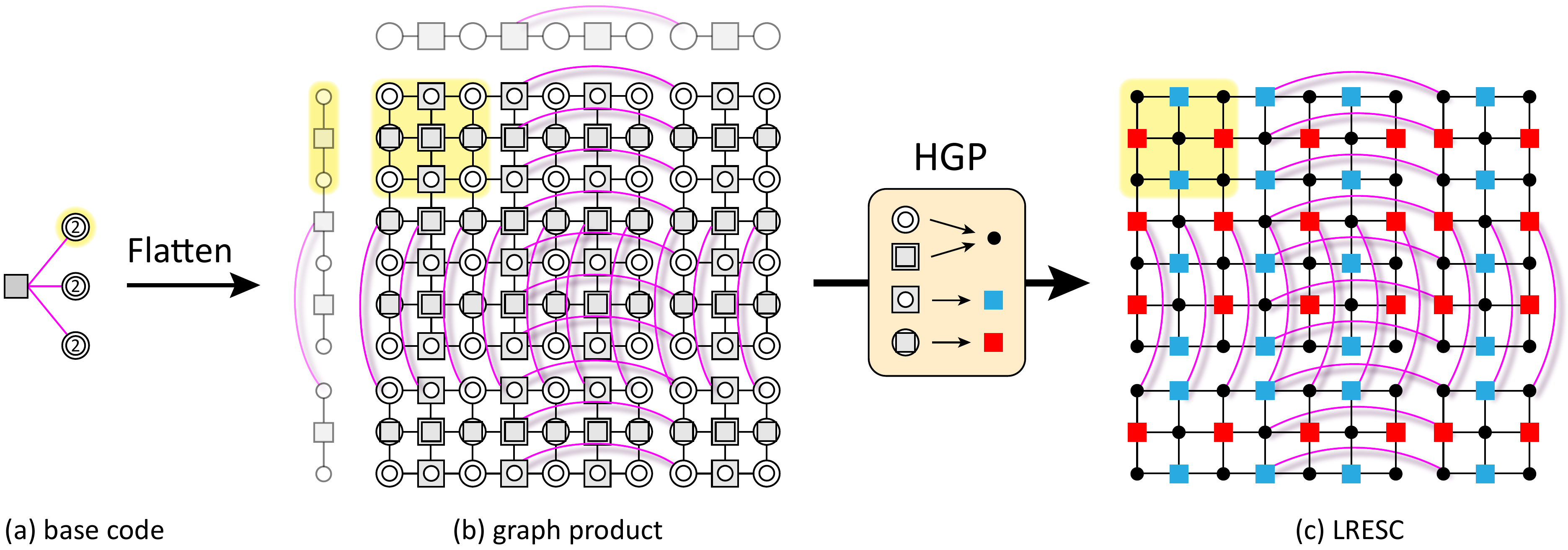}
\caption{The steps to construct a $\llbracket 52,4,4 \rrbracket$ HGP code based on a parent $[3(2),2,2(2)]$ concatenated code are illustrated. (a) The Tanner graph of the base $[3,2,2]$ code is depicted where the square and circles represents the parity check and bits respectively. The double concentric circle with a ``2" in the middle means each bit in this base code is actually the logical bit of a length-2 inner repetition code after concatenation. (b) The Tanner graph of the concatenated code is flattened in 1D, and then a Euclidean graph product is taken which produces four types of vertices in a 2D embedding. (c) The HGP procedure transforms the product graph into a CSS Tanner graph. 20 long-range interactions (magenta curves) of range 4 are required for this code. A $k=4$, $d=4$ surface code of the same layout will require $n \geq 100$ physical qubits.}
\label{fig:HGP construction detailed}
\end{figure*}

Due to the spatial locality of a surface code in two spatial dimensions, it is highly desirable for experimentalists; nearly all platforms, including atoms in optical tweezers \cite{Saffman2010, Saffman2016, kaufman2021quantum, wu2022erasure, cong2022hardware, evered2023high,ma2023high}, trapped ions~\cite{PhysRevLett.74.4091, Bruzewicz_2019, kihwan_2010, Britton_2012, Barreiro_2011}, or superconducting qubits \cite{PhysRevLett.89.117901, PhysRevLett.79.2328, Krantz_2019, Brooks_2013, Gu_2017}, can realize geometrically local interactions in two spatial dimensions.  Unfortunately, quantum computation with $O(10^3)$ logical qubits in a surface code architecture with typical error rates of $O(10^{-3})$ may require an architecture with $O(10^7)$ physical qubits \cite{Gidney_2021}, which could be prohibitively difficult to build in the near term.

An exciting alternative are quantum low-density parity-check (qLDPC) codes, which can achieve $k\sim n$: the overhead for encoding logical information is finite. At the same time, the stabilizers are few-body just like the surface code (but not necessarily spatially local), meaning they can in principle be measured efficiently using few-qubit operations.  The first qLDPC construction with a finite rate ($k\sim n$) and large distance ($d\sim \sqrt{n}$) was the hypergraph product (HGP) \cite{HGP}; a series of improvements \cite{TP_codes, BP_codes, LP_codes} eventually led to ``good" codes with $k\sim d \sim n$ \cite{Panteleev_2022, qTanner_codes, dinur2022good}. 

Spatial locality constrains the implementation of qLDPC codes in quantum hardware. Suppose that each physical qubit is arranged in a two-dimensional grid, and qubits can only interact with other qubits a finite distance away.  Then one can prove \cite{BPT_2010} that $kd^2 \lesssim n$: there is an unavoidable tradeoff between robustness to error ($d$) and number of logical qubits ($k$), given a fixed number of physical qubits ($n$). Conversely, it is known \cite{Baspin_2022} that to implement a qLDPC in 2D, at least $\Omega\left(\sqrt{\frac{k}{n}}d\right)$ interactions of range $\Omega\left(\sqrt{\frac{k}{\sqrt{n}}}\right)$ are necessary.  If we only ask for $d\sim\sqrt{n}$ as in the surface code, the bounds of \cite{Baspin_2022} alone admit the prospect of $k\sim \sqrt{n}$ using interactions of $O(1)$ range.  Since \cite{BPT_2010} proves that these finite-range interactions only allow $k=O(1)$, the cost of nonlocality in qLDPC codes is even  higher than implied by \cite{Baspin_2022}. Further challenges to qLDPC implementation in 2D were discussed in \cite{delfosse2021, baspin2023improved}. It is thus of crucial interest to know: how many nonlocal stabilizers are needed to add logical qubits to a surface code, while keeping $d$ and $n$ fixed?   If we find a code that uses the least nonlocality to add logical qubits to the surface code, is it realizable in any near-term quantum  hardware?

This paper answers these questions.  We present \emph{long-range-enhanced surface codes} (LRESCs): an interpolating family of hypergraph product codes that bridges the surface code with constant-rate qLDPC codes.   These codes: (\emph{1}) have as few nonlocal stabilizers as possible in the $k=O(1)$ limit, (\emph{2}) maintain the code distance $d$ of the surface code while adding logical qubits, i.e. increasing $k$, (\emph{3}) have lower logical failure rates compared to a surface code under measurement noise, and (\emph{4}) enable fault-tolerant gadgets to be inherited from those of classical codes.  The simplest realization of the LRESC has a ``hierarchical" structure similar to a recent construction \cite{pattison2023}; however, unlike \cite{pattison2023}, LRESCs are LDPC stabilizer codes, employing as little nonlocality as possible.   Moreover, as we will explain, these codes are well suited for implementation using multiple different architectures for quantum computation, as the specific form of nonlocality required by LRESCs is far more efficient to implement than a generic qLDPC code.


\section{The LRESC}

\subsection{Construction}

We begin by summarizing intuitively the structure of LRESCs; technical details are provided in appendices.  Our construction consists of three parts, visualized in Fig. \ref{fig:HGP construction detailed}.

(\emph{1}) First, begin with a classical base code. For instance, one can pick a ``good" classical LDPC (cLDPC) code \cite{gallager_1962,random_ldpc}, which uses $L_0$ classical bits to store $\Theta(L_0)$ logical bits with $\Theta(L_0)$ distance.  In this paper, we will focus on relatively small code sizes where $L_0\sim 3-10$, both for pedagogy and near-term relevance.   Note that a good cLDPC code will require nonlocal parity checks, with $\Omega(L_0)$ range in general, between the classical bits to ensure constant rate.  Appendix \ref{app:cLDPC} overviews classical codes.

(\emph{2}) Next, we increase the number of classical bits: $L_0 \rightarrow cL_0 = L$, while proportionally increasing the distance of the code to $\Theta(L)$ and keeping the number of logical bits fixed as $\Theta(L_0)$.  We can do so by replacing the bits of our starting code with another classical code, such as the repetition code, which stores a single logical bit in $c$ physical bits, with codewords $0\rightarrow 0\cdots 0$ and $1\rightarrow 1 \cdots 1$.  The repetition code has spatially local parity checks when bits are laid out in one-dimension: the parity checks demand that the parity of two nearest-neighbor bits agree.  We thus build a \emph{concatenated} code by replacing the cLDPC ``physical bits" with repetition codes of length $c$.  There is no code with fewer nonlocal edges in one spatial dimension that has $\Theta(L_0)$ logical bits and $\Theta(L)$ code distance (see Appendix \ref{app:cLDPC}).  Decoding this concatenated code can be done in two steps: we first decode each repetition code, and then decode the size-$L_0$ LDPC using the state of each repetition code as an effective ``physical bit".

(\emph{3}) We now build a quantum code by taking the \emph{hypergraph product} (HGP) of this classical concatenated code with itself.  A formal definition of the HGP is technical and relegated to Appendix \ref{app:HGP}; Fig. \ref{fig:HGP construction detailed} sketches the idea.  We lay out two copies of the classical code of length $L$ along the $x$ and $y$ directions in the plane.  Based on the connections between checks and physical bits of the classical code, we lay out physical qubits and $X$ and $Z$ type stabilizers of the quantum code in two dimensions. Note that the hypergraph product of two classical repetition codes is the quantum surface code.  Since our classical codes contain repetition code segments, our quantum code consists of two-dimensional surface code patches.  Long-range parity checks from step (\emph{1}) induce stabilizers with range $\Omega(L)$ that connect distant patches in the code, while ensuring that each stabilizer itself has low-weight (is a product of $O(1)$ $X$s or $Z$s).

These are the LRESCs. The total number of physical qubits is $n\sim L^2$, the quantum code distance is $d\sim L$, and the number of logical qubits is $k\sim L_0^2$.  Alternatively, we have constructed a code with $d\sim \sqrt{n}$, just like the surface code, but where we have added $k$ logical qubits at the cost of adding $O(L\sqrt{k})$ long-range stabilizers.


\subsection{Bounding nonlocality}

\begin{figure}[t]
\centering
\includegraphics[width=0.4\textwidth]{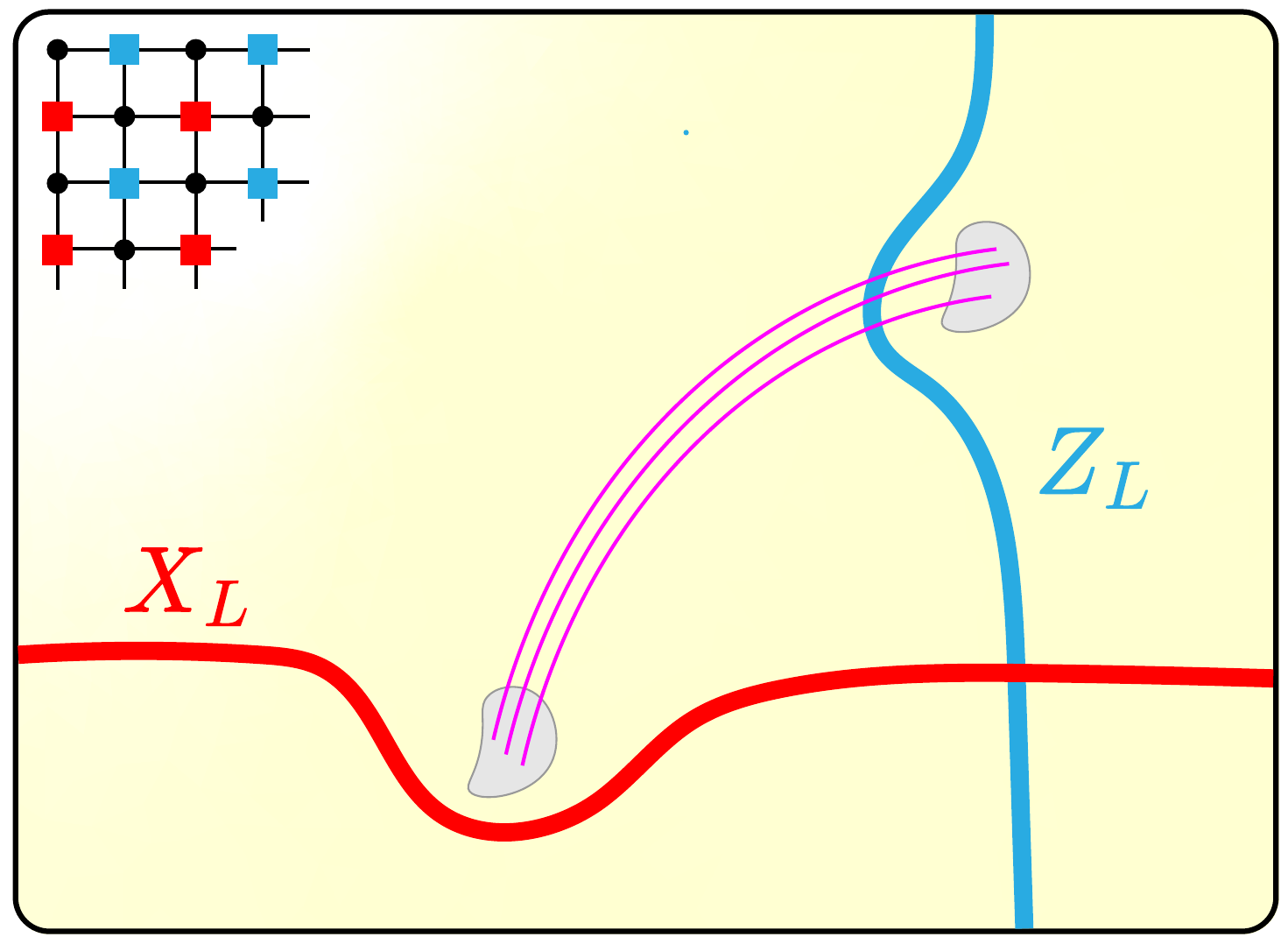}
\caption{A geometrically local 2D stabilizer code (surface code depicted) is modified with $\ell \ll d$ long-range interactions (magenta curves). For this modified code, all logical operators (two shown as thick red and blue curves) can be cleaned to exist completely outside of the support of the long-range checks (gray regions).}
\label{fig:2D few LR edges}
\end{figure}

\begin{figure*}[t]
\centering
\includegraphics[width=0.48\textwidth]{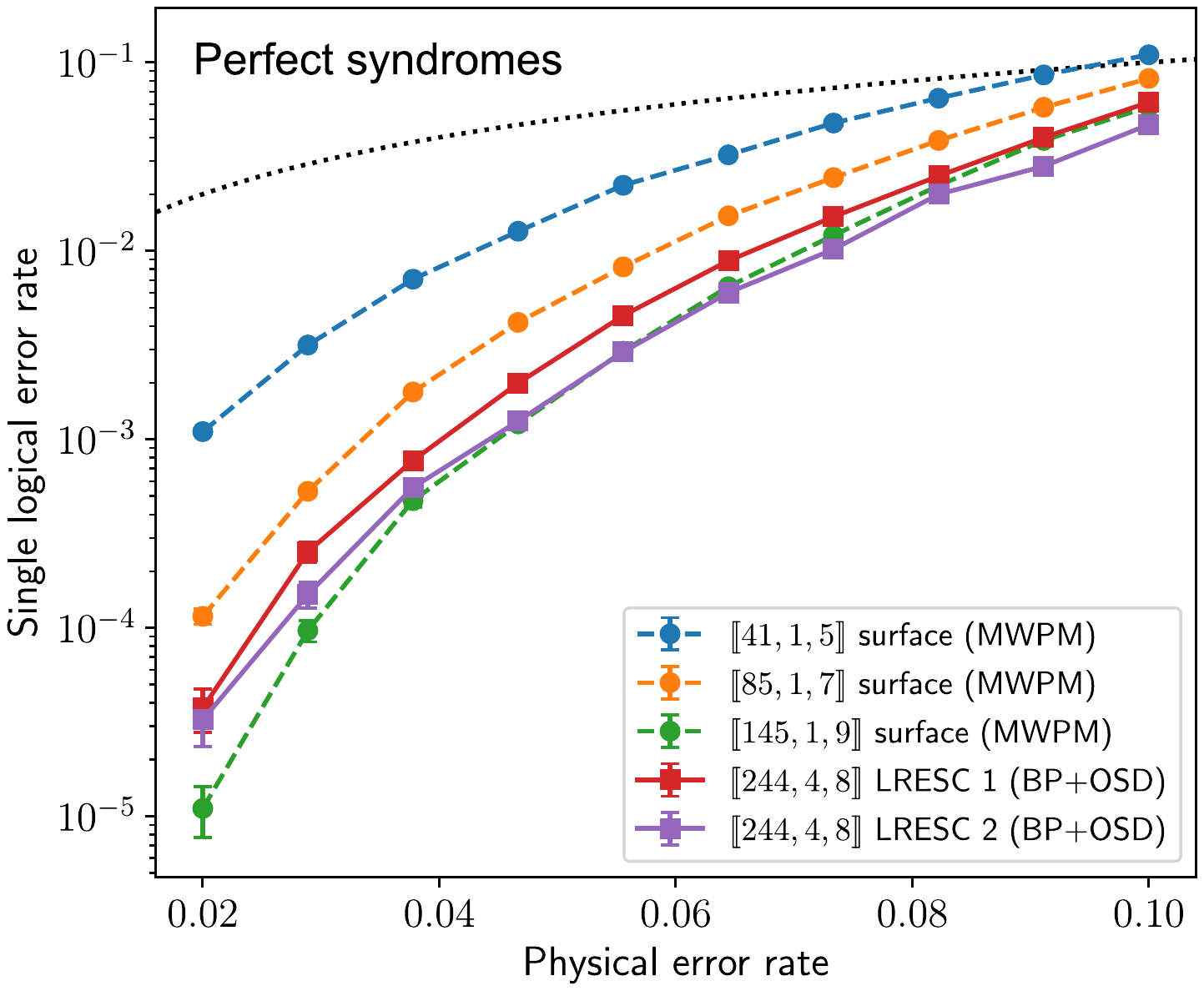} \hspace{1em}
\includegraphics[width=0.48\textwidth]{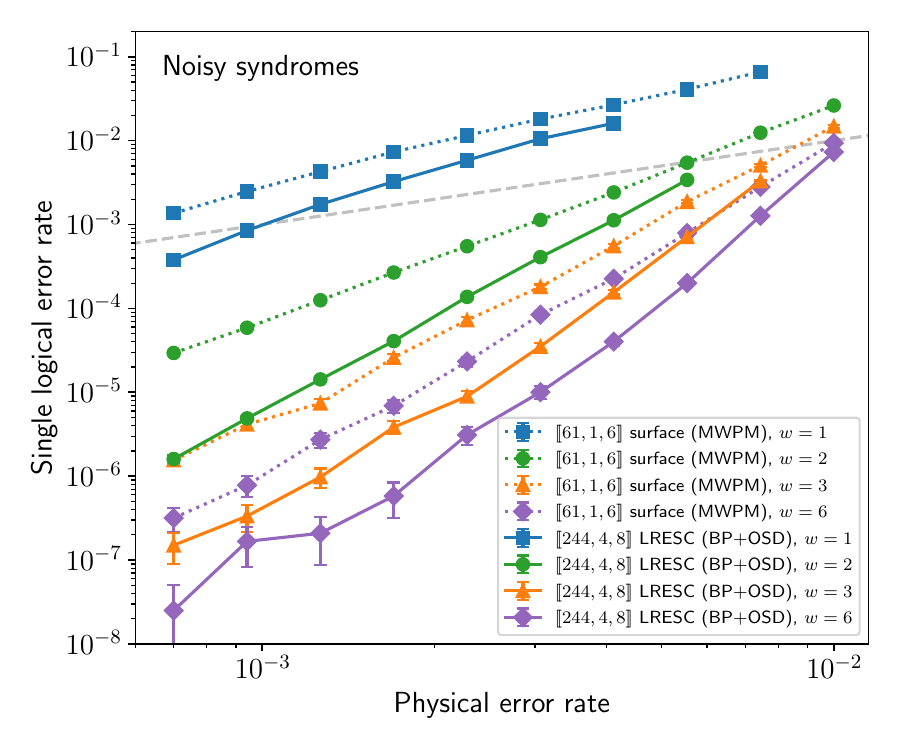}
\caption{QEC performance is numerically estimated using MWPM decoding for three surface codes with increasing distance as well as BP+OSD for LRESCs from Table \ref{tab:LRESCs}.  Left: $\sim 10^5$ clean QEC cycles are averaged per data point. Right: A comparison between a $d=6$ surface code and LRESC 2 is shown for $\sim 10^4$ samples of 100 noisy QEC cycles. Decoding is performed using a sliding window technique with variable window size $w$. Uncertainties are given by standard errors. A round of noiseless decoding is performed internally after each noisy cycle to probe the residual errors. The break-even line is plotted in dashed gray.}
\label{fig:HGP vs surface}
\end{figure*}

There is no code that has parametrically fewer long-range stabilizers in 2D than an LRESC, while maintaining the same code distance $d\sim \sqrt{n}$ and adding O(1) additional qubits beyond the bound of \cite{BPT_2010}.  To see why, start with a 2D code $\mathcal{C}_0$ whose checks have support only on qubits within a ball of radius $r_0 = O(1)$ and whose distance satisfies $d_0 \sim \sqrt{n}$. BPT \cite{BPT_2010} tells us that there is at most a finite $k_0 = O(1)$ number of logical qubits, such that $k^{\vps}_0 d^2_0 \le Kn$ for some $K = O(1)$.  To increase the number of logical qubits from $k^{\vps}_0$, we will need to add longer-range interactions:  how many are required?  Suppose that by modifying $\ell \ll d_0$ of our local stabilizers to be spatially nonlocal (but still low weight, i.e. few Paulis) we obtain a new code $\mathcal{C}_1$ with modified parameters $\llbracket n, k_1, d_1 \rrbracket$. Further suppose that we have $d_1 \sim d_0$. Then, since the new code is still LDPC, the size of the region $R$ which contains the long-range checks is $O(\ell) \ll d_1$ and hence a correctable region. The Cleaning lemma \cite{Bravyi_2009} ensures that we may choose all logical operators such that they are supported only on sites outside of the support of these long-range checks, as shown in Fig. \ref{fig:2D few LR edges}. Since $\mathcal{C}_0$ and $\mathcal{C}_1$ share the same local checks outside of region $R$, and we have only cleaned the logical operators off of at most $O(r_0^2 \ell) \ll d_0$ sites, the cleaned logical operators are also valid logical operators for the original local code $\mathcal{C}_0$; the number of such logical operators for inequivalent logical qubits is hence bounded by BPT: we can have no more than $k_0$ logical qubits.

The only way around the above argument is relax the condition that the long-range region is correctable so that we cannot clean out all logical operators. As a consequence, there would then exist a logical operator supported entirely within this region, and so we would have $d_1 \ll d_0$. Thus, to add logical qubits to a surface code without sacrificing distance, $d\sim \sqrt{n}\sim L$ long-range stabilizers are required. The LRESC achieves this scaling, and is parametrically as local as possible for finite $k>k_0$. Interestingly, adding logical qubits beyond the $\ell=\Omega(d)$ restriction need not require additional long-range interactions, but rather a ``rewiring" of them. For example, consider a $L\times L$ surface code and add a line of $L$ qubits encoded in a good $\llbracket L,\Theta(L),\Theta(L) \rrbracket$ quantum LDPC code. Then the number of logical qubits is $k \sim L$ with $\ell \sim L$ long-range interactions.

Note that the Cleaning lemma is crucial to the above argument. It has no direct classical analogue, which is why we could improve the classical code dimension with only $\ell \ll d$ long-range interactions while maintaining the scaling of the distance. The physical intuition for the Cleaning lemma is that due to the unitarity of quantum mechanics, a correctable region must not reveal any encoded information; otherwise it may be possible to violate the No-Cloning theorem. As a consequence, a small correctable region is effectively ``invisible" to the logical codespace. For classical codes this is not true: in the repetition code the value of a single physical bit reveals the value of the logical bit.


\subsection{Quantum error correction}

Quantum error correction (QEC) for stabilizer codes is typically done by extracting the eigenvalues of all stabilizers, which can be deduced by measuring a set of generators called the check set; the outcomes of these measurements comprise the error syndrome. Decoding then proceeds by finding a suitable correction operator according to the syndrome. The combination of the original error and the correction then either leaves the codespace unchanged (success) or enacts an undesirable logical operation (fail). For some codes, such as the surface code, when syndrome measurements can be faulty, it is necessary to perform several rounds of syndrome measurements and collectively decode over a ``spacetime" decoding graph \cite{Dennis_2002}.

We conduct numerical simulations of QEC, using both a code-capacity (clean syndromes) and a weighted phenomenological (noisy syndromes) noise model under a local, stochastic depolarization channel (single-qubit $X$, $Y$ or $Z$ errors are equally likely) with probability $p$: see Fig. \ref{fig:HGP vs surface}. For the phenomenological noise model we scale both the physical error and the syndrome measurement error rates according the degree distribution of the Tanner graph: a qubit participating in $v$ checks has an error rate of $vp$, and a check with weight $w$ is incorrectly measured with probability $wp$; this mimics the experimental way that such syndromes are measured, as we will explain later. We implement a single-stage decoder utilizing belief propagation with ordered-statistics \cite{Fossorier_1995, Panteleev_2021} post-processing (BP+OSD), which has been previously shown to have favorable performance as a general-purpose qLDPC decoder. We use the ``min-sum" and ``combination-sweep" $(\lambda=30)$ variants of BP and OSD from open-source software \cite{Roffe_LDPC_Python_tools_2022}. Syndrome errors are accounted for by adding an additional variable node for each check node in the Tanner graph \cite{Kuo_2021, PRXQuantum.4.020332}. For the phenomenological model, we perform 100 noisy QEC cycles for each Monte Carlo trial. When syndrome errors are present, decoding is not perfect, and there will often be a ``residual error" which is carried onto the next QEC cycle. In order to ensure that this residual error is not detrimental, we perform noiseless decoding after each noisy cycle as an internal flag to ensure that residual errors are successfully controlled; the QEC simulation outputs a failure if the residual error cannot be properly decoded. We decode using a ``sliding window" technique \cite{Dennis_2002}, where the correction at time $t$ is determined using the syndrome information from times $t$ to $t+w$ with window size $w$. Recently, it was shown that this technique considerably reduced logical error rates for several classes of qLDPC codes \cite{huang2023improved} beyond the single-shot regime ($w=1$).

\begin{table}[t]
\centering
\begin{tabular}{c|c|c|c}
LRESC & \;edges\; & \;LR edges (optimized)\; & \;LR ratio \;\\ \hline
$\llbracket 244,4,8 \rrbracket$\; & 924 & 60 (20) & 6.5\% (2.2\%) \\
$\llbracket 244,4,8 \rrbracket$\; & 1056 & 528 (176) & 50\% (16.7\%)
\end{tabular}
\caption{The number of long-range interactions vs total number of pairwise interactions are displayed for three LRESCs. Optimized embeddings of the Tanner graphs lower the number of long-range interactions (in parentheses).}
\label{tab:LRESCs}
\end{table}

To construct the LRESCs, we use parent codes (1) $[3(4),2,2(4)]$ and (2) $[6(2),2,4(2)]$, where $[n'(c),k',d'(c)]$ is short for an outer $[n',k',d']$ code concatenated with an inner $[c,1,c]$ repetition code. The corresponding LRESCs have mutual parameters $\llbracket 244,4,8 \rrbracket$ with rate $k/n \approx 1.64\%$ (61 physical qubits per logical qubit). The second LRESC contains more long-range interactions than the first: see Table \ref{tab:LRESCs}. The performance of BP+OSD decoding on these LRESCs is compared with that of minimum-weight perfect-matching (MWPM) on $d=5,7,9$ surface codes. For the MWPM simulations we use the open-source software PyMatching \cite{pymatching}. We observe that the first two LRESCs perform similarly to surface codes of similar distance under the code-capacity model. The benefits of the long-range interactions are revealed when syndrome noise is taken into consideration. Under the phenomenological noise model, we observe that the second LRESC performs considerably better than a surface code with the same rate for window sizes $w = 1,2,3,6$.

With an encoding rate of 61 physical qubits per logical qubit, the LRESCs begin to significantly outperform the surface codes of similar rate. With hundreds of physical qubits, an LRESC can surpass the break-even point -- where the collective logical qubit is more stable than a single isolated qubit -- once one- and two-qubit operations are achieved with $\gtrsim 99.5\%$ fidelity, which in many platforms is near-term~\cite{evered2023high, ma2023high} or within reach~\cite{leu_fast_2023, ding2023high, zhang2023tunable}.



\subsection{Logical operators} To understand why the LRESC not only stores more logical qubits, but also has reduced logical error rates, we need to understand how LRESCs encode logical qubits. As hinted at previously, since logical operators \emph{locally} look like repetition codes in the concatenated cLDPC codes (Step 2 above), in the HGP, logical operators \emph{locally} look like surface code logicals, which are strings of Pauli $X$ or $Z$ stretching across a surface code patch.  What differs from the usual surface code is the global structure of the logical operator: i.e. how strings in different patches are joined together.  A sketch is shown in Fig. \ref{fig:logical operators}, with technical details in Appendix \ref{app:anyon}.   In a nutshell: the simplest logical operator in an LRESC corresponds to strings in $O(\sqrt{k})$ of the surface code patches, corresponding to an analogous logical codeword of our cLDPC from Step 1 above.  

\begin{figure}[t]
\centering
\includegraphics[width=0.4\textwidth]{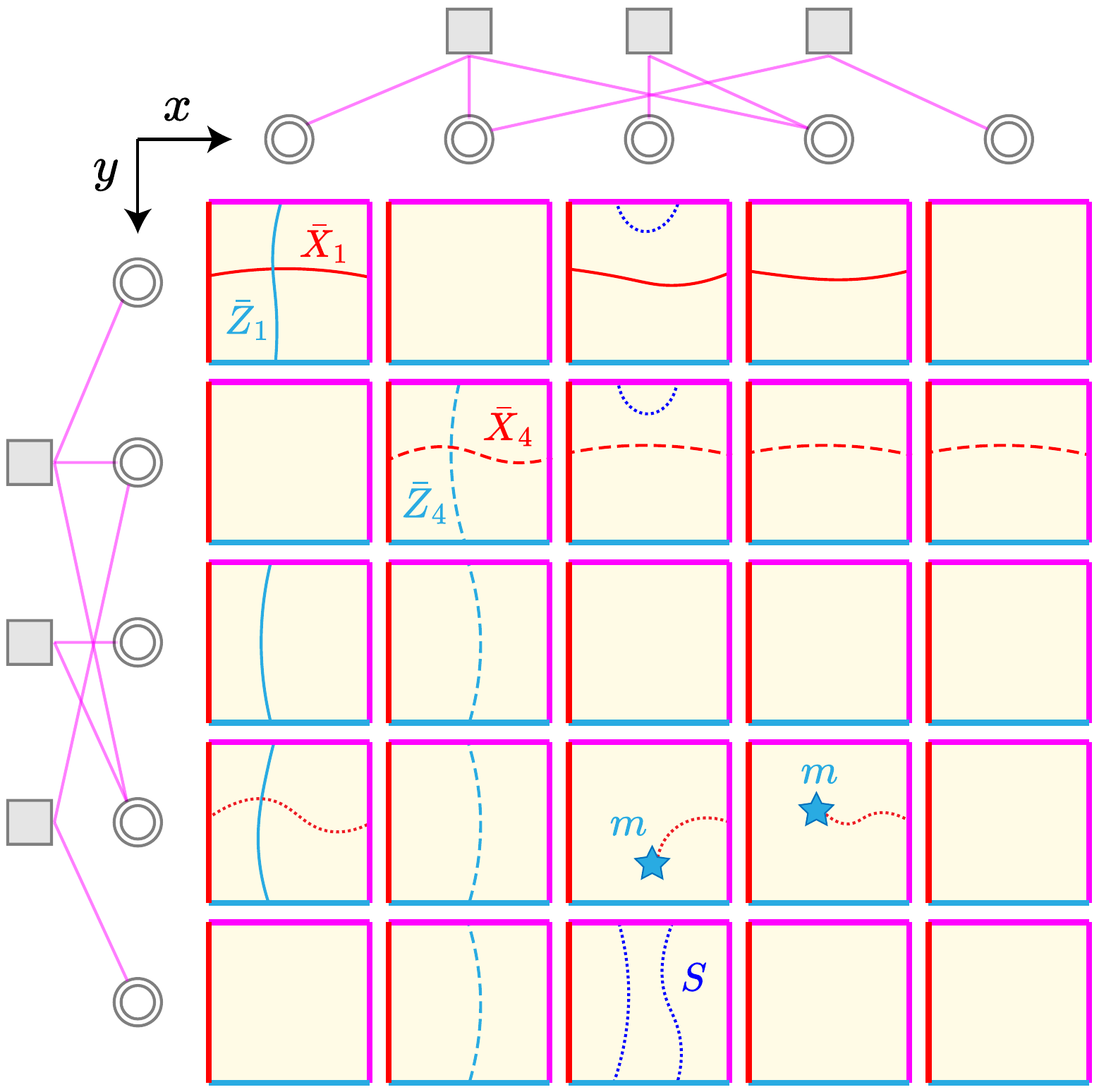}
\caption{A 2D layout of a LRESC with an outer $[5,2,3]$ parent code is shown. Smooth ($X$-type) and rough ($Z$-type) boundaries are located on the left (thick red) and bottom sides (thick blue) of each patch respectively, with long-range boundaries on the other (magenta) sides. Four logical operators ($\bar{X}_1, \bar{Z}_1, \bar{X}_4, \bar{Z}_4$) corresponding to two logical qubits are drawn (solid and dashed curves). A $Z$-type stabilizer and an $X$-type error string also shown (dotted red and blue).}
\label{fig:logical operators}
\end{figure}

We can intuitively understand why LRESCs are more effective at protecting logical information by showing that no matter how a logical error forms via local processes, during the formation of the error, we always violate more check operators than in an ordinary surface code.  Since more checks are violated, we have more opportunities to catch the physical qubit errors before they introduce a logical error.  In the surface code, we can create a logical error by introducing a physical error near one boundary and then causing a cascade of additional errors on adjacent sites, i.e. growing a logical string in Fig. \ref{fig:logical operators}.  At any step during this process, for the ordinary surface code, only one check is violated, meaning the error is almost undetected.  In condensed matter physics, we can interpret this as an \emph{anyonic} particle that is free to diffuse around the system.   In the LRESC, we can similarly grow an error through a single patch; however, when the error hits the long-range boundary, it will flip \emph{multiple} checks in adjacent patches (anyons are not conserved across the long-range boundaries of the LRESC).  The rules for anyon splitting are discussed in Appendix \ref{app:anyon}.  Since the error must grow across multiple patches to constitute a logical, we must inevitably flip more checks during the formation of the logical error, implying that it is easier to detect.

\subsection{Logical gates}
\label{sec:logical gates}

\begin{figure*}[t]
\centering
\includegraphics[width=0.75\textwidth]{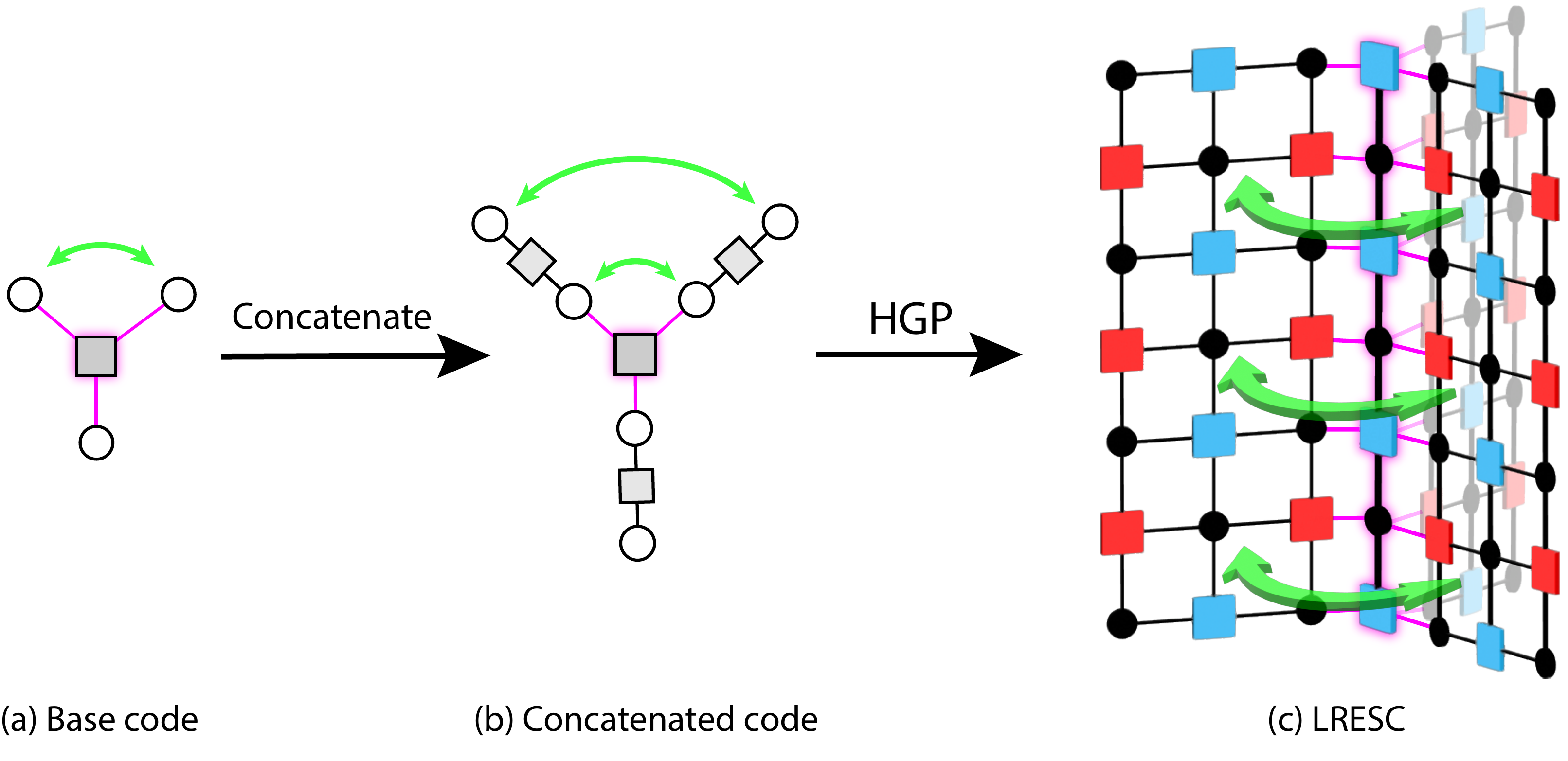}
\caption{The inheritance of the SWAP-CNOT logical gadget from the $[3,2,2]$ base code to the corresponding LRESC is illustrated. (a) The Tanner graph of the $[3,2,2]$ is shown, where circles and squares represent bits and checks respectively. A logical CNOT is performed by swapping the bits according to the green arrow. (b) Upon concatenation with a repetition code, the bits in the base code now become segments of repetition code. The single SWAP in the base code now becomes a segment-transversal SWAP between the respective repetition codes. (c) The hypergraph product code using the previous concatenated code with a repetition code of length 4 is depicted. The long-range components are organized into a single central column to which the surface-code patches are attached at their boundaries. The logical CNOT is implemented by a patch-transversal SWAP between the corresponding patches of surface codes.}
\label{fig:[3,2,2] SWAP-CNOT}
\end{figure*}

Implementing one- and two-qubit logical gates on an LRESC is (in principle) quite simple due to its similarly to a surface code. By readily organizing one of our logical qubits into a contiguous surface-code patch (e.g. moving surface-code patches in Fig. \ref{fig:logical operators} so that a logical string becomes ``continuous" and adjacent to the global boundaries), we can apply generalized lattice surgery techniques to perform Clifford gates \cite{Cohen_2022}. Note that one will require surface code patches of $O(n)$ physical qubits to implement logical gates on $O(1)$ logical qubits. Alternatively, one can use the same lattice surgery to teleport a logical qubit onto a surface code \cite{xu2023constantoverhead}. Once a logical qubit is in an ordinary surface code patch, standard methods \cite{Horsman_2012, Brown_2017} can then be used to apply all logical Clifford operations in a fault tolerant way. This process can be repeated to pass multiple qubits into surface code patches, onto which two-qubit gates can be fault-tolerantly applied. As an alternative, some work has been done on applying logical Clifford gates without the need for lattice surgery by generalizing hole-braiding in the surface code \cite{HGP_FTgates}.

Transversal gates have both constant spatial and temporal cost owing to their inherent parallelization at the physical level, but a hypergraph product code has yet to be found which can transversally implement the entire Clifford group on all its logical qubits. Some progress on transversal gates has been previously made, which can fill the Clifford group when supplemented with more expensive techniques such as code switching and state injection \cite{quintavalle2022}. Because the previous techniques apply to generic hypergraph product codes, they apply to LRESCs as well.

We now extend upon these previous results by presenting additional transversal gates in specific LRESCs, inherited from those of specially structured parent codes, with the amount of fault tolerance, quantified by transversality, tuned by adjusting the size of the local surface-code patches (i.e. the concatenation parameter $c$). We showcase an example using the $[3,2,2]$ parity code, with details regarding the general procedure in Appendix \ref{app:logical gates}. The $[3,2,2]$ code is the dual of the 3-bit repetition code with a single parity check $H = (1\;1\;1)$ and two logical bits with codewords $\bar{1}\bar{0} = 101$ and $\bar{0}\bar{1} = 011$. The logical $\mathrm{CNOT}_{\bar{1}\rightarrow \bar{2}}$ between the first (control) and second (target) logical bits maps $\bar{1}\bar{0} \rightarrow \bar{1}\bar{1}$ and $\bar{1}\bar{1} \rightarrow \bar{1}\bar{0}$ while leaving the other two logical bitstrings invariant. This transformation can be realized by physically swapping the second and third bits. The complementary logical $\mathrm{CNOT}_{\bar{2}\rightarrow\bar{1}}$ with control and target switched is realized by physically swapping the first and third bits. Note that the single 111 parity check remains invariant under both of these physical SWAPs.

When we concatenate with a 1D repetition code of length $c$, we obtain a $[3(c),2,2(c)]$ code. The codewords for this new concatenated code mimic those of the original base code but each 0 and 1 now become strings of 0s and 1s of length $c$. Because the values of the bits along each repetition-code segment are simply copies of the original bits in the $[3,2,2]$ base code, any transformations of the base code now become \emph{segment-transversal}  in the concatenated code: we simply apply the same transformation in parallel to all physical bits in the corresponding repetition codes. See Fig. \ref{fig:[3,2,2] SWAP-CNOT}b for an illustration. 

Taking the hypergraph product of the above concatenated $[3(c),2,2(c)]$ code with a 1D repetition code results in a quantum CSS code which can be arranged as three surface-code patches connected by a shared central boundary: see Fig. \ref{fig:[3,2,2] SWAP-CNOT}c. One can quickly verify by inspection that the stabilizer checks remain invariant upon swapping any two surface-code patches due to the ``fold" symmetry about the central column. In addition, like the segment-transversal implementation of the concatenated parent code, this patch swap can be implemented in a \emph{patch-transversal} manner: apply parallel SWAPs between qubits paired under the ``fold" symmetry. Choosing our ``horizontal" logical $\bar{X}$ operators to mimic the structure of the codewords of the parent $[3(c),2,2(c)]$ code, we can ensure that swapping surface-code patches enacts the correct CNOT transformation on the logical $\bar{X}$ operators. It suffices to verify the SWAP action on the logical $\bar{Z}$ operators. Choose $\bar{Z}_1$ and $\bar{Z}_2$ to be ``vertical" strings living in the first and second patches respectively. Then swapping the second and third patches leaves $\bar{Z}_1$ invariant while moving $\bar{Z}_2$ from the second to the third patch, which becomes stabilizer equivalent to $\bar{Z}_1\bar{Z}_2$. Similarly, swapping the first and third patches leaves $\bar{Z}_2$ invariant while transforming $\bar{Z}_1\rightarrow \bar{Z}_1\bar{Z}_2$. So we see that the patch-transversal SWAPs successfully implement the desired logical CNOT gates.

We conclude this section with some comments regarding the general procedure for logical gate inheritance and implementations for non-Clifford gates. The remarkable feature of performing entangling gates at the logical level with non-entangling gates at the physical level stems from the long-range interactions of the LRESC: the codespace contains additional entanglement from the long-range interactions and simply rearranging this entanglement is enough to couple logical qubits. In the same example, we could have also taken the hypergraph product of the $[3(c),2,3(c)]$ code with itself to obtain a LRESC with four logical qubits. We would then have four logical gadgets comprised of simultaneous CNOTs between $(\bar{1}\bar{2}, \bar{3}\bar{4})$ and $(\bar{1}\bar{3}, \bar{2}\bar{4})$ with either left or right logical qubits as control or target. In general, we will only be able to perform simultaneous logical gates along ``rows" and ``columns" of logical qubits. Performing arbitrary two-qubit logical CNOTs in this setting may require the code-switching techniques of \cite{quintavalle2022}. The $[3,2,2]$ code is also the smallest code in the family of (shortened) Hadamard, or equivalently dual Hamming, codes with parameters $[2^k-1,k,2^{k-1}]$. This family of codes is equidistant: all codewords have weight $d=2^{k-1}$. As a consequence, the SWAP-CNOT gadget for the $[3,2,2]$ code generalizes to this entire code family, and any two-bit logical CNOT gate can be implemented by physical SWAPs. However, these codes are not LDPC and so may be difficult to implement in a fault-tolerant setting. In addition, in the corresponding hypergraph product codes, one also needs to track the action of the SWAPs on the parity-check matrices (detailed in Appendix \ref{app:logical gates}), which will in general not be invariant under these SWAPs. So some level of concatenation with repetition codes (thus creating LRESCs) may be necessary to maintain fault tolerance. Nonetheless, for small base code sizes, the Hadamard codes provide many new additional CNOT gadgets available to their corresponding LRESCs, which can reduce the need for code switching or lattice surgery. The SWAP-CNOT gadget is also very amenable for near-term experiments where the dominant source of error is from two-qubit entangling gates, as we will explain in Sec. \ref{sec:experiment}.

To achieve a universal gate set, the Clifford group needs to be supplemented with a non-Clifford gate, such as the $T$ gate ($\pi/8$ rotation). Since logical operators of HGP codes can be chosen to be perpendicular ``strips" intersecting on only one qubit (recall Fig. \ref{fig:logical operators}), we do not expect non-Clifford gates to be transversally implementable \cite{Bravyi_2013}. Nonetheless, as previously mentioned, one can teleport logical qubits onto ordinary surface codes \cite{xu2023constantoverhead}, from which magic state distillation \cite{knill2004, Bravyi_2005} can subsequently be applied, though \cite{bombin2015gauge, kubica, Brown_2020} provide alternatives.

\subsection{Weight balancing}\label{sec:weight bal}

\begin{figure}[t]
\centering
\includegraphics[width=0.48\textwidth]{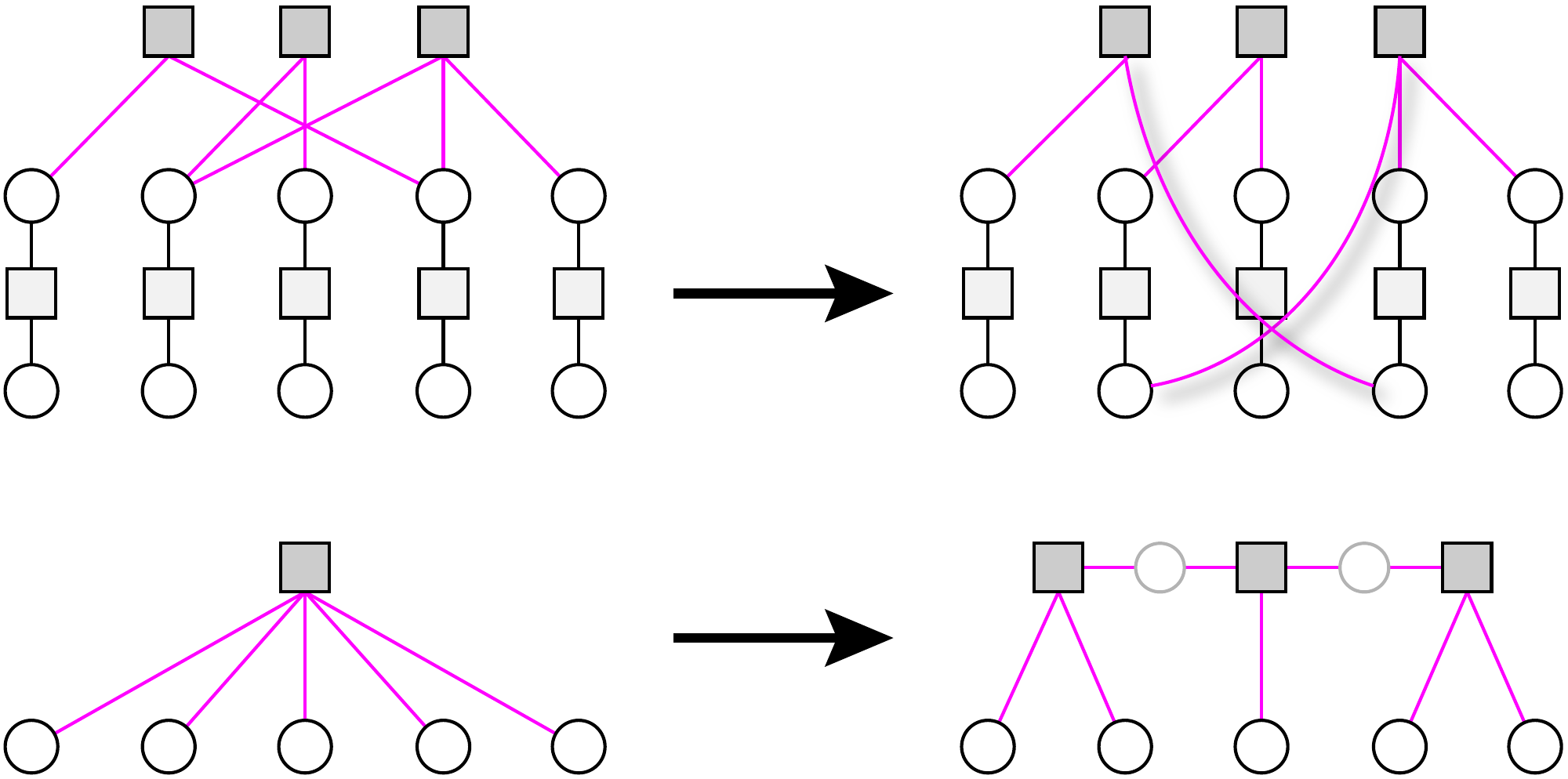}
\caption{The two different weight-balancing procedures are depicted. Top: a $[5(2),2,3(2)]$ code is modified so that all physical bits participate in at most one long-range parity check. Bottom: A weight-5 parity check is decomposed into three weight-3 checks with two additional auxiliary bits (gray circles).}
\label{fig:weight balancing}
\end{figure}

There are two simple, but practical enhancements to the LRESC described thus far by modifying the parent codes. In Step 2 of the LRESC construction, notice that each ``physical bit" of the cLDPC from Step 1 consists of a repetition code, but we assigned all of the ``long-range" parity checks to a single bit.  We can instead evenly distribute these parity checks to different bits inside of the repetition code: see Fig. \ref{fig:weight balancing} -- so long as $c$ is larger than the maximal number of parity checks per bit of the cLDPC (Step 1), this will mean that each physical bit is involved in at most one long-range parity check in Step 2.

The second modification is to introduce auxiliary bits into the parent codes in order to reduce the weight of each long-range parity check. The parity-check constraints of a classical code can be reformulated as a boolean satisfiability problem (SAT). It is well known in computer science that any SAT problem can be decomposed into conjunctions of smaller SATs of maximum size three (3-SAT), with the potential of introducing some auxiliary bits. Moreover, this SAT $\rightarrow$ 3-SAT decomposition can be performed in polynomial time \cite{SATto3SAT}; for our linear constraints, this decomposition takes a particularly simple form, see Fig. \ref{fig:weight balancing}. When we apply this decomposition to the parity checks of a classical code, we obtain new parity checks with bounded weights $\leq 3$ acting on the combination of our original physical bits and some new auxiliary bits. Importantly, the code distance remains unchanged, though the relative distance may decrease by an $O(1)$ factor if this method is applied to all checks. At the quantum level, this decomposition bears resemblance to a measurement-only version of Shor's cat-state syndrome extraction circuit \cite{shor1997faulttolerant}, where we have included the cat-state ancillas and measured operators as auxiliary qubits and new stabilizer checks respectively.

The modified parent codes will now have at most weight-3 parity checks with each physical bit participating in at most one long-range interaction. Furthermore, by arranging each long-range parity check to be adjacent to an endpoint of a repetition-code segment, we can always localize at least one of its long-range edges. In turn, the LRESCs will contain at most weight-6 stabilizer checks with each physical and ancilla qubit participating in at most 4 long-range interactions. A major caveat of the weight-balancing procedure is that we will lose the concatenated structure for logical gate inheritance described in Sec. \ref{sec:logical gates}. Nonetheless, if we simply desire to use an LRESC as a quantum memory block, then the two ``weight-balancing" procedures will be particularly advantageous for experimental implementations, as we will discuss in Sec. \ref{sec:experiment}.

\section{Experimental implementations} 
\label{sec:experiment}

Typically, experimental design of quantum hardware has been strongly limited by the choice of QEC code and its resulting requirements on circuit connectivity. LRESCs imply that one can exploit the tunable addition of nonlocality on top of the most local of codes, the surface code, once improvements in physical error rates, or increases in physical qubit number, have been exhausted. Such a theoretical advance offers a timely new tool for improving the performance of state-of-the-art platforms, including super-conducting qubits \cite{PhysRevLett.89.117901, PhysRevLett.79.2328, Krantz_2019, Brooks_2013, Gu_2017}, trapped-ions \cite{PhysRevLett.74.4091, Bruzewicz_2019, kihwan_2010, Britton_2012, Barreiro_2011} and neutral-atom arrays \cite{Saffman2010, Saffman2016, kaufman2021quantum, wu2022erasure, cong2022hardware, evered2023high,ma2023high} since, as we now explain, the specific type of nonlocality needed for the LRESC is relatively mild in multiple experimental platforms.  

In superconducting circuits, novel circuit graphs have simulated many-body physics in novel geometries \cite{Kollar_2019}.  To realize the LRESC, one must use multiple planes of wiring \cite{Rosenberg_2017}, and we expect that this construction is doable for modest values of $k$ (i.e. not encoding too many logical qubits). For devices with larger values of $k$, we can also employ fault-tolerant quantum repeater networks \cite{Fowler_2010, Rozpedek_2021} to teleport ancilla qubits down a strictly two-dimensional architecture.  The number of such quantum repeater rounds is constrained by the requirement that we cannot pass two logical qubits ``through each other".   This latter construction is quite similar to the ``hierarchical codes" recently discussed in \cite{pattison2023}. 

While a superconducting-qubit-based quantum computer may take advantage of LRESCs or hierarchical codes, we believe that the LRESC is significantly more optimized for architectures capable of non-local gates and reconfigurability, namely, trapped ions and neutral atom arrays. Using the race-track geometry, trapped-ion quantum computers have shown all-to-all connectivity with high fidelity gates and mid-circuit operations, including measurement and reset, which have been exploited for state-of-the-art demonstrations of quantum error correction~\cite{moses_race_2023}. Meanwhile, recent advances in neutral-atom quantum computers using reconfigurable tweezer arrays have enabled demonstrations of the circuit encodings for a variety of quantum error correction codes~\cite{bluvstein2022quantum, Bluvstein_2023}. Importantly, while both platforms fundamentally allow for all-to-all connectivity, the qubit moves required for this are up to a few orders of magnitude slower than other operations, such as single and two-qubit gates~\cite{moses_race_2023,bluvstein2022quantum, Bluvstein_2023}.
Accordingly, in order to develop optimally performing quantum computers --- constrained by hardware level limitations associated with qubit number, circuit encoding time, operational fidelities, etc. --- quantum error-correcting codes are desired that allow for flexible optimization within the space of these constraints. The LRESC construction allows for precisely this flexibility within these two platforms, by balancing physical qubit number against nonlocality and computation time in order to optimize encoding capacity. The proposed implementations below reflect this perspective.

\subsection{Trapped ions}

The Quantum Charge-Coupled Device (QCCD) approach \cite{kielpinski_architecture_2002} to quantum computation with trapped ions has the potential to realize LRESCs in the near future.
This architecture relies on a trap device capable of confining multiple one-dimensional arrays of ions. Within these so-called ``ion crystals", multi-qubit operations are achieved through laser- or microwave-induced spin-motion couplings. To facilitate interactions between ions initially residing in separate ion crystals, the architecture requires real-time shuttling, splitting, and merging operations of ion crystals that occur on faster timescales compared to the coherence time of the data qubits. This dynamic control over system connectivity is made possible through precise manipulation of electric fields that generate the trapping potentials.

The high operational fidelities (up to 99.9999\% single-qubit fidelity \cite{harty_high-fidelity_2014} and 99.94\% two-qubit fidelity \cite{clark_high-fidelity_2021}) allowed fault-tolerant demonstrations of quantum error-correcting codes encoding a single logical qubit in small-scale quantum processors \cite{ryan-anderson_implementing_2022, postler_demonstration_2022}.
As systems with 100s of controllable qubits become available in the near future it will be feasible to incorporate LRESCs in the QCCD architecture.
In particular, if state-of-the-art fidelities can be maintained for a large-scale device, then LRESC 2 from Fig. \ref{fig:HGP vs surface} significantly surpasses the break-even point with 244 physical qubits, and a similar number of ancilla qubits, assuming two-qubit fidelity from \cite{clark_high-fidelity_2021}.

\begin{figure}[t]
\centering
\includegraphics[width=0.47\textwidth]{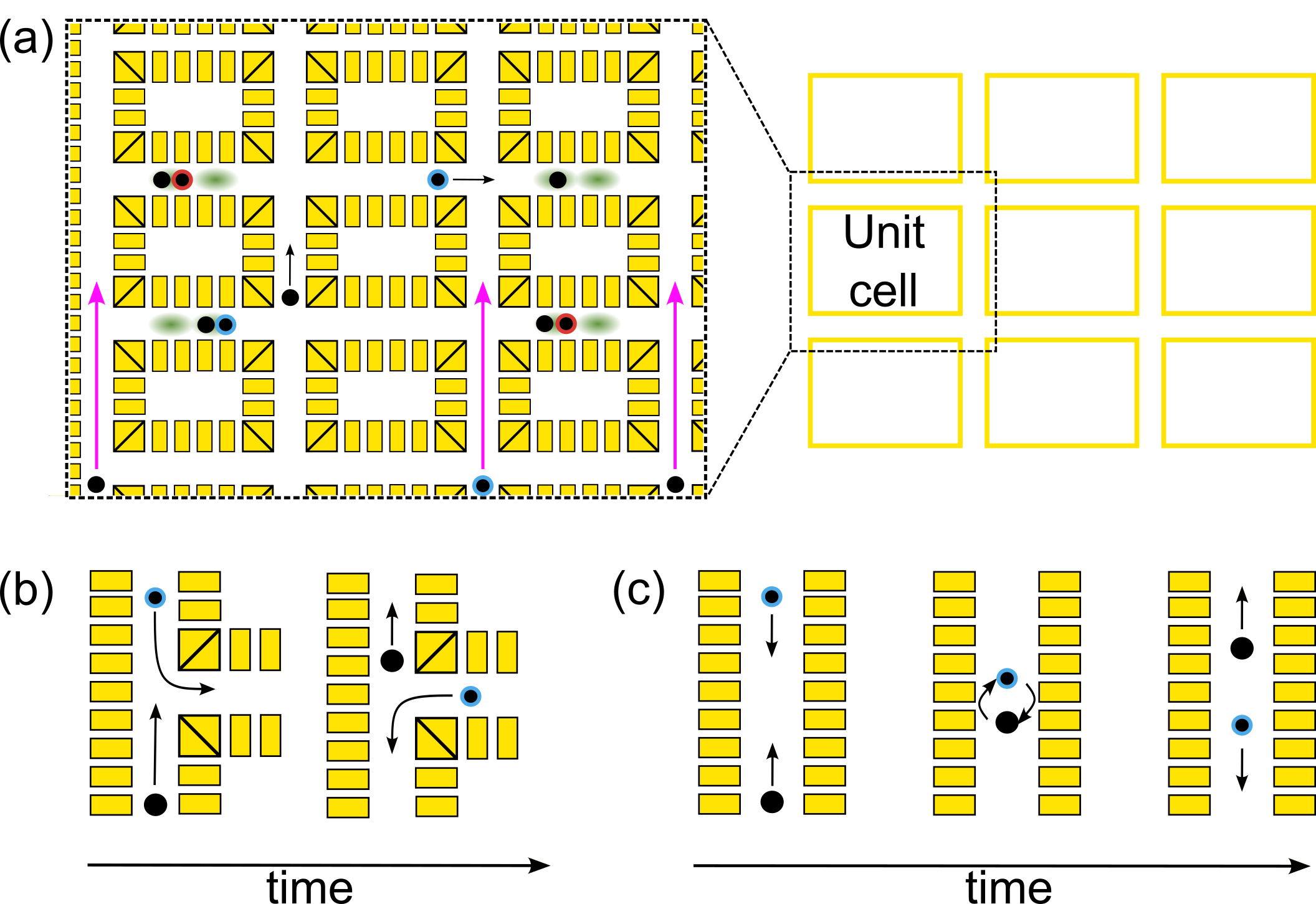}
\caption{LRESC implementation using a trapped-ion Quantum Charge-Coupled Device architecture. (a) A possible quantum processor is structured into multiple unit cells, each representing a surface code tile (yellow tile in Fig. \ref{fig:HGP construction detailed}). Each unit cell contains the necessary data qubits (black dots) and measure qubits for parity checks (blue and red rims). Ions are transported across different interaction regions (green-shaded areas) by precise control of the voltage applied to the trap electrodes (yellow boxes) to perform the necessary local operations. Each cell also contains additional ancilla qubits (not shown) used for re-cooling operations after transport operations. Sparse nonlocal operations required by LRESC are performed via long-range transport of qubits across different cells (pink arrows). (b)-(c) two possible ways to perform qubit permutation within the QCCD architecture. }
\label{fig:ion_proposal}
\end{figure}

A possible implementation of LRESCs with trapped ions is shown in Fig. \ref{fig:ion_proposal}. 
The envisioned architecture is structured into multiple unit cells, each representing a surface code tile (yellow tile in Fig. \ref{fig:HGP construction detailed}). Within each unit cell, multiple interaction regions are designed to facilitate parallel single- and two-qubit gates. Every unit cell contains both the data qubits necessary for surface code operations and the necessary ancilla ions. Data qubits are transported between interaction regions to perform the necessary two-qubit gates. During transport, unwanted motional excitations may arise due to imperfect control of applied fields. To maintain high two-qubit fidelities, ancilla ions are then used to sympathetic re-cool an ion crystal following a transport operation. 
While qubits primarily move within a unit cell, an LRESC requires sparse long-range operations. 
The QCCD architecture in Fig. \ref{fig:ion_proposal} efficiently facilitates the parallel transport of multiple ions to different unit cells, thus minimizing the time associated with nonlocal operations.

The main complexity lies in the optimal scheduling of gates and transport operations, as well as the  delivery of the laser light used to coherently control the qubits in each interaction region.
For trapped ions, scalable laser light delivery based on free-space optics can be challenging due to the need for tightly focused beams for single-qubit addressing and the presence of nearby trap electrodes. The complexity is further increased if multiple wavelengths of light are needed in each region. A promising approach to address the issue of light delivery is using optical waveguides integrated into the structure of the ion trap. This approach allows compact routing of light to the various interaction regions and the generation of tightly focused beams using grating couplers \cite{dalgoutte_thin_1975, mehta_integrated_2016}. Ion traps with integrated waveguides have been successfully employed in devices controlling a single region \cite{mehta_integrated_2016, niffenegger_integrated_2020, mehta_integrated_2020} as well as multiple ones \cite{kwon_multi-site_2023,mordini_multi-zone_2024}. Furthermore, experiments demonstrated the use of integrated waveguides with multiple wavelengths ranging from violet to infrared \cite{niffenegger_integrated_2020} as well as schemes for performing all qubit control with longer wavelengths of light \cite{lindenfelser_cooling_2017, hendricks_doppler_2008}, thus simplifying the waveguide requirements.
To mitigate the challenges associated with ion transport and the scheduling of gate operation, a possible solution is a system that allows the manipulation of ion crystals with more than two data qubits. Such a system not only would reduce scheduling complexity but is also likely to reduce the unit cell's size, as fewer interaction regions are needed. Consequently, it would also decrease the overall execution time, since the time required to transport and re-cool a crystal is generally longer than those of two-qubit gates in medium size ion crystals \cite{pino_demonstration_2021, moses_race_2023}.
However, working with large ion crystals can add extra control challenges. State-of-the-art two-qubit gates between ions in long ion chains are generally slower than gates on two-qubit ion crystals and also yield lower fidelities \cite{landsman_two-qubit_2019, wang_high-fidelity_2020}. Furthermore, multi-qubit gates mediated by normal modes of motion cannot be easily executed in parallel. 
Therefore, we speculate that a likely optimal architecture that implements LRESCs will compromise the advantages offered by the QCCD architecture and those offered by the manipulation of medium-size ion chains.

Depending on the details of the experimental apparatus (i.e. physical size of the quantum processor, qubit coherence time, maximum achievable transport speed and re-cooling times), long-range transport may cause an increased physical error rate due to the finite qubit coherence time and the longer time required for long-range ion shuttling. To mitigate this issue teleportation of the qubit state can be employed to replace long-distance transport. This approach requires generating entangled Bell pairs between two distant regions of the quantum processor using schemes for remote entanglement generation \cite{monroe_large-scale_2014, stephenson_2020, Krutyanskiy_2019}. This scheme would also be compatible with a modular ion-trap architecture \cite{monroe_large-scale_2014} composed of multiple interlinked small devices each with a limited number of qubits and correspondingly little computational power \cite{monroe_large-scale_2014, campbell_distributed_2007}.

\subsection{Neutral atom arrays}

Reconfigurable atom arrays manipulated with optical tweezers are also well-suited to reap the benefits of LRESC~\cite{Saffman2016,Browaeys2020,kaufman2021quantum,bluvstein2022quantum, Bluvstein_2023}. In particular, scaling to 100s of controllable qubits has already been demonstrated~\cite{scholl2021quantum, ebadi2021quantum, young2023atomic}, while scaling to 1000s is a near-term prospect~\cite{tao2023high}; two-qubit gate fidelities of $>98.5\%$ have been shown in multiple atomic species, with the state-of-the-art performance at $99.5\%$~\cite{ma2023high,evered2023high}. Accordingly, this platform lies within an order of magnitude of the break-even point of an LRESC (see Fig. \ref{fig:HGP vs surface}). As important, the optical methods used for atomic reconfigurability enable parallelism that is well-suited to the surface code and LRESCs~\cite{bluvstein2022quantum,wu2022erasure}. 

\begin{figure}[t]
\centering
\includegraphics[width=0.5\textwidth]{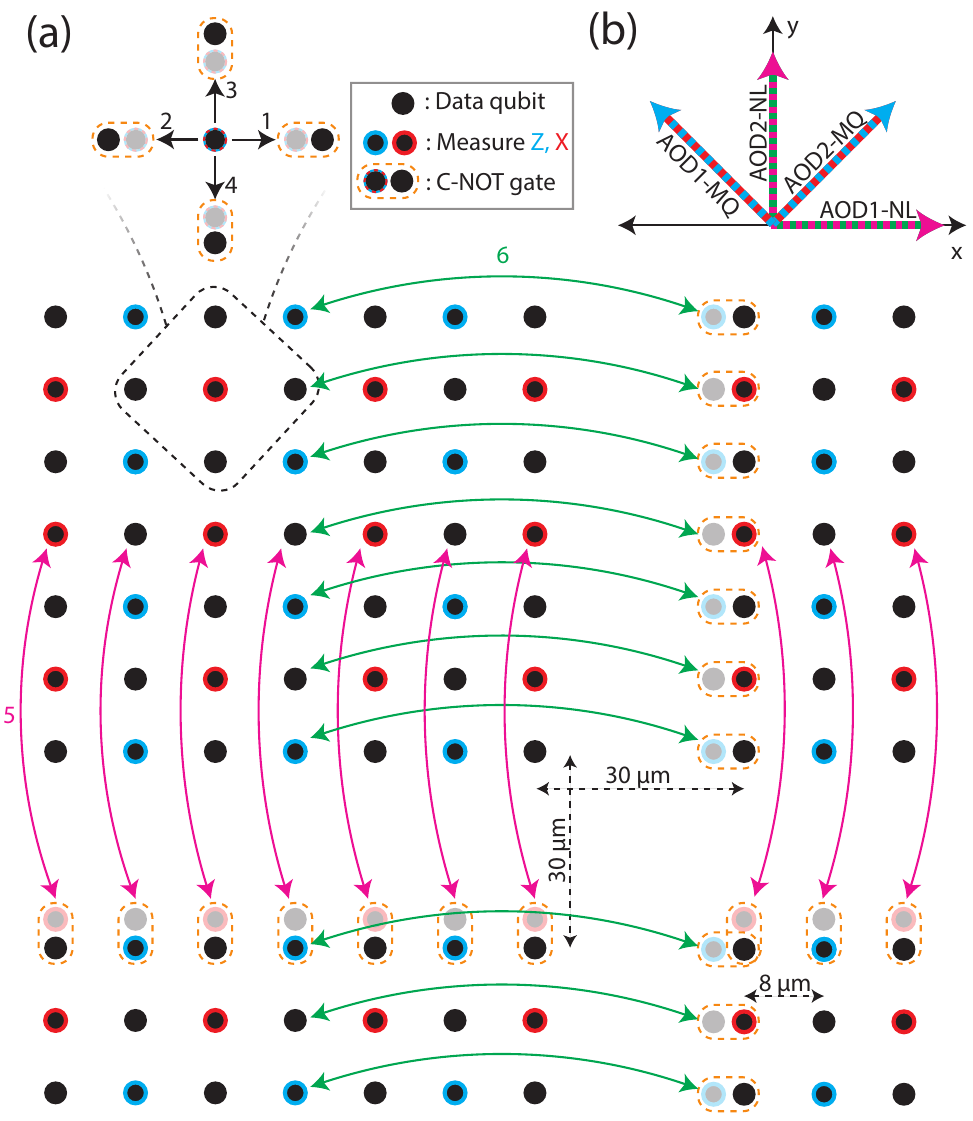}
\caption{LRESC implementation on a neutral-atom-based processor using Rydberg-mediated interactions. (a) Data qubits (black circles) and measure qubits (blue and red rims) are initialized in a static ordered 2D array generated by a spatial light modulator or optical lattice. Local parity checks are performed with sequential two-qubit gates (orange-dashed lines) performed on all measure/data qubits in parallel, where each measure qubit is transported in close proximity with a neighbor qubit (step 1-4) using fast crossed acousto-optic deflectors (AODs). Another pair of crossed AODs is used to perform nonlocal operations by transporting data and measure qubits between different locations of the quantum processor. (b) Two different pairs of crossed AODs are used for short-range and long-range atom transport, respectively labeled as MQ and NL. Arrows represent the transport direction for a varying radio-frequency offset in each AOD.}
\label{fig:atoms}
\end{figure}

Fig.~\ref{fig:atoms} illustrates a possible implementation of an LRESC using atom arrays. A static array --- formed with a spatial light modulator or optical lattice~\cite{scholl2021quantum,ebadi2021quantum, bluvstein2022quantum, eckner2023realizing, young2023atomic} --- holds atomic data qubits. The measure qubits that yield $X$ and $Z$ parity checks (red and blue rims) sit on a grid of traps rotated 45 degrees from the $x$/$y$ axes. This array of traps is formed with crossed acoustic-optic deflectors (AOD1-MQ, AOD2-MQ in Fig.~\ref{fig:atoms}b) driven with a comb of radio-frequencies. This entire array can be moved by adding an overall offset frequency to the comb of tones inside each deflector, allowing any rigid array translation in the $x$-$y$ plane. Such moves are used to bring all measure qubits into proximity with the appropriate neighbor, in order to exploit short-range Rydberg-mediated interactions for a two-qubit gate (orange-dashed lines) for parallelized two-qubit gates~\cite{bluvstein2022quantum, wu2022erasure}. Due to the short distance scales and the use of AODs, each stepwise move of the SC (top of Fig.~\ref{fig:atoms}) can be executed in  $\lesssim 10\mu$s. 

The nonlocal gates that underlie LRESCs likewise can be implemented in a straight-forward fashion, with one adjustment. A pair of crossed AODs (Fig.~\ref{fig:atoms}b), AOD1-NL and AOD2-NL, can be used for row translations along the $y$ direction (step 5 in Fig.~\ref{fig:atoms}a), as well as column translations along the $x$ direction (step 6). For the nonlocal gates, both measure and data qubits are moved, which necessitates qubit transfers between different optical potentials. Such methods have been demonstrated and can be done while preserving  coherence~\cite{trotzky2008time,kaufman2015entangling, Bluvstein_2023}, yet come at the price of longer timescales ($\sim100\mu \textrm{s}$) to mitigate motional heating. 

In addition to allowing the core components of LRESCs, the atom array platform is compatible with other more general needs of QEC. Initialization of the qubit array into the set of optical potentials discussed can be accomplished with atomic rearrangement~\cite{barredo2016atom, endres2016atom, eckner2023realizing}. Parity checks on the measure qubits will require mid-circuit readout and reset. This can be done in-situ using qubit shelving methods, as recently demonstrated for $^{171}$Yb or by using mixed atomic species~\cite{singh2023mid, lis2023mid, norcia2023mid} --- this circumvents the need for large moves and zoned read-out~\cite{pino_demonstration_2021,bluvstein2022quantum}. Lossless state detection of neutral atoms can be slow (at best, a few milliseconds~\cite{lis2023mid}); this timescale can be improved using destructive state detection ~\cite{ma2023high,scholl2023erasure, Bluvstein_2023}, that is then paired with a qubit reservoir for rapid replenishment~\cite{pause2023reservoir,norcia2023mid}. High fidelity single- and two-qubit gates can be accomplished at low cross-talk with the qubit separations illustrated, using a combination of tightly-focused and laser beams and globally-addressing fields~\cite{xia2015randomized, wang2016single,sheng2018high, jenkins2022ytterbium,ma2022universal,lis2023mid, evered2023high, ma2023high}. Qubit loss --- a prevalent error channel during two-qubit gates and measurement --- can be mitigated using syndrome extraction circuits and three-outcome measurements~\cite{suchara2015leakage}.

Finally, the weight-balancing procedures described earlier (Sec.~\ref{sec:weight bal}), which allow for reducing the number of qubits per check (and checks per qubit), are relevant for the implementation of LRESCs in atom arrays. So long as each qubit participates in at most 4 long-range interactions, a single physical qubit will be involved in at most 8 rounds of row and column swaps -- 4 local and 4 nonlocal -- to couple all corresponding physical and ancilla qubits during syndrome extraction for one round of QEC.  Using a single AOD each for long-range row and column permutations, this may require $O(\sqrt{k})$ sequential swaps.  These swaps could be further parallelized by carefully arranging the long-range edges, or by adding additional AODs, though we leave further optimization for future work.


\section{Outlook}
We have described the LRESC: a minimal generalization of the surface code capable of encoding multiple logical qubits without sacrificing code  distance. We show how long-range interactions can (i) improve code overhead, (ii) improve code performance and (iii) enable new kinds of fault-tolerant gadgets. The LRESC is well-suited for near-term hardware, where we anticipate that our fault-tolerant code might be realizable within the next few years. 

An immediate direction for future work is to design a better decoder for LRESCs. Depending on qubit shuttling times, a more sophisticated two-stage decoder could be designed as follows. (1) Perform multiple rounds of local syndrome measurements in the surface code patches while waiting for the long-range syndrome measurements to complete \cite{Berthusen:2023lpm}. (2) Use one's favorite standard decoder (e.g. MWPM \cite{Dennis_2002} or Union-Find \cite{Delfosse_2021}) for the multi-round syndromes within the surface-code patches and feed the output decisions into a single-stage BP+OSD decoder for global decoding. In this manner, one can strike a balance between the ``fast" (but less robust) checks of the surface code and the ``slower" (but more robust) long-range checks. Another related avenue is the construction of additional fault-tolerant gadgets, taking advantage of specially structured base codes. While the more compact encoding of a LRESC facilitates the implementation of certain multi-qubit logical gates, single-qubit logical gates become harder for the same reason. In an LRESC, the support of logical operators overlap, and it is difficult to only manipulate one logical qubit without affecting others.

The construction of the LRESC also opens possible avenues to investigate new quantum phases of matter.  In particular, it suggests new ``topological phases" are enabled using only a small density of long-range interactions, and can thus be investigated in experiment. In the longer term, a large-scale LRESC may also be the foundation for an autonomous self-correcting quantum memory. Indeed, our proposed architecture may well represent a more convenient strategy for passive error correction versus a four-dimensional toric code \cite{Dennis_2002}.  It may also be more amenable to single-shot error correction than three-dimensional single-shot codes \cite{Bombin_2015, Kubica_2022, stahl2023} due to its lower overhead.



\section*{Acknowledgements} We thank Evan Wickenden and Charles Stahl for useful discussions, and Jeff Thompson for a careful reading of the manuscript. This work was supported by the Office of Naval Research via Grant N00014-23-1-2533 (YH, AMK, AL), the Alfred P. Sloan Foundation via Grant FG-2020-13795 (AL), NIST (AMK) and the Swiss National Science Foundation under grant 211072 (MM).

\begin{appendix}
\renewcommand{\thesubsection}{\thesection.\arabic{subsection}}

\section{Classical LDPC codes}\label{app:cLDPC}

\begin{figure}[t]
\centering
\includegraphics[width=0.45\textwidth]{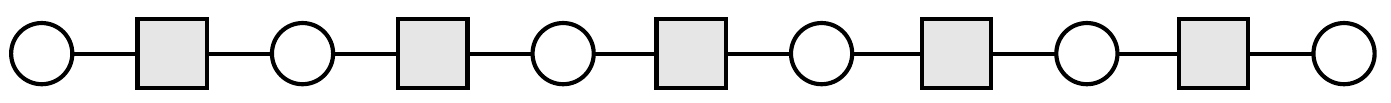}
\caption{The Tanner graph of a $n=6$ repetition code is illustrated. The circles and squares represent physical bits and parity checks, respectively.}
\label{fig:repetition tanner graph}
\end{figure}

We begin by reviewing classical low-density parity-check (cLDPC) codes \cite{gallager_1962}, which play an important role in our construction.  A classical linear code $\mathcal{C}$ is specified by a set of constraints called parity checks and a set of codeword generators satisfying those constraints. The state of the system can be represented as an element of $\mathbb{F}^{n}_2$, where $\mathbb{F}^{\vps}_2=\lbrace0,1\rbrace$, and in $\mathbb{F}_2$, $1+1=0$. We often represent the parity checks as rows of an $\mathbb{F}_2$-valued \emph{parity-check matrix} $H$ and  logical codewords as rows of a matrix $G$. The statement that the codewords satisfy the parity-check constraints becomes $H G^\transpose = 0$. The dual code $\mathcal{C}^\perp$ is defined as the code where $G$ and $H$ are swapped. We say a linear code is LDPC if its parity-check matrix $H$ is sparse: the number of ones per row and column are bounded by a constant irrespective of $n$. The code is useful if $G$ is not sparse: the code distance $d$ is the smallest number of 1s in a codeword. We can represent any linear code as a bipartite \emph{Tanner graph}, drawing an edge between a ``variable node" $v$ and a ``check node" $c$ if the corresponding element of $H$ is non-zero: $H_{cv}=1$. The Tanner graph of a repetition code is depicted in Fig. \ref{fig:repetition tanner graph}. All linear codes satisfy the \emph{Singleton bound}:
\begin{align}\label{eq:Singleton bound}
    k \leq n - \abs{C}
\end{align}
where $C$ is a correctable region satisfying $\abs{C} \geq d - 1$. Correctable here means that all codewords can be successfully recovered upon erasure of $C$.

\begin{figure}[t]
\centering
\includegraphics[width=0.49\textwidth]{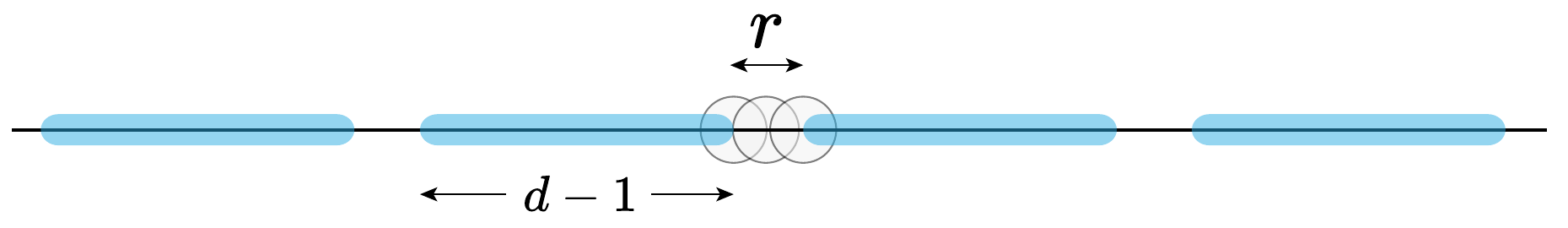}
\caption{A 1D chain is partitioned into disconnected, correctable regions (blue) of length $\approx d$ with separation $\approx r$ such all checks (circles) have support in at most one region.}
\label{fig:1d chain partition}
\end{figure}

A code generated by a random sparse $H$ has both $k = \Theta(n)$ and $d = \Theta(n)$ with high probability \cite{random_ldpc}; its corresponding Tanner graph is an asymptotically good expander \cite{Richardson2008}. However, if we arrange the variable nodes locally in one dimension, such a code will necessarily involve checks $c$ that are nonlocal. If we enforce geometric locality in $D$-dimensional Euclidean space, then the code parameters must satisfy \cite{BPT_2010}
\begin{align}\label{eq:classical BPT}
    k d^{1/D} = O(n) \, .
\end{align}
The sketch of the proof in $D=1$ is as follows. The idea is to partition the 1D chain into disjoint, correctable regions $C^{\vps}_i$ of length $\abs{C^{\vps}_i} \approx d$ where the separation between each region is large enough (say $r$) so that no parity check acts in more than one region: see Fig. \ref{fig:1d chain partition}. Since all the correctable regions do not share any checks, their union is entirely correctable \cite{Bravyi_2009}. The Singleton bound \eqref{eq:Singleton bound} then imposes that $k \leq n - \abs{C} = \abs{\bar{C}} = O(rn/d)$, and we thus arrive at \eqref{eq:classical BPT} for $r = O(1)$ and $D=1$. Now suppose we add in $\ell$ long-range connections to surpass \eqref{eq:classical BPT}. We can simply avoid the long-range edges and partition the rest of the chain as before, arriving at $\abs{C} \rightarrow \abs{C} - O(\ell)$ and thus $k \rightarrow k + O(\ell)$. Hence, the number of logical bits $k$ can scale at most linearly with the number of long-range connections $\ell$.

We now saturate the asymptotic constraints above with a cLDPC code of $d = \Theta(n)$, $k$ logical bits, and $\Theta(k)$ long-range checks.  A $[n'c,k',d'c]$ code is produced from the concatenation of an ``outer" $[n',k',d']$ code with an ``inner" $[c,1,c]$ repetition code of variable length $c$ (denoted $[n'(c),k',d'(c)]$): see Step 1 of Fig. \ref{fig:HGP construction detailed}. Concatenation means that we connect a single bit of each inner repetition code to the parity checks of the outer $[n',k',d']$ code. This concatenating procedure can also be interpreted as first cutting up a 1D repetition code into disconnected segments and then reconnecting these segments with long-range interactions. The only long-range checks come from the outer code, and if it is a ``good" $[n',k',d'] = [\Theta(k'),k',\Theta(k')]$ cLDPC code, the concatenated code has parameters $[\Theta(ck'),k',\Theta(ck')]$ with $\Theta(k')$ long-range connections, which is parametrically optimal. Since we are allowed to attach the long-range edges to \emph{any} bits of the inner repetition codes, we have some flexibility in designing the long-range couplings (recall Sec. \ref{sec:weight bal}). This concatenation procedure can be considered as a ``dual" variant to the edge-augmentation construction of \cite{Roffe_2020}: instead of having the repetition codes live on the edges of a cLDPC code, we attach them to the variable nodes themselves. For a cLDPC with average vertex degree $\bar{w}$, the concatenated construction reduces the number of surface-code patches by a factor of $\bar{w}^2$ compared to the approach in \cite{Roffe_2020}. As we will later see, the ``hierarchical" structure of concatenated codes also lends the dynamics to be factorized in a systematic manner: we can analyze the dynamics within the inner and outer codes separately.


\section{Hypergraph product codes}\label{app:HGP}

\begin{figure*}[t]
\centering
\includegraphics[width=0.8\textwidth]{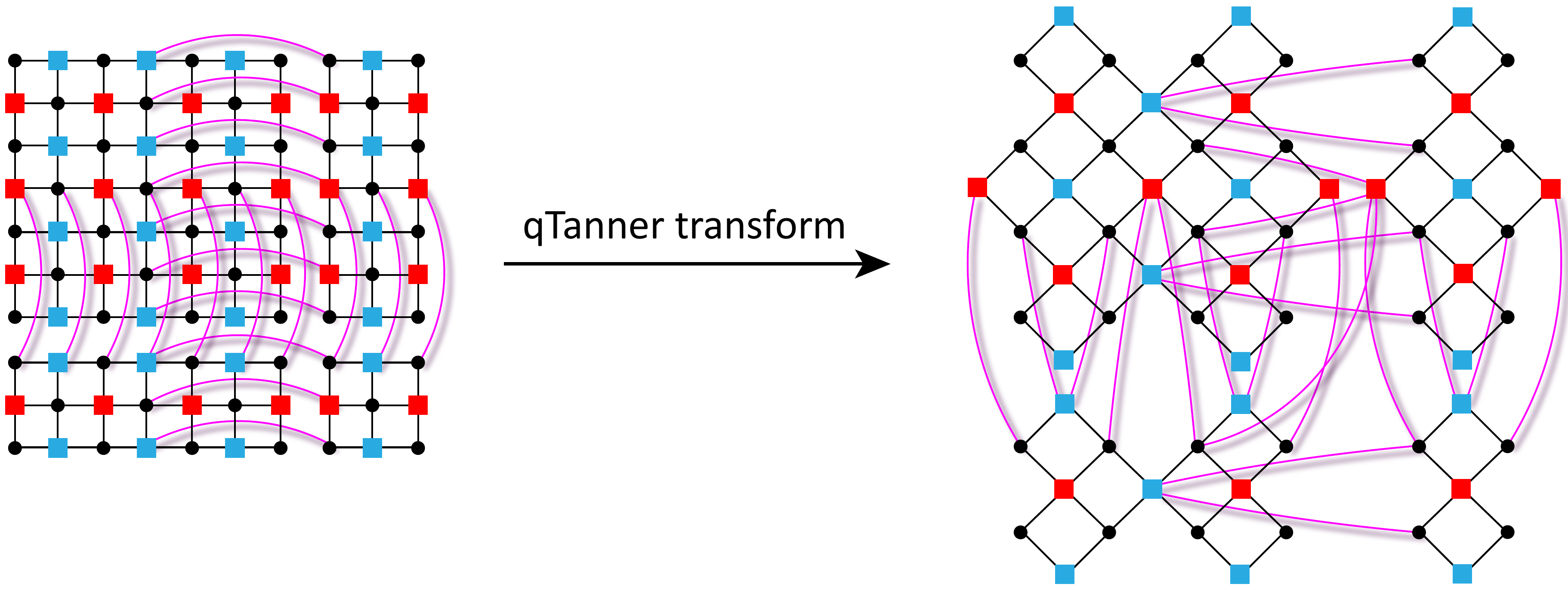}
\caption{The transformation of a $\llbracket 52,4,4 \rrbracket$ HGP code into a $\llbracket 36,4,4 \rrbracket$ quantum Tanner code is shown. Solid black dots represent physical qubits, and red (blue) squares represent $X$ ($Z$) checks. 21 long-range interactions (magenta curves) are required. A $k=4$, $d=4$ surface code of the same layout will require $n \geq 64$ physical qubits.}
\label{fig:[3(2),2,2(2)] qTanner code}
\end{figure*}

Using an $\mathbb{F}^{2n}_2$ representation for Paulis, the stabilizer checks of a CSS code can be represented by the parity-check matrix
\begin{align}
    H = \begin{pmatrix}
        H^{\vps}_X & 0 \\
        0 & H^{\vps}_Z
    \end{pmatrix}
\end{align}
where commutativity requires $H^{\vps}_Z H^{\transpose}_X = 0$. We use the hypergraph product (HGP) \cite{HGP} to construct a quantum CSS code from two classical linear codes. Specifically, suppose we have two classical codes with parameters $[n^{\vps}_1, k^{\vps}_1, d^{\vps}_1]$, $[n^{\vps}_2, k^{\vps}_2, d^{\vps}_2]$ and parity-check matrices $H^{\vps}_1$, $H^{\vps}_2$ with $m_1$, $m_2$ rows respectively. The associated HGP code has parity-check matrices defined as
\begin{subequations}
\begin{align}
    H^{\vps}_X &= \big(\, H^{\vps}_1 \otimes \ident_{n_2} \;\;|\;\; \ident_{m_1} \otimes H^\transpose_2 \,\big) \\
    H^{\vps}_Z &= \big(\, \ident_{n_1} \otimes H^{\vps}_2 \;\;|\;\; H^\transpose_1 \otimes \ident_{m_2} \,\big) \, .
\end{align}
\end{subequations}
By construction, the orthogonality constraint $H^{\vps}_Z H^{\transpose}_X = 2(H^{\vps}_1 \otimes H^\transpose_2) = 0$ is automatically satisfied. The $\llbracket N,K,D \rrbracket$ parameters of the HGP code are given by
\begin{align}\label{eq:HGP params generic}
\begin{aligned}
    N &= n_1 n_2 + m_1 m_2 \\
    K &= k^{\vps}_1 k^{\vps}_2 + k^\transpose_1 k^\transpose_2 \\
    D &= \min \left( d^{\vps}_1, d^{\vps}_2, d^\transpose_1, d^\transpose_2 \right) \, ,
\end{aligned}
\end{align}
where $k^\transpose$ and $d^\transpose$ are the usual $k$ and $d$ for the transpose code. For the rest of the appendix, we will use lower-case letters for classical code parameters and upper-case letters for quantum code parameters. If $H_1$ and $H_2$ have full rank (no redundant parity checks), then their transpose codes are trivial, and the above HGP code parameters \eqref{eq:HGP params generic} simplify to
\begin{align}\label{eq:HGP params full rank}
\begin{aligned}
    N &= n_1 n_2 + m_1 m_2 \\
    K &= k^{\vps}_1 k^{\vps}_2 \\
    D &= \min \left( d^{\vps}_1, d^{\vps}_2 \right) \, .
\end{aligned}
\end{align}
Geometrically, the Tanner graph of the HGP code takes the form of a Cartesian graph product between those of the two classical parent codes. Given two graphs $\mathcal{G}^{\vps}_1 = (V^{\vps}_1, E^{\vps}_1)$ and $\mathcal{G}^{\vps}_2 = (V^{\vps}_2, E^{\vps}_2)$, the product graph $\mathcal{G}^{\vps}_1 \times \mathcal{G}^{\vps}_2$ is a graph with vertices labeled by pairs $(x,y)$ where $x \in V^{\vps}_1$ and $y \in V^{\vps}_2$. Two vertices $(x,y)$, $(x',y')$ are connected by an edge if either $x=x'$ and $\{y,y'\} \in E^{\vps}_2$ or $y=y'$ and $\{x,x'\} \in E^{\vps}_1$. The steps to convert this product graph into a CSS Tanner graph are:
\begin{enumerate}
    \item If the vertex of the product graph is of the form (node, node) or (factor, factor), then that vertex becomes a node representing a physical qubit.
    \item If the vertex of the product graph is of the form (node, factor), then that vertex becomes a factor representing an $X$ stabilizer.
    \item If the vertex of the product graph is of the form (factor, node), then that vertex becomes a factor representing a $Z$ stabilizer.
\end{enumerate}
Importantly, if the two parent codes are LDPC, then so is the resultant HGP code. If the two parent codes can be locally embedded in $D^{\vps}_1$ and $D^{\vps}_2$ spatial dimensions, then the HGP code can in $D^{\vps}_1 + D^{\vps}_2$ dimensions. The surface code is the HGP of two 1D repetition codes.

The LRESC is simply the HGP of the classical concatenated code defined earlier with itself. Its parameters are $\llbracket \Theta(c^2k'^2), k'^2, \Theta(ck') \rrbracket$ with $\Theta(ck'^2)$ long-range interactions. Denoting $L \equiv ck'$ and $K \equiv k'^2$, the code parameters simplify as $\llbracket \Theta(L^2), K, \Theta(L) \rrbracket$ with $\Theta\left( L\sqrt{K} \right) $ long-range interactions.
For $K \ll N$, the 2D layout of this HGP code can be understood as patches of surface code of length $c$, whose boundaries are connected by long-range stabilizers: see Fig. \ref{fig:HGP construction detailed}. The graph product structure arranges these long-range interactions as parallel row and column couplings.

In the surface code, the quantum Tanner transformation \cite{leverrier2022efficient, qTanner_rotated_surface} can reduce $N = D^2 + (D-1)^2$ to $N = D^2$ while maintaining the same distance, producing the so-called ``rotated surface code" with parameters $\llbracket D^2,1,D \rrbracket$. Examining \ref{fig:HGP construction detailed}, we see that the qubits of the HGP code can be partitioned into two sublattices corresponding to node-node (primary) and check-check (secondary) vertices of the graph product. The idea of the quantum Tanner transform is to multiply adjacent checks of the same type in order to produce new checks which commute when restricted to the primary sublattice; the secondary sublattice can then be discarded: see Fig. \ref{fig:[3(2),2,2(2)] qTanner code}. Applied to a HGP code, the transformation will reduce $N = n_1n_2 + m_1m_2$ to $N = n_1n_2$. The Tanner transform of a LRESC will unfortunately introduce a ``diagonal" interaction for every long-range 4-cycle in the original Tanner graph. If the parent code has $O(k')$ long-range edges, then the HGP code will contain $O(k'^2) = O(K)$ additional ``diagonal" interactions in a 2D layout. For small platforms, the factor of $\approx\,$2 reduction in overhead may still be advantageous despite the increase in the routing complexity needed to implement the long-range checks. In practice, one will also need to worry about decreasing the effective code distance under circuit-level noise. For traditional HGP codes, it has been shown that the usual methods for syndrome extraction maintain the code distance \cite{manes2023}. After performing a quantum Tanner transformation, a specially engineered syndrome extraction circuit may be required to maintain this effective code distance (c.f. surface code vs rotated surface code syndrome extraction circuits).


\begin{figure*}[t]
\centering
\includegraphics[width=0.85\textwidth]{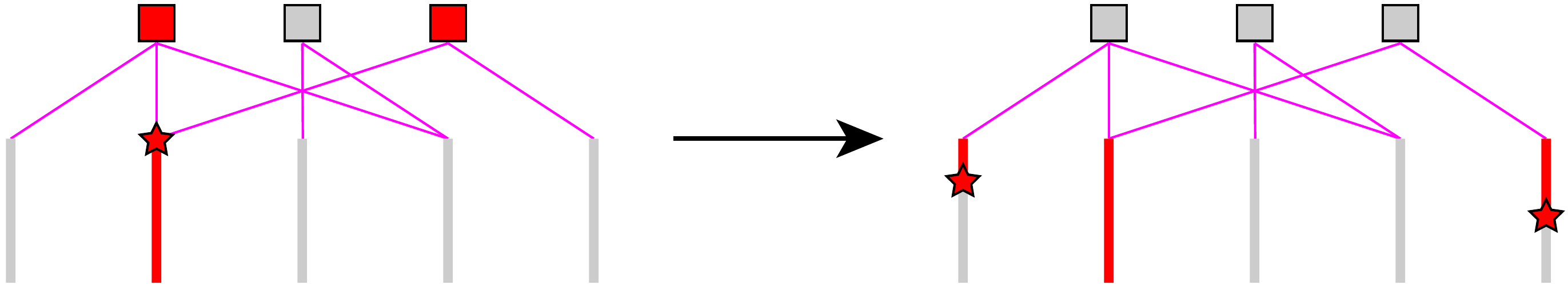}
\caption{The classical dynamics of long-range boundaries is depicted for a $[5(c),2,3(c)]$ concatenated code. A domain-wall excitation (star) is created in an inner repetition code and is transported across to the long-range boundaries (magenta lines). Left: Upon reaching this boundary, the long-range checks of the outer code will be violated (red squares). Right: Additional $d'-1=2$ excitations must appear amongst the other connected surface-code patches in order to complete a logical operation (codeword 11001).}
\label{fig:domain wall transport}
\end{figure*}

\section{Anyons, long-range boundaries and confinement}\label{app:anyon}

In this section, we characterize the structure of logical operators in LRESCs using concepts from condensed-matter physics. We show how anyon transport properties in the LRESC are related to domain-wall dynamics in the classical parent codes. Finally, we describe how the long-range boundaries in an LRESC can lead to anyon confinement and improved single-shot decoding.

\subsection{Logical operators and boundary dynamics}\label{app:boundary dynamics}

Interpreting the parity checks of the 1D repetition code (Fig. \ref{fig:repetition tanner graph}) as energetic terms in a Hamiltonian, we arrive at the 1D Ising model. A local bit flip in the 1D Ising model creates a pair of domain walls separating 1s and 0s. When these domain walls move via additional bit flips, the number of violated checks remains constant, and we say that the domain walls are ``deconfined". These domain walls can then travel to opposite endpoints of the chain, flipping all physical bits in the process. Thus, a logical error in the 1D repetition code can be enacted with local processes while violating only $O(1)$ checks.

The concatenated codes mentioned earlier consist of a $[n',k',d']$ base code and an inner repetition code. We now describe how the structure of the base code dictates the dynamics of propagating domain walls. As a concrete example, suppose our base code was a $[5,2,3]$ code with
\begin{align}\label{eq:[5,2,3] matrices}
    H = \begin{pmatrix}
        1 & 1 & 0 & 1 & 0 \\
        0 & 1 & 0 & 0 & 1 \\
        0 & 0 & 1 & 1 & 0
    \end{pmatrix} \quad,\quad
    G = \left(\begin{array}{cc|ccc}
        1 & 0\, & 1 & 1 & 0 \\
        0 & 1\, & 1 & 1 & 1
    \end{array}\right)
\end{align}
where we have expressed $G$ in reduced row echelon (standard) form.
Domain walls can freely propagate within each inner repetition code. However, upon hitting the long-range boundaries, these domain walls will excite the long-range checks of the base code: see Fig. \ref{fig:domain wall transport}. Satisfying the long-range checks requires locally exciting domain walls on other repetition code segments according to the codewords generated by $G$. When a domain wall reaches a long-range check, we examine the codewords which contain a 1 at the position of its corresponding repetition-code segment. The other 1s in the codeword label the other segments which can spawn the additional domain walls, thereby satisfying all long-range parity checks. The minimum number of additional domain walls is $d'-1$, where $d'$ is the distance of the outer code. By using a good cLDPC code as the base code, expansion properties guarantee that this domain-wall splitting scales with the size of the base code.

\begin{figure*}[t]
\centering
\includegraphics[width=0.85\textwidth]{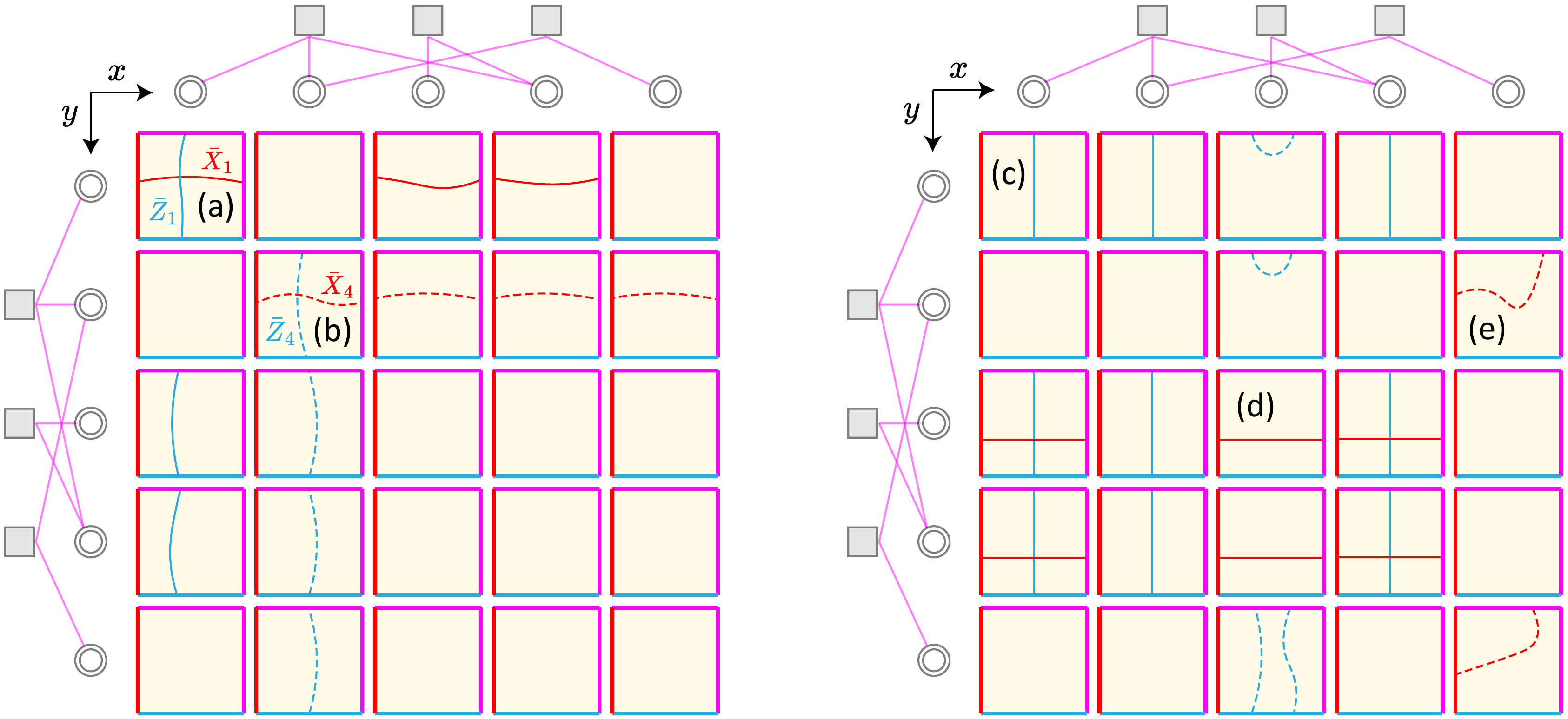}
\caption{Some logical (left diagram) and stabilizer (right diagram) operators are depicted for a LRESC with a parent $[5,2,3]$ outer code \eqref{eq:[5,2,3] matrices}. (a) The $\bar{X}_1$ and $\bar{Z}_1$ logical operators are constructed using the codeword $10110 \in G$. (b) The $\bar{X}_4$ and $\bar{Z}_4$ logical operators are constructed using the codeword $01111 \in G$. (c) A $Z$-type stabilizer is constructed using $x: 11010 \in H$ and $y: 10110 \in G$. (d) An $X$-type stabilizer is constructed using $x: 10110 \in G$ and $y: 00110 \in H$. (e) An $X$-type stabilizer is constructed from a ``contractible loop" through the long-range boundaries according to $y: 01001 \in H$. Logical operators may be deformed through the long-range boundaries by multiplying appropriate stabilizers.}
\label{fig:logicals and stabs}
\end{figure*}

The HGP will produce two types of horizontal and vertical long-range boundaries: an $X$-type and a $Z$-type. For every long-range edge connecting a node and a check, the graph product will produce a long-range edge connecting a (node, node) $\rightarrow$ qubit to a (check, node) $\rightarrow$ $Z$-check or a (node, check) $\rightarrow$ $X$-check to a (check, check) $\rightarrow$ qubit. We denote an excitation of a $X$($Z$)-check as a $e$ ($m$) anyon. Using these conventions, we can now analyze anyon transport through the long-range boundaries. Suppose we try and move an $e$ particle through an $X$-type long-range boundary (by growing its ``error" string of $Z$s). The combination of the original and newly emerging strings must overlap on an even number of sites with each long-range $X$-check. This constraint is satisfied precisely by the codewords generated by $G$. If the code distance of the outer code is $d'$, then the $e$ must split into at least $d'-1$ additional $e$ particles. Now if we try to move an $e$ particle through a $Z$-type long-range boundary, we can simply multiply the error string by long-range $Z$-checks, which will extend the support of this error to additional surface-code patches given by $H$. Growing the error strings in these other patches will create additional $e$ particles. The rules for $m$ particle follow analogously by switching the roles of $X$ and $Z$. For each surface-code patch labeled $(x,y)$ with the origin at the upper-left, we can arrange the rough ($e$ absorbing), smooth ($m$ absorbing) and long-range boundaries as depicted in Fig. \ref{fig:logicals and stabs}. Rough boundaries are present at the bottom and smooth boundaries on the left. The top and right boundaries are the long-range boundaries. The anyon transport rules through the long-range boundaries can now be summarized as
\begin{itemize}
    \item $e$ anyons tunnel through horizontal (vertical) boundaries according to $G$ ($H$).
    \item $m$ anyons tunnel through horizontal (vertical) boundaries according to $H$ ($G$).
\end{itemize}
So the tunneling of $e$ ($m$) anyons through horizontal (vertical) boundaries are analogous to that of domain walls in the classical parent code: the codewords generated by $G$ label the $y$ ($x$) coordinates of surface-code patches where additional anyons can appear. The tunneling rules in the other directions are analogous but using the dual codewords generated by $H$. Because $e$ and $m$ anyons behave differently through long-range boundaries ($G \neq H$ in general), we have lost the usual $e\leftrightarrow m$ duality that is present in the ordinary surface code. However, if we use a \emph{self-dual} code where $G \simeq H$ (e.g. $[8,4,4]$ extended Hamming), then this duality is restored.

We can use the above tunneling rules to construct our logical operators. We choose the standard form of $G$ ($\ident^{\vps}_{k'}$ on the left) as a canonical basis for our logical operators like in \eqref{eq:[5,2,3] matrices}. Starting on each surface-code patch $(1\leq x \leq k', 1\leq y \leq k')$ in the upper-left corner, the $X$($Z$)-type logical operators are horizontal (vertical) lines spanning the surface-code patches given by the $x$($y$)-th row of $G$ with the other coordinate fixed. The $X$($Z$)-type logical strings can be interpreted as dragging a single $m$ ($e$) from a smooth (rough) boundary where they are condensed, transporting it to the long-range boundary on the opposite side, and then transporting all tunneled anyons across the new surface-code patches and absorbing them at opposing smooth (rough) boundaries. Using this procedure, we successfully construct $X$ and $Z$ logical operators for all $K=k'^2$ logical qubits. Since $G$ is in standard form, these logical operators only intersect once inside the patches in the upper-left $k'\times k'$ corner.

\subsection{Anyon confinement and single-shot decoding}

The presence of syndrome measurement errors is detrimental for surface code decoding, often lowering the error threshold by an order of magnitude. The intuitive reason is that error strings are only detectable at their endpoints, and so if both endpoints have a syndrome measurement error, then that string becomes undetectable. The usual scheme to account for syndrome measurement errors is to perform multiple rounds of syndrome measurements and use the global space-time history for decoding. However, the number of measurement rounds per QEC cycle will scale with the system size \cite{Dennis_2002}. If a decoder is able to account for these measurement errors with a only a small overhead, then we say this decoder is capable of \emph{single-shot} correction. For stabilizer codes, the relation between confinement and single-shot ability has been well established \cite{Bombin_2015, Quintavalle_2021}. Confinement implies that enacting a logical operation via local moves will necessarily violate an increasing amount of stabilizers, and so even if a few measurements are faulty, there still exists a sufficient number of violated stabilizers to undo most of the error such that any residual error remains controlled over subsequent QEC cycles.

The Tanner graphs of good cLDPC codes are typically constructed from bipartite expander graphs. Expander graphs have the property that the boundary of a small subset of vertices scales proportionally to the size of that subset. In particular, we say that a (regular) bipartite graph $G = (B \cup C,E)$ of size $\{\abs{B},\abs{C}\} = \{n,m\}$ and degrees $\Delta_B, \Delta_C$ is $(\gamma,\delta)$ left-expanding if for any subset $S\subset B$ with volume $\abs{S} \leq \gamma n$, the size of the boundary, the number of connected checks, obeys $\abs{\partial S} \geq (1-\delta)\Delta_B \abs{S}$. The definition of right-expansion follows analogously by switching the roles of $B$ and $C$ above. On a Tanner graph, the boundary of a subset of nodes is an upper bound on the number of violated parity checks for an error supported on that subset. For Tanner graphs that are left-expanding with $\delta < 1/2$, one can show a linear code distance $d \geq \gamma n$ by counting the number of \emph{unique} neighbors, a lower bound on the syndrome weight, of small subsets of nodes. One can further show that the syndrome weight $\abs{\mathbf{s}} \geq (1-2\delta) \Delta_B \abs{\mathbf{e}}$ for any error with weight $\abs{\mathbf{e}} \leq \gamma n$, with $\gamma n < d$ \cite{sipser1996}. In other words, a cLDPC code is guaranteed to exhibit \emph{linear} confinement if the underlying Tanner graph exhibits sufficient left-expansion. As a consequence, effecting a logical error via local bit flips, or equivalently moving a domain-wall excitation, will necessarily violate a growing number of checks in the process. Reformulating the parity checks as multi-spin interactions in a classical Hamiltonian, we can reinterpret the previous statement as the existence of macroscopic energy barriers between different ground states.

The notion of confinement for cLDPC codes has been generalized to HGP codes \cite{Leverrier_2015, Fawzi_2020}\footnote{In these references, confinement is referred to as robustness.}. We summarize the relevant results as follows. For HGP codes to achieve $D = \Theta(\sqrt{N})$, we require the Tanner graph of the parent cLDPC codes to be both left- \emph{and} right-expanding with $\delta<1/2$ so that both the distance and transpose distance are linear in the system size. In quantum codes, due to degeneracy, there are many errors related by stabilizer elements which produce the same error syndrome. It suffices to examine the so-called ``reduced weight" of a given error, which is defined as the minimum Hamming weight over its stabilizer group orbit. If the parent expansion satisfies $\delta<1/6$, then it has been shown that the syndrome weight obeys $\abs{\mathbf{s}} \geq \frac{1}{3}\abs{\mathbf{e}}_{\rm red}$ for errors in the HGP code with reduced weight $\abs{\mathbf{e}}_{\rm red} \leq \min(\gamma n,\gamma m)$ with $\gamma \min(n,m) \leq \min(d,d^\transpose)$ \cite{quintavalle2022}. For a HGP code to provably exhibit linear confinement, its parent code requires a larger expansion ($\delta<1/6$) than what is needed for the classical analogue ($\delta<1/2$). 

Finding explicit constructions of bipartite graphs with the necessary expansion parameters above is often difficult. Fortunately, a random regular bipartite graph with degrees $\Delta_B,\Delta_C > 1/\delta$ exhibits $(\gamma=\Omega(1),\delta)$ expansion with high probability \cite{Richardson2008}. We also note that in practice, one can often get away with smaller expansions compared to the theoretical guarantees \cite{grospellier2019}.

To construct the parent code of a LRESC, we begin with a good cLDPC code and concatenate a repetition code of variable length $c$. If the Tanner graph of the outer cLDPC code exhibits $\delta<1/6$ expansion, then we know that $\abs{\mathbf{s}} \geq \frac{2}{3} \Delta_B \abs{\mathbf{e}}$. The concatenation with a repetition code simply decreases the confinement to $\abs{\mathbf{s}} \geq \frac{2}{3c} \Delta_B \abs{\mathbf{e}}$, since the structure of the outer code is unchanged but each physical outer bit can now host $c$ inner bits. An analogy holds for the associated LRESC due to the product structure of the HGP. The long-range checks in both horizontal and vertical directions mimic those of the non-concatenated HGP code, but now the surface codes can host an additional $\approx 2c^2$ physical qubits. Thus, the confinement in the LRESC can be quantified as $\abs{\mathbf{s}} \geq \frac{1}{6c^2} \abs{\mathbf{e}}_{\rm red}$. Let us now analyze the scaling of confinement with the number of long-range interactions. Let $[n',\Theta(n'),\Theta(n')]$ be the parameters of the outer cLDPC code constructed using the expander graph methods previously mentioned; this code will necessarily contain $\Theta(n')$ long-range interactions. After concatenating with a length-$c$ repetition code, the parameters become $[n'c,\Theta(n'),\Theta(n'c)]$. The confinement of this concatenated code scales as $\abs{\mathbf{s}} = \Omega(n'/n \cdot \abs{\mathbf{e}})$. Suppose the number of long-range interactions scales as $n' = n^b$ for some $0\leq b<1$. Then small errors of weight $\abs{\mathbf{e}} < n^{1-b}$ have no confinement because they can be chosen to live on a single repetition-code segment. For large errors with weight $\abs{\mathbf{e}} = \Omega(n)$, the confinement scales as $\abs{\mathbf{s}} = \Omega(n^b)$. Taking the HGP of this code with itself yields a LRESC with $\llbracket \Theta(n'^2c^2),\Theta(n'^2),\Theta(n'c) \rrbracket$ code parameters, $\abs{\mathbf{s}} = \Omega(N^{b-1} \cdot \abs{\mathbf{e}}_{\rm red})$ confinement, and $O(n'\sqrt{N})$ long-range interactions. For a $b>0$ scaling of the long-range interactions in the parent code, the LRESC satisfies the ``good confinement" definition of \cite{Quintavalle_2021} and is provably single-shot decodable under adversarial noise. A sustainable threshold under local stochastic noise has been proven for $b=1$ (linear confinement and fully nonlocal limit), but it remains an open problem as to whether this threshold can exist for $b<1$, though numerical evidence suggests an affirmative for certain families of 3D homological product codes \cite{Quintavalle_2021}. Because LRESCs can systematically vary their density of long-range interactions, they provide tunable qLDPC codes to numerically benchmark sustainable thresholds for $0<b<1$. We leave such studies to future work.

We also emphasize that in practice, the existence of a threshold may not be as important when dealing with finite-size overheads, as supported by the numerical simulations in Fig. \ref{fig:HGP vs surface}.


\section{Logical gate inheritance}\label{app:logical gates}

In this section, we elaborate on how a HGP code can inherit certain logical gates from its parent classical codes. In the parent codes, we show how non-fault-tolerant operations can be made increasingly transversal upon concatenating with a repetition code. We then demonstrate that this transversality is inherited by the associated LRESC, using the parent $[3(c),2,2(c)]$ code as a guiding example.

\subsection{Linear transformations on the codespace}

Suppose we have a classical linear code with parity-check matrix $H \in \mathbb{F}^{m\times n}_2$ and generator matrix $G \in \mathbb{F}^{k\times n}_2$. We examine transformations on the code of the form
\begin{align}\label{eq:linear code transform}
    H \longrightarrow HU \quad,\quad G \longrightarrow GU \, ,
\end{align}
where $U \in \mathbb{F}^{n\times n}_2$ is an orthogonal matrix obeying $U^\transpose U = U U^\transpose = \ident$. The columns of $U$ represent \emph{linear} transformations on the columns (physical bits) of $H$ and $G$. In particular, since $U$ is invertible, we can always decompose this linear transformation into a series of elementary ones corresponding to adding $(A)$ and swapping $(S)$ columns.
Each column swap corresponds to a physical SWAP, while each column addition can be implemented by a CNOT gate on the physical bits:
\begin{align}
    \mathrm{CNOT}_{i\rightarrow j} \implies \begin{array}{cc}
         H_{\star i} \longrightarrow H_{\star i} + H_{\star j}  \\
         G_{\star i} \longrightarrow G_{\star i} + G_{\star j} \, ,
    \end{array}
\end{align}
where $H_{\star i}$ denotes the $i$th column of $H$. In order for $U$ to be a valid transformation on the codespace, we require the row space of $G$ (i.e. $\ker(H)$) to remain invariant, which is satisfied if and only if 
\begin{align}\label{eq:G row reduction}
    GU = VG \, ,
\end{align}
for some invertible $V \in \mathbb{F}^{m\times m}_2$. Since $U$ is orthogonal, we have $(HU)(GU)^\transpose = HUU^\transpose G^\transpose = HG^\transpose = 0$. At the same time, $H(GU)^\transpose = H(VG)^\transpose = HG^\transpose V^\transpose = 0$. Hence $\ker(H) = \ker(HU)$, which implies that $H$ and $HU$ are row-equivalent; i.e. there exists some invertible $W \in \mathbb{F}^{m\times m}_2$ such that
\begin{align}\label{eq:H row reduction}
    HU = WH \, .
\end{align}

As an example, let's take the 3-bit parity code with
\begin{align}
    H = \begin{pmatrix}
        1 & 1 & 1
    \end{pmatrix} \quad,\quad
    G = \begin{pmatrix}
        1 & 0 & 1 \\
        0 & 1 & 1
    \end{pmatrix} \, .
\end{align}
A logical CNOT gate between the first logical qubit (first row of $G$, control) and the second logical qubit (second row of $G$, target) results in the transformation
\begin{subequations}\label{eq:CNOT row transform}
\begin{align}
    g_1 &\longrightarrow g_1 + g_2 \\
    g_2 &\longrightarrow g_2 \, ,
\end{align}
\end{subequations}
which becomes $101\rightarrow 110$ and $011\rightarrow 011$ explicitly for the 3-bit parity code. This logical operation can be achieved by swapping the second and third bits (columns of $G$) given by the permutation matrix
\begin{align}\label{eq:1<->3 swap matrix}
    U = \begin{pmatrix}\,
        1 & 0 & 0 \\
        0 & 0 & 1 \\
        0 & 1 & 0
    \,\end{pmatrix} \, .
\end{align}
Similarly, the logical CNOT with control and target roles reversed is achieved by swapping bits 1 and 3. The column swaps results in the desired transformation \eqref{eq:linear code transform} where $U$ is a permutation, and thereby orthogonal, matrix. The logical CNOTs correspond to row addition on $G$, and so \eqref{eq:G row reduction} is also satisfied. Because $H$ is permutation symmetric, we have $W=\ident$ in \eqref{eq:H row reduction}.

\subsection{Segment-transversality with repetition}

We now show how concatenation with a repetition code maintains the same circuit depth to perform logical operations. Intuitively, this is because the repetition codes simply clone the value of the physical bits of the outer code. Any operations performed on the outer bits now become \emph{parallel} (transversal) operations amongst the repetition code segments. Since the code distance is increased in the process, this concatenation will increase the fault tolerance of the original gadgets.

Suppose the $m'\times n'$ parity-check matrix of the outer code is $H$. Then the parity-check matrix $H^{(c)}$ of the concatenated code can be written as

\begin{align}\label{eq:concatenated H}
    H^{(c)} = \left(
    \begin{array}{c}
        H \otimes \mathbf{v}  \\
        \hline
        \ident_{n'} \otimes H^{(c)}_{\rm rep}
    \end{array}
    \right) \, ,
\end{align}
where $\otimes$ denotes the kronecker product, $\mathbf{v} = (1\;0\;0\;\dots)$ is a vector of length $c$ with 1 in the first entry and 0 elsewhere, and $H^{(c)}_{\rm rep}$ is the parity-check matrix of the 1D repetition code of length $c$ with dimensions $(c-1) \times c$, e.g.
\begin{align}
    H^{(3)}_{\rm rep} = \begin{pmatrix}
        1 & 1 & 0 \\
        0 & 1 & 1
    \end{pmatrix} \, .
\end{align}
The generator matrix $G^{(c)}$ takes the form
\begin{align}\label{eq:concatenated G}
    G^{(c)} = G \otimes \mathbf{1}_{1\times c}
\end{align}
where $\mathbf{1}_{1\times c} = (\begin{array}{cccc} 1&1&1&\dots \end{array})$ is the nonzero codeword of the repetition code of length $c$. The orthogonal column transformation \eqref{eq:linear code transform} on the concatenated code takes the transversal form
\begin{align}\label{eq:concatenated U}
    U^{(c)} = U \otimes \ident_c \, ,
\end{align}
where the kronecker product can be physically interpreted as performing $U$ transversally between the repetition code segments; we accordingly describe this physical operation as \emph{segment-transversal}. It is straightforward to verify that the right-action of \eqref{eq:concatenated U} on \eqref{eq:concatenated H} and \eqref{eq:concatenated G} reproduces the concatenated analogue of \eqref{eq:H row reduction} and \eqref{eq:G row reduction} respectively with
\begin{subequations}
\begin{align}
    V^{(c)} &= V  \label{eq:concatenated V} \\
    W^{(c)} &= \mathrm{diag}\left( W,\, U \otimes \ident_{c-1} \right) \, . \label{eq:concatenated W}
\end{align}
\end{subequations}
Since $V^{(c)} = V$, we conclude that the right-action of $U^{(c)}$ \eqref{eq:concatenated U} enacts the same logical transformation as the original $U$.

\subsection{Patch-transversality in LRESCs}

Suppose now that we use the previous concatenated code as the parent code in the hypergraph product. Recall that the parity-check matrices for a HGP code with parent code $H^{(c)}$ take the form
\begin{subequations}\label{eq:LRESC H transversal}
\begin{align}\label{eq:H_X transversal}
    H^{\vps}_X &= \big(\, H^{(c)} \otimes \ident_{n} \;\;|\;\; \ident_{m} \otimes H^{(c)\transpose} \,\big) \\
    H^{\vps}_Z &= \big(\, \ident_{n} \otimes H^{(c)} \;\;|\;\; H^{(c)\transpose} \otimes \ident_{m} \,\big) \, .
\end{align}
\end{subequations}
For simplicity, we will focus on codewords induced by $G^{(c)}$ on the primary lattice (node-node qubits); the analysis of the transpose codewords on the secondary lattice (check-check qubits) follows analogously. Mirroring the structure of the parity checks, we can construct generator matrices for the $K=k^2$ logical qubits as follows:
\begin{subequations}\label{eq:LRESC G transversal}
\begin{align}
    G_Z &= \big(\, G^{(c)} \otimes \ident_{n} \;\;|\;\; 0 \;\big)  \label{eq:LRESC logical Z} \\
    G_X &= \big(\, \ident_{n} \otimes G^{(c)} \;\;|\;\; 0 \;\big)  \label{eq:LRESC logical X} \, .
\end{align}
\end{subequations}
It is easy to verify that $H^{\vps}_X G^\transpose_Z = H^{\vps}_Z G^\transpose_X = 0$. For quantum CSS codes, we can choose either $X$-type or $Z$-type logical operators to inherit our classical transformations. Suppose we want to enact our desired logical transformation on the logical $\bar{Z}$ operators \eqref{eq:LRESC logical Z}. Define the following right-action on $H_X$:
\begin{align}\label{eq:U_X transversal}
    U_X^{(\bar{Z})} = \mathrm{diag}\big(\, U^{(c)} \otimes \ident_{n} \,,\, W^{(c)} \otimes \ident_{m} \,\big) \, ,
\end{align}
which physically corresponds to applying $U^{(c)}$ simultaneously to all primary rows and $W^{(c)}$ to all secondary rows in the 2D layout of Fig. \ref{fig:HGP construction detailed}. Examining the structure of $U^{(c)}$ \eqref{eq:concatenated U} and $W^{(c)}$ \eqref{eq:concatenated W}, we see that the above operation can be interpreted as \emph{transversal} operations amongst surface-code patches given by the hypergraph product of the repetition code segments; we accordingly describe this physical operation as \emph{patch-transversal}. For ease of notation, we will henceforth drop the superscript $(\bar{Z})$; unless stated otherwise, $U_X$ corresponds to the choice \eqref{eq:U_X transversal}. Note that the physical action of $U_X$ does \emph{not} transform $H_Z$ in the same way. Nonetheless, we can easily compute the corresponding $U_Z$ by decomposing $U_X$ into elementary matrices:
\begin{align}
    U_X = S_6 A_5 S_4 A_3 \cdots \, .
\end{align}
The swap operations $S$ correspond to physical SWAP gates and so remain the same for $U_Z$. However, the addition operations become transposed because CNOT reverses the roles of control and target for $X\leftrightarrow Z$. So the corresponding $U_Z$ is given by
\begin{align}\label{eq:U_Z transversal}
    U^{(\bar{Z})}_Z &= S_6 A^\transpose_5 S_4 A^\transpose_3 \cdots = \left( U^{-1}_X \right)^\transpose \equiv U^{-\transpose}_X  \notag \\
    &= \mathrm{diag}\big(\, U^{(c)} \otimes \ident_{n} \,,\, (W^{(c)})^{-\transpose} \otimes \ident_{m} \,\big) \, ,
\end{align}
where we have used the fact that $S^2=A^2=\ident$ over $\mathbb{F}_2$ in the first line and $U^\transpose = U^{-1}$ in the second line. The orthogonality condition $H^{\vps}_X H^\transpose_Z \rightarrow H^{\vps}_X U^{\vps}_X U^\transpose_Z H^\transpose_Z = H^{\vps}_X U^{\vps}_X U^{-1}_X H^\transpose_Z = H^{\vps}_X H^\transpose_Z = 0$ is thus preserved, which is expected since we know that physical unitary operations preserve commutativity.

To ensure that the stabilizer group remains invariant, it suffices to verify that $(H_X U_X) H^\transpose_Z = 0$. Acting $U_X$ \eqref{eq:U_X transversal} to the right of $H_X$ \eqref{eq:H_X transversal} gives
\begin{align}
    H_X U_X &= \big(\, H^{(c)} U^{(c)} \otimes \ident_{n} \;\;|\;\; W^{(c)} \otimes H^{(c)\transpose} \,\big)  \notag \\
    &= \big(\, W^{(c)} H^{(c)} \otimes \ident_{n} \;\;|\;\; W^{(c)} \otimes H^{(c)\transpose} \,\big)  \notag \\
    &= \big( W^{(c)} \otimes \ident_{n} \big)\, H_X  \notag \\
    &\equiv W_X H_X \, ,
\end{align}
from which the orthogonality condition $(H_X U_X) H^\transpose_Z = W_X H_X H^\transpose_Z = 0$ follows. Since $W_X$ is invertible, we conclude that $H_X U_X$ is row-equivalent to the original $H_X$. A similar calculation also shows row-equivalence between $H_Z U_Z$ and $H_Z$. Thus, the patch-transversal operation corresponding to \eqref{eq:U_X transversal} and \eqref{eq:U_Z transversal} preserves the stabilizer group.

\begin{figure}[t]
\centering
\includegraphics[width=0.32\textwidth]{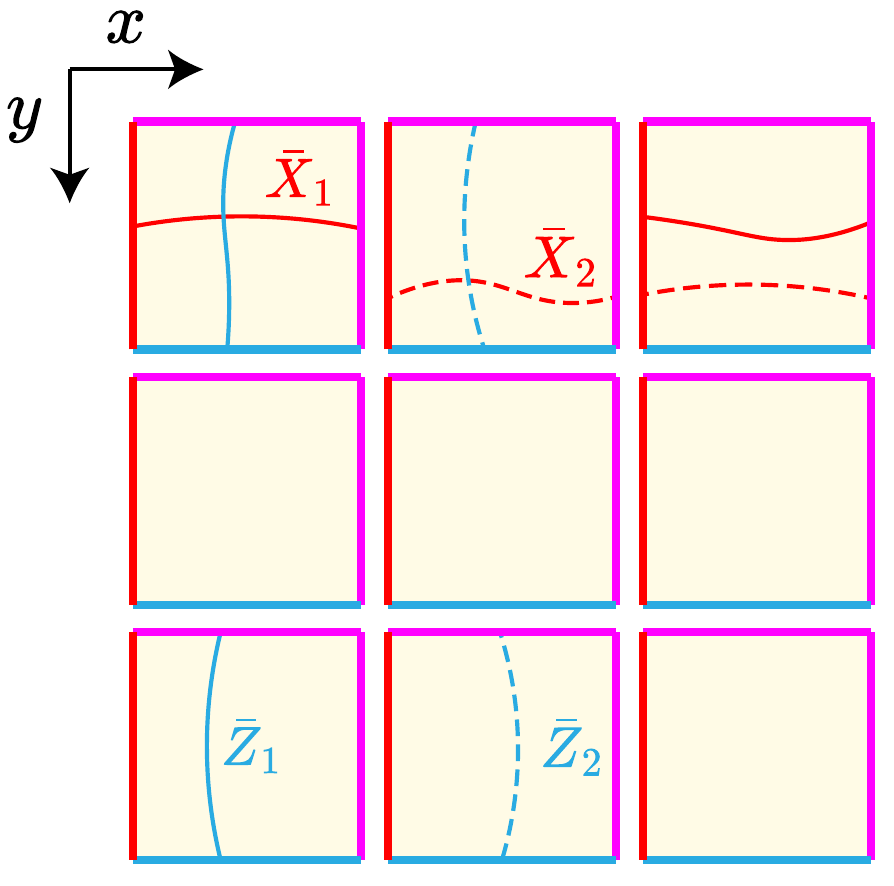}
\caption{The layout of the local surface-code patches of the $[3,2,2]$ LRESC family is illustrated. The patches can be labeled by coordinates $(x,y)$ where $x,y=1,2,3$. The logical $\bar{X}$ and $\bar{Z}$ operators for the first two logical qubits are also shown.}
\label{fig:[3,2,2] LRESC logicals}
\end{figure}

The action on the codespace is given by
\begin{subequations}
\begin{align}
    G_Z U_Z &= \big(\, G^{(c)} U^{(c)} \otimes \ident_{n} \;\;|\;\; 0 \;\big)  \notag \\
    &= \big(\, V G^{(c)} \otimes \ident_{n} \;\;|\;\; 0 \;\big)  \\
    G_X U_X &= \big(\, U^{(c)} \otimes G^{(c)} \;\;|\;\; 0 \;\big)  \label{eq:G_X U_X} \, ,
\end{align}
\end{subequations}
where in the second line we have used \eqref{eq:concatenated V} and \eqref{eq:G row reduction}. We see that the patch-transversal operation enacts parallel logical operations along the $n'=n/c$ columns of surface-code patches, transforming the logical $\bar{Z}$ operators in the same manner as in the classical code \eqref{eq:G row reduction}. We can also enact patch-transversal operations between columns of surface-code patches, accordingly transforming the logical $\bar{X}$ operators, using the choice
\begin{subequations}\label{eq:LRESC U transform}
\begin{align}
    U^{(\bar{X})}_X &= \mathrm{diag}\big(\, \ident_n \otimes U^{(c)} \,,\, \ident_m \otimes (W^{(c)})^{-\transpose} \,\big)  \\
    U^{(\bar{X})}_Z &= \mathrm{diag}\big(\, \ident_n \otimes U^{(c)} \,,\, \ident_m \otimes W^{(c)} \,\big) \, .
\end{align}
\end{subequations}
The analogous operations for the transpose code follows similarly by switching the roles of the primary and secondary lattices (left and right sides of \eqref{eq:LRESC H transversal} and \eqref{eq:LRESC G transversal}). In general, from the form of \eqref{eq:LRESC U transform}, we see that both codewords and transpose codewords are transformed simultaneously. Another thing to check is whether the complementary transformation on the other type of logical \eqref{eq:G_X U_X} is the desired one.

Now let's see how the 3-bit parity code fits into the above machinery. Its associated LRESC family has four logical qubits living amongst nine surface-code patches arranged in a $3\times 3$ layout in the manner of Fig. \ref{fig:[3,2,2] LRESC logicals}. In the base $[3,2,2]$ code, the $U$ which implements the logical $\overline{\mathrm{CNOT}}_{1\rightarrow 2}$ is given by \eqref{eq:1<->3 swap matrix}. In the LRESC, the corresponding transformation \eqref{eq:LRESC U transform} involves exchanging the $x=2$ column of patches with those of $x=3$. We will now show that this transformation enacts $\overline{\mathrm{CNOT}}_{1\rightarrow 2} \cdot \overline{\mathrm{CNOT}}_{3\rightarrow 4}$, focusing on $\overline{\mathrm{CNOT}}_{1\rightarrow 2}$ since the analysis of $\overline{\mathrm{CNOT}}_{3\rightarrow 4}$ follows identically by examining the second ($y=2$) row instead of the first ($y=1$). It is easy to see that this column swap maps $\bar{X}_1 \rightarrow \bar{X}_1\bar{X}_2$ while keeping $\bar{X}_2$ and $\bar{Z}_1$ unchanged, up to stabilizer equivalence. To complete the (operator) CNOT truth table, it suffices to verify that $\bar{Z}_2 \rightarrow \bar{Z}_1\bar{Z}_2$. Notice that $\bar{Z}_2$ gets mapped from the second $(x=2)$ to the third column $(x=3)$; denote the transformed operator as $\bar{Z}'_2$. Using the tunneling rules of Appendix \ref{app:boundary dynamics}, we can construct a $Z$-type stabilizer element consisting of vertical strings living in patches indexed by $h_1 = (1\;1\;1)$ for the $x$ coordinate and $g_1 = (1\;0\;1)$ for the $y$ coordinate. This $Z$-stabilizer element is precisely given by
\begin{align}
    S_Z = \bar{Z}_1\bar{Z}_2\bar{Z}'_2 \, .
\end{align}
We thus verify that $\bar{Z}'_2 \simeq \bar{Z}'_2 S_Z = \bar{Z}_1 \bar{Z}_2$ for the logical $\overline{\mathrm{CNOT}}_{1\rightarrow 2}$, completing the truth table.

For larger codes in the Hadamard code family, we will in general lose the permutation symmetry of $H$ in the parent code, leading to a nontrivial $W$ in \eqref{eq:H row reduction}. As a consequence, from \eqref{eq:concatenated W} and \eqref{eq:U_X transversal}, we see that additional transversal gates may need to be applied between the long-range boundaries of the associated LRESCs. For suitably small base codes and presentations of $H$, the additional SWAP-CNOT gadgets may still be advantageous despite the added gate costs.

\end{appendix}

\bibliography{thebib}

\begin{thebibliography}{129}%
\makeatletter
\providecommand \@ifxundefined [1]{%
 \@ifx{#1\undefined}
}%
\providecommand \@ifnum [1]{%
 \ifnum #1\expandafter \@firstoftwo
 \else \expandafter \@secondoftwo
 \fi
}%
\providecommand \@ifx [1]{%
 \ifx #1\expandafter \@firstoftwo
 \else \expandafter \@secondoftwo
 \fi
}%
\providecommand \natexlab [1]{#1}%
\providecommand \enquote  [1]{``#1''}%
\providecommand \bibnamefont  [1]{#1}%
\providecommand \bibfnamefont [1]{#1}%
\providecommand \citenamefont [1]{#1}%
\providecommand \href@noop [0]{\@secondoftwo}%
\providecommand \href [0]{\begingroup \@sanitize@url \@href}%
\providecommand \@href[1]{\@@startlink{#1}\@@href}%
\providecommand \@@href[1]{\endgroup#1\@@endlink}%
\providecommand \@sanitize@url [0]{\catcode `\\12\catcode `\$12\catcode
  `\&12\catcode `\#12\catcode `\^12\catcode `\_12\catcode `\%12\relax}%
\providecommand \@@startlink[1]{}%
\providecommand \@@endlink[0]{}%
\providecommand \url  [0]{\begingroup\@sanitize@url \@url }%
\providecommand \@url [1]{\endgroup\@href {#1}{\urlprefix }}%
\providecommand \urlprefix  [0]{URL }%
\providecommand \Eprint [0]{\href }%
\providecommand \doibase [0]{http://dx.doi.org/}%
\providecommand \selectlanguage [0]{\@gobble}%
\providecommand \bibinfo  [0]{\@secondoftwo}%
\providecommand \bibfield  [0]{\@secondoftwo}%
\providecommand \translation [1]{[#1]}%
\providecommand \BibitemOpen [0]{}%
\providecommand \bibitemStop [0]{}%
\providecommand \bibitemNoStop [0]{.\EOS\space}%
\providecommand \EOS [0]{\spacefactor3000\relax}%
\providecommand \BibitemShut  [1]{\csname bibitem#1\endcsname}%
\let\auto@bib@innerbib\@empty
\bibitem [{\citenamefont {Gottesman}(1997)}]{gottesman1997stabilizer}%
  \BibitemOpen
  \bibfield  {author} {\bibinfo {author} {\bibfnamefont {Daniel}\ \bibnamefont
  {Gottesman}},\ }\href@noop {} {\enquote {\bibinfo {title} {Stabilizer codes
  and quantum error correction},}\ } (\bibinfo {year} {1997}),\ \Eprint
  {http://arxiv.org/abs/quant-ph/9705052} {arXiv:quant-ph/9705052 [quant-ph]}
  \BibitemShut {NoStop}%
\bibitem [{\citenamefont {Calderbank}\ and\ \citenamefont
  {Shor}(1996)}]{Calderbank_1996}%
  \BibitemOpen
  \bibfield  {author} {\bibinfo {author} {\bibfnamefont {A.~R.}\ \bibnamefont
  {Calderbank}}\ and\ \bibinfo {author} {\bibfnamefont {Peter~W.}\ \bibnamefont
  {Shor}},\ }\bibfield  {title} {\enquote {\bibinfo {title} {Good quantum
  error-correcting codes exist},}\ }\href {\doibase 10.1103/physreva.54.1098}
  {\bibfield  {journal} {\bibinfo  {journal} {Physical Review A}\ }\textbf
  {\bibinfo {volume} {54}},\ \bibinfo {pages} {1098--1105} (\bibinfo {year}
  {1996})}\BibitemShut {NoStop}%
\bibitem [{\citenamefont {Steane}(1996)}]{Steane_1996}%
  \BibitemOpen
  \bibfield  {author} {\bibinfo {author} {\bibfnamefont {Andrew}\ \bibnamefont
  {Steane}},\ }\bibfield  {title} {\enquote {\bibinfo {title}
  {Multiple-particle interference and quantum error correction},}\ }\href
  {\doibase 10.1098/rspa.1996.0136} {\bibfield  {journal} {\bibinfo  {journal}
  {Proceedings of the Royal Society of London. Series A: Mathematical, Physical
  and Engineering Sciences}\ }\textbf {\bibinfo {volume} {452}},\ \bibinfo
  {pages} {2551--2577} (\bibinfo {year} {1996})}\BibitemShut {NoStop}%
\bibitem [{\citenamefont {Kitaev}(2003)}]{Kitaev_2003}%
  \BibitemOpen
  \bibfield  {author} {\bibinfo {author} {\bibfnamefont {A.Yu.}\ \bibnamefont
  {Kitaev}},\ }\bibfield  {title} {\enquote {\bibinfo {title} {Fault-tolerant
  quantum computation by anyons},}\ }\href {\doibase
  10.1016/s0003-4916(02)00018-0} {\bibfield  {journal} {\bibinfo  {journal}
  {Annals of Physics}\ }\textbf {\bibinfo {volume} {303}},\ \bibinfo {pages}
  {2–30} (\bibinfo {year} {2003})}\BibitemShut {NoStop}%
\bibitem [{\citenamefont {Bravyi}\ and\ \citenamefont
  {Kitaev}(1998)}]{bravyi1998}%
  \BibitemOpen
  \bibfield  {author} {\bibinfo {author} {\bibfnamefont {S.~B.}\ \bibnamefont
  {Bravyi}}\ and\ \bibinfo {author} {\bibfnamefont {A.~Yu.}\ \bibnamefont
  {Kitaev}},\ }\href@noop {} {\enquote {\bibinfo {title} {Quantum codes on a
  lattice with boundary},}\ } (\bibinfo {year} {1998}),\ \Eprint
  {http://arxiv.org/abs/quant-ph/9811052} {arXiv:quant-ph/9811052 [quant-ph]}
  \BibitemShut {NoStop}%
\bibitem [{\citenamefont {Fowler}\ \emph {et~al.}(2012)\citenamefont {Fowler},
  \citenamefont {Mariantoni}, \citenamefont {Martinis},\ and\ \citenamefont
  {Cleland}}]{Fowler_2012}%
  \BibitemOpen
  \bibfield  {author} {\bibinfo {author} {\bibfnamefont {Austin~G.}\
  \bibnamefont {Fowler}}, \bibinfo {author} {\bibfnamefont {Matteo}\
  \bibnamefont {Mariantoni}}, \bibinfo {author} {\bibfnamefont {John~M.}\
  \bibnamefont {Martinis}}, \ and\ \bibinfo {author} {\bibfnamefont
  {Andrew~N.}\ \bibnamefont {Cleland}},\ }\bibfield  {title} {\enquote
  {\bibinfo {title} {Surface codes: Towards practical large-scale quantum
  computation},}\ }\href {\doibase 10.1103/physreva.86.032324} {\bibfield
  {journal} {\bibinfo  {journal} {Physical Review A}\ }\textbf {\bibinfo
  {volume} {86}} (\bibinfo {year} {2012}),\
  10.1103/physreva.86.032324}\BibitemShut {NoStop}%
\bibitem [{\citenamefont {Bluvstein}\ \emph {et~al.}(2022)\citenamefont
  {Bluvstein}, \citenamefont {Levine}, \citenamefont {Semeghini}, \citenamefont
  {Wang}, \citenamefont {Ebadi}, \citenamefont {Kalinowski}, \citenamefont
  {Keesling}, \citenamefont {Maskara}, \citenamefont {Pichler}, \citenamefont
  {Greiner} \emph {et~al.}}]{bluvstein2022quantum}%
  \BibitemOpen
  \bibfield  {author} {\bibinfo {author} {\bibfnamefont {Dolev}\ \bibnamefont
  {Bluvstein}}, \bibinfo {author} {\bibfnamefont {Harry}\ \bibnamefont
  {Levine}}, \bibinfo {author} {\bibfnamefont {Giulia}\ \bibnamefont
  {Semeghini}}, \bibinfo {author} {\bibfnamefont {Tout~T}\ \bibnamefont
  {Wang}}, \bibinfo {author} {\bibfnamefont {Sepehr}\ \bibnamefont {Ebadi}},
  \bibinfo {author} {\bibfnamefont {Marcin}\ \bibnamefont {Kalinowski}},
  \bibinfo {author} {\bibfnamefont {Alexander}\ \bibnamefont {Keesling}},
  \bibinfo {author} {\bibfnamefont {Nishad}\ \bibnamefont {Maskara}}, \bibinfo
  {author} {\bibfnamefont {Hannes}\ \bibnamefont {Pichler}}, \bibinfo {author}
  {\bibfnamefont {Markus}\ \bibnamefont {Greiner}},  \emph {et~al.},\
  }\bibfield  {title} {\enquote {\bibinfo {title} {A quantum processor based on
  coherent transport of entangled atom arrays},}\ }\href@noop {} {\bibfield
  {journal} {\bibinfo  {journal} {Nature}\ }\textbf {\bibinfo {volume} {604}},\
  \bibinfo {pages} {451--456} (\bibinfo {year} {2022})}\BibitemShut {NoStop}%
\bibitem [{\citenamefont {AI}(2023)}]{Google_SC}%
  \BibitemOpen
  \bibfield  {author} {\bibinfo {author} {\bibfnamefont {Google~Quantum}\
  \bibnamefont {AI}},\ }\bibfield  {title} {\enquote {\bibinfo {title}
  {Suppressing quantum errors by scaling a surface code logical qubit},}\
  }\href {\doibase 10.1038/s41586-022-05434-1} {\bibfield  {journal} {\bibinfo
  {journal} {Nature}\ }\textbf {\bibinfo {volume} {614}},\ \bibinfo {pages}
  {676--681} (\bibinfo {year} {2023})}\BibitemShut {NoStop}%
\bibitem [{\citenamefont {Bluvstein}\ \emph {et~al.}(2023)\citenamefont
  {Bluvstein}, \citenamefont {Evered}, \citenamefont {Geim}, \citenamefont
  {Li}, \citenamefont {Zhou}, \citenamefont {Manovitz}, \citenamefont {Ebadi},
  \citenamefont {Cain}, \citenamefont {Kalinowski}, \citenamefont {Hangleiter},
  \citenamefont {Bonilla~Ataides}, \citenamefont {Maskara}, \citenamefont
  {Cong}, \citenamefont {Gao}, \citenamefont {Sales~Rodriguez}, \citenamefont
  {Karolyshyn}, \citenamefont {Semeghini}, \citenamefont {Gullans},
  \citenamefont {Greiner}, \citenamefont {Vuletić},\ and\ \citenamefont
  {Lukin}}]{Bluvstein_2023}%
  \BibitemOpen
  \bibfield  {author} {\bibinfo {author} {\bibfnamefont {Dolev}\ \bibnamefont
  {Bluvstein}}, \bibinfo {author} {\bibfnamefont {Simon~J.}\ \bibnamefont
  {Evered}}, \bibinfo {author} {\bibfnamefont {Alexandra~A.}\ \bibnamefont
  {Geim}}, \bibinfo {author} {\bibfnamefont {Sophie~H.}\ \bibnamefont {Li}},
  \bibinfo {author} {\bibfnamefont {Hengyun}\ \bibnamefont {Zhou}}, \bibinfo
  {author} {\bibfnamefont {Tom}\ \bibnamefont {Manovitz}}, \bibinfo {author}
  {\bibfnamefont {Sepehr}\ \bibnamefont {Ebadi}}, \bibinfo {author}
  {\bibfnamefont {Madelyn}\ \bibnamefont {Cain}}, \bibinfo {author}
  {\bibfnamefont {Marcin}\ \bibnamefont {Kalinowski}}, \bibinfo {author}
  {\bibfnamefont {Dominik}\ \bibnamefont {Hangleiter}}, \bibinfo {author}
  {\bibfnamefont {J.~Pablo}\ \bibnamefont {Bonilla~Ataides}}, \bibinfo {author}
  {\bibfnamefont {Nishad}\ \bibnamefont {Maskara}}, \bibinfo {author}
  {\bibfnamefont {Iris}\ \bibnamefont {Cong}}, \bibinfo {author} {\bibfnamefont
  {Xun}\ \bibnamefont {Gao}}, \bibinfo {author} {\bibfnamefont {Pedro}\
  \bibnamefont {Sales~Rodriguez}}, \bibinfo {author} {\bibfnamefont {Thomas}\
  \bibnamefont {Karolyshyn}}, \bibinfo {author} {\bibfnamefont {Giulia}\
  \bibnamefont {Semeghini}}, \bibinfo {author} {\bibfnamefont {Michael~J.}\
  \bibnamefont {Gullans}}, \bibinfo {author} {\bibfnamefont {Markus}\
  \bibnamefont {Greiner}}, \bibinfo {author} {\bibfnamefont {Vladan}\
  \bibnamefont {Vuletić}}, \ and\ \bibinfo {author} {\bibfnamefont
  {Mikhail~D.}\ \bibnamefont {Lukin}},\ }\bibfield  {title} {\enquote {\bibinfo
  {title} {Logical quantum processor based on reconfigurable atom arrays},}\
  }\href {\doibase 10.1038/s41586-023-06927-3} {\bibfield  {journal} {\bibinfo
  {journal} {Nature}\ }\textbf {\bibinfo {volume} {626}},\ \bibinfo {pages}
  {58–65} (\bibinfo {year} {2023})}\BibitemShut {NoStop}%
\bibitem [{\citenamefont {Ataides}\ \emph {et~al.}(2021)\citenamefont
  {Ataides}, \citenamefont {Tuckett}, \citenamefont {Bartlett}, \citenamefont
  {Flammia},\ and\ \citenamefont {Brown}}]{Bonilla_Ataides_2021}%
  \BibitemOpen
  \bibfield  {author} {\bibinfo {author} {\bibfnamefont {J.~Pablo~Bonilla}\
  \bibnamefont {Ataides}}, \bibinfo {author} {\bibfnamefont {David~K.}\
  \bibnamefont {Tuckett}}, \bibinfo {author} {\bibfnamefont {Stephen~D.}\
  \bibnamefont {Bartlett}}, \bibinfo {author} {\bibfnamefont {Steven~T.}\
  \bibnamefont {Flammia}}, \ and\ \bibinfo {author} {\bibfnamefont
  {Benjamin~J.}\ \bibnamefont {Brown}},\ }\bibfield  {title} {\enquote
  {\bibinfo {title} {The {XZZX} surface code},}\ }\href {\doibase
  10.1038/s41467-021-22274-1} {\bibfield  {journal} {\bibinfo  {journal}
  {Nature Communications}\ }\textbf {\bibinfo {volume} {12}} (\bibinfo {year}
  {2021}),\ 10.1038/s41467-021-22274-1}\BibitemShut {NoStop}%
\bibitem [{\citenamefont {Saffman}\ \emph {et~al.}(2010)\citenamefont
  {Saffman}, \citenamefont {Walker},\ and\ \citenamefont
  {M{\o}lmer}}]{Saffman2010}%
  \BibitemOpen
  \bibfield  {author} {\bibinfo {author} {\bibfnamefont {M.}~\bibnamefont
  {Saffman}}, \bibinfo {author} {\bibfnamefont {T.~G.}\ \bibnamefont {Walker}},
  \ and\ \bibinfo {author} {\bibfnamefont {K.}~\bibnamefont {M{\o}lmer}},\
  }\bibfield  {title} {\enquote {\bibinfo {title} {{Quantum information with
  Rydberg atoms}},}\ }\href {\doibase 10.1103/RevModPhys.82.2313} {\bibfield
  {journal} {\bibinfo  {journal} {Rev. Mod. Phys.}\ }\textbf {\bibinfo {volume}
  {82}},\ \bibinfo {pages} {2313--2363} (\bibinfo {year} {2010})}\BibitemShut
  {NoStop}%
\bibitem [{\citenamefont {Saffman}(2016)}]{Saffman2016}%
  \BibitemOpen
  \bibfield  {author} {\bibinfo {author} {\bibfnamefont {M}~\bibnamefont
  {Saffman}},\ }\bibfield  {title} {\enquote {\bibinfo {title} {Quantum
  computing with atomic qubits and rydberg interactions: progress and
  challenges},}\ }\href {\doibase 10.1088/0953-4075/49/20/202001} {\bibfield
  {journal} {\bibinfo  {journal} {Journal of Physics B: Atomic, Molecular and
  Optical Physics}\ }\textbf {\bibinfo {volume} {49}},\ \bibinfo {pages}
  {202001} (\bibinfo {year} {2016})}\BibitemShut {NoStop}%
\bibitem [{\citenamefont {Kaufman}\ and\ \citenamefont
  {Ni}(2021)}]{kaufman2021quantum}%
  \BibitemOpen
  \bibfield  {author} {\bibinfo {author} {\bibfnamefont {Adam~M}\ \bibnamefont
  {Kaufman}}\ and\ \bibinfo {author} {\bibfnamefont {Kang-Kuen}\ \bibnamefont
  {Ni}},\ }\bibfield  {title} {\enquote {\bibinfo {title} {Quantum science with
  optical tweezer arrays of ultracold atoms and molecules},}\ }\href@noop {}
  {\bibfield  {journal} {\bibinfo  {journal} {Nature Physics}\ }\textbf
  {\bibinfo {volume} {17}},\ \bibinfo {pages} {1324--1333} (\bibinfo {year}
  {2021})}\BibitemShut {NoStop}%
\bibitem [{\citenamefont {Wu}\ \emph {et~al.}(2022)\citenamefont {Wu},
  \citenamefont {Kolkowitz}, \citenamefont {Puri},\ and\ \citenamefont
  {Thompson}}]{wu2022erasure}%
  \BibitemOpen
  \bibfield  {author} {\bibinfo {author} {\bibfnamefont {Yue}\ \bibnamefont
  {Wu}}, \bibinfo {author} {\bibfnamefont {Shimon}\ \bibnamefont {Kolkowitz}},
  \bibinfo {author} {\bibfnamefont {Shruti}\ \bibnamefont {Puri}}, \ and\
  \bibinfo {author} {\bibfnamefont {Jeff~D}\ \bibnamefont {Thompson}},\
  }\bibfield  {title} {\enquote {\bibinfo {title} {Erasure conversion for
  fault-tolerant quantum computing in alkaline earth rydberg atom arrays},}\
  }\href@noop {} {\bibfield  {journal} {\bibinfo  {journal} {Nature
  communications}\ }\textbf {\bibinfo {volume} {13}},\ \bibinfo {pages} {4657}
  (\bibinfo {year} {2022})}\BibitemShut {NoStop}%
\bibitem [{\citenamefont {Cong}\ \emph {et~al.}(2022)\citenamefont {Cong},
  \citenamefont {Levine}, \citenamefont {Keesling}, \citenamefont {Bluvstein},
  \citenamefont {Wang},\ and\ \citenamefont {Lukin}}]{cong2022hardware}%
  \BibitemOpen
  \bibfield  {author} {\bibinfo {author} {\bibfnamefont {Iris}\ \bibnamefont
  {Cong}}, \bibinfo {author} {\bibfnamefont {Harry}\ \bibnamefont {Levine}},
  \bibinfo {author} {\bibfnamefont {Alexander}\ \bibnamefont {Keesling}},
  \bibinfo {author} {\bibfnamefont {Dolev}\ \bibnamefont {Bluvstein}}, \bibinfo
  {author} {\bibfnamefont {Sheng-Tao}\ \bibnamefont {Wang}}, \ and\ \bibinfo
  {author} {\bibfnamefont {Mikhail~D}\ \bibnamefont {Lukin}},\ }\bibfield
  {title} {\enquote {\bibinfo {title} {Hardware-efficient, fault-tolerant
  quantum computation with rydberg atoms},}\ }\href@noop {} {\bibfield
  {journal} {\bibinfo  {journal} {Physical Review X}\ }\textbf {\bibinfo
  {volume} {12}},\ \bibinfo {pages} {021049} (\bibinfo {year}
  {2022})}\BibitemShut {NoStop}%
\bibitem [{\citenamefont {Evered}\ \emph {et~al.}(2023)\citenamefont {Evered},
  \citenamefont {Bluvstein}, \citenamefont {Kalinowski}, \citenamefont {Ebadi},
  \citenamefont {Manovitz}, \citenamefont {Zhou}, \citenamefont {Li},
  \citenamefont {Geim}, \citenamefont {Wang}, \citenamefont {Maskara} \emph
  {et~al.}}]{evered2023high}%
  \BibitemOpen
  \bibfield  {author} {\bibinfo {author} {\bibfnamefont {Simon~J}\ \bibnamefont
  {Evered}}, \bibinfo {author} {\bibfnamefont {Dolev}\ \bibnamefont
  {Bluvstein}}, \bibinfo {author} {\bibfnamefont {Marcin}\ \bibnamefont
  {Kalinowski}}, \bibinfo {author} {\bibfnamefont {Sepehr}\ \bibnamefont
  {Ebadi}}, \bibinfo {author} {\bibfnamefont {Tom}\ \bibnamefont {Manovitz}},
  \bibinfo {author} {\bibfnamefont {Hengyun}\ \bibnamefont {Zhou}}, \bibinfo
  {author} {\bibfnamefont {Sophie~H}\ \bibnamefont {Li}}, \bibinfo {author}
  {\bibfnamefont {Alexandra~A}\ \bibnamefont {Geim}}, \bibinfo {author}
  {\bibfnamefont {Tout~T}\ \bibnamefont {Wang}}, \bibinfo {author}
  {\bibfnamefont {Nishad}\ \bibnamefont {Maskara}},  \emph {et~al.},\
  }\bibfield  {title} {\enquote {\bibinfo {title} {High-fidelity parallel
  entangling gates on a neutral atom quantum computer},}\ }\href@noop {}
  {\bibfield  {journal} {\bibinfo  {journal} {arXiv preprint arXiv:2304.05420}\
  } (\bibinfo {year} {2023})}\BibitemShut {NoStop}%
\bibitem [{\citenamefont {Ma}\ \emph {et~al.}(2023)\citenamefont {Ma},
  \citenamefont {Liu}, \citenamefont {Peng}, \citenamefont {Zhang},
  \citenamefont {Jandura}, \citenamefont {Claes}, \citenamefont {Burgers},
  \citenamefont {Pupillo}, \citenamefont {Puri},\ and\ \citenamefont
  {Thompson}}]{ma2023high}%
  \BibitemOpen
  \bibfield  {author} {\bibinfo {author} {\bibfnamefont {Shuo}\ \bibnamefont
  {Ma}}, \bibinfo {author} {\bibfnamefont {Genyue}\ \bibnamefont {Liu}},
  \bibinfo {author} {\bibfnamefont {Pai}\ \bibnamefont {Peng}}, \bibinfo
  {author} {\bibfnamefont {Bichen}\ \bibnamefont {Zhang}}, \bibinfo {author}
  {\bibfnamefont {Sven}\ \bibnamefont {Jandura}}, \bibinfo {author}
  {\bibfnamefont {Jahan}\ \bibnamefont {Claes}}, \bibinfo {author}
  {\bibfnamefont {Alex~P}\ \bibnamefont {Burgers}}, \bibinfo {author}
  {\bibfnamefont {Guido}\ \bibnamefont {Pupillo}}, \bibinfo {author}
  {\bibfnamefont {Shruti}\ \bibnamefont {Puri}}, \ and\ \bibinfo {author}
  {\bibfnamefont {Jeff~D}\ \bibnamefont {Thompson}},\ }\bibfield  {title}
  {\enquote {\bibinfo {title} {High-fidelity gates with mid-circuit erasure
  conversion in a metastable neutral atom qubit},}\ }\href@noop {} {\bibfield
  {journal} {\bibinfo  {journal} {arXiv preprint arXiv:2305.05493}\ } (\bibinfo
  {year} {2023})}\BibitemShut {NoStop}%
\bibitem [{\citenamefont {Cirac}\ and\ \citenamefont
  {Zoller}(1995)}]{PhysRevLett.74.4091}%
  \BibitemOpen
  \bibfield  {author} {\bibinfo {author} {\bibfnamefont {J.~I.}\ \bibnamefont
  {Cirac}}\ and\ \bibinfo {author} {\bibfnamefont {P.}~\bibnamefont {Zoller}},\
  }\bibfield  {title} {\enquote {\bibinfo {title} {Quantum computations with
  cold trapped ions},}\ }\href {\doibase 10.1103/PhysRevLett.74.4091}
  {\bibfield  {journal} {\bibinfo  {journal} {Phys. Rev. Lett.}\ }\textbf
  {\bibinfo {volume} {74}},\ \bibinfo {pages} {4091--4094} (\bibinfo {year}
  {1995})}\BibitemShut {NoStop}%
\bibitem [{\citenamefont {Bruzewicz}\ \emph {et~al.}(2019)\citenamefont
  {Bruzewicz}, \citenamefont {Chiaverini}, \citenamefont {McConnell},\ and\
  \citenamefont {Sage}}]{Bruzewicz_2019}%
  \BibitemOpen
  \bibfield  {author} {\bibinfo {author} {\bibfnamefont {Colin~D.}\
  \bibnamefont {Bruzewicz}}, \bibinfo {author} {\bibfnamefont {John}\
  \bibnamefont {Chiaverini}}, \bibinfo {author} {\bibfnamefont {Robert}\
  \bibnamefont {McConnell}}, \ and\ \bibinfo {author} {\bibfnamefont
  {Jeremy~M.}\ \bibnamefont {Sage}},\ }\bibfield  {title} {\enquote {\bibinfo
  {title} {Trapped-ion quantum computing: Progress and challenges},}\ }\href
  {\doibase 10.1063/1.5088164} {\bibfield  {journal} {\bibinfo  {journal}
  {Applied Physics Reviews}\ }\textbf {\bibinfo {volume} {6}},\ \bibinfo
  {pages} {021314} (\bibinfo {year} {2019})}\BibitemShut {NoStop}%
\bibitem [{\citenamefont {Kim}\ \emph {et~al.}(2010)\citenamefont {Kim},
  \citenamefont {Chang}, \citenamefont {Korenblit}, \citenamefont {Islam},
  \citenamefont {Edwards}, \citenamefont {Freericks}, \citenamefont {Lin},
  \citenamefont {Duan},\ and\ \citenamefont {Monroe}}]{kihwan_2010}%
  \BibitemOpen
  \bibfield  {author} {\bibinfo {author} {\bibfnamefont {Kihwan}\ \bibnamefont
  {Kim}}, \bibinfo {author} {\bibfnamefont {M-S}\ \bibnamefont {Chang}},
  \bibinfo {author} {\bibfnamefont {Simcha}\ \bibnamefont {Korenblit}},
  \bibinfo {author} {\bibfnamefont {R.}~\bibnamefont {Islam}}, \bibinfo
  {author} {\bibfnamefont {E}~\bibnamefont {Edwards}}, \bibinfo {author}
  {\bibfnamefont {J}~\bibnamefont {Freericks}}, \bibinfo {author}
  {\bibfnamefont {G-D}\ \bibnamefont {Lin}}, \bibinfo {author} {\bibfnamefont
  {L-M}\ \bibnamefont {Duan}}, \ and\ \bibinfo {author} {\bibfnamefont
  {C}~\bibnamefont {Monroe}},\ }\bibfield  {title} {\enquote {\bibinfo {title}
  {Quantum simulation of frustrated ising spins with trapped ions},}\ }\href
  {\doibase 10.1038/nature09071} {\bibfield  {journal} {\bibinfo  {journal}
  {Nature}\ }\textbf {\bibinfo {volume} {465}},\ \bibinfo {pages} {590--3}
  (\bibinfo {year} {2010})}\BibitemShut {NoStop}%
\bibitem [{\citenamefont {Britton}\ \emph {et~al.}(2012)\citenamefont
  {Britton}, \citenamefont {Sawyer}, \citenamefont {Keith}, \citenamefont
  {Wang}, \citenamefont {Freericks}, \citenamefont {Uys}, \citenamefont
  {Biercuk},\ and\ \citenamefont {Bollinger}}]{Britton_2012}%
  \BibitemOpen
  \bibfield  {author} {\bibinfo {author} {\bibfnamefont {Joseph~W.}\
  \bibnamefont {Britton}}, \bibinfo {author} {\bibfnamefont {Brian~C.}\
  \bibnamefont {Sawyer}}, \bibinfo {author} {\bibfnamefont {Adam~C.}\
  \bibnamefont {Keith}}, \bibinfo {author} {\bibfnamefont {C.-C.~Joseph}\
  \bibnamefont {Wang}}, \bibinfo {author} {\bibfnamefont {James~K.}\
  \bibnamefont {Freericks}}, \bibinfo {author} {\bibfnamefont {Hermann}\
  \bibnamefont {Uys}}, \bibinfo {author} {\bibfnamefont {Michael~J.}\
  \bibnamefont {Biercuk}}, \ and\ \bibinfo {author} {\bibfnamefont {John~J.}\
  \bibnamefont {Bollinger}},\ }\bibfield  {title} {\enquote {\bibinfo {title}
  {Engineered two-dimensional ising interactions in a trapped-ion quantum
  simulator with hundreds of spins},}\ }\href {\doibase 10.1038/nature10981}
  {\bibfield  {journal} {\bibinfo  {journal} {Nature}\ }\textbf {\bibinfo
  {volume} {484}},\ \bibinfo {pages} {489--492} (\bibinfo {year}
  {2012})}\BibitemShut {NoStop}%
\bibitem [{\citenamefont {Barreiro}\ \emph {et~al.}(2011)\citenamefont
  {Barreiro}, \citenamefont {Müller}, \citenamefont {Schindler}, \citenamefont
  {Nigg}, \citenamefont {Monz}, \citenamefont {Chwalla}, \citenamefont
  {Hennrich}, \citenamefont {Roos}, \citenamefont {Zoller},\ and\ \citenamefont
  {Blatt}}]{Barreiro_2011}%
  \BibitemOpen
  \bibfield  {author} {\bibinfo {author} {\bibfnamefont {Julio~T.}\
  \bibnamefont {Barreiro}}, \bibinfo {author} {\bibfnamefont {Markus}\
  \bibnamefont {Müller}}, \bibinfo {author} {\bibfnamefont {Philipp}\
  \bibnamefont {Schindler}}, \bibinfo {author} {\bibfnamefont {Daniel}\
  \bibnamefont {Nigg}}, \bibinfo {author} {\bibfnamefont {Thomas}\ \bibnamefont
  {Monz}}, \bibinfo {author} {\bibfnamefont {Michael}\ \bibnamefont {Chwalla}},
  \bibinfo {author} {\bibfnamefont {Markus}\ \bibnamefont {Hennrich}}, \bibinfo
  {author} {\bibfnamefont {Christian~F.}\ \bibnamefont {Roos}}, \bibinfo
  {author} {\bibfnamefont {Peter}\ \bibnamefont {Zoller}}, \ and\ \bibinfo
  {author} {\bibfnamefont {Rainer}\ \bibnamefont {Blatt}},\ }\bibfield  {title}
  {\enquote {\bibinfo {title} {An open-system quantum simulator with trapped
  ions},}\ }\href {\doibase 10.1038/nature09801} {\bibfield  {journal}
  {\bibinfo  {journal} {Nature}\ }\textbf {\bibinfo {volume} {470}},\ \bibinfo
  {pages} {486--491} (\bibinfo {year} {2011})}\BibitemShut {NoStop}%
\bibitem [{\citenamefont {Martinis}\ \emph {et~al.}(2002)\citenamefont
  {Martinis}, \citenamefont {Nam}, \citenamefont {Aumentado},\ and\
  \citenamefont {Urbina}}]{PhysRevLett.89.117901}%
  \BibitemOpen
  \bibfield  {author} {\bibinfo {author} {\bibfnamefont {John~M.}\ \bibnamefont
  {Martinis}}, \bibinfo {author} {\bibfnamefont {S.}~\bibnamefont {Nam}},
  \bibinfo {author} {\bibfnamefont {J.}~\bibnamefont {Aumentado}}, \ and\
  \bibinfo {author} {\bibfnamefont {C.}~\bibnamefont {Urbina}},\ }\bibfield
  {title} {\enquote {\bibinfo {title} {Rabi oscillations in a large
  josephson-junction qubit},}\ }\href {\doibase 10.1103/PhysRevLett.89.117901}
  {\bibfield  {journal} {\bibinfo  {journal} {Phys. Rev. Lett.}\ }\textbf
  {\bibinfo {volume} {89}},\ \bibinfo {pages} {117901} (\bibinfo {year}
  {2002})}\BibitemShut {NoStop}%
\bibitem [{\citenamefont {Nakamura}\ \emph {et~al.}(1997)\citenamefont
  {Nakamura}, \citenamefont {Chen},\ and\ \citenamefont
  {Tsai}}]{PhysRevLett.79.2328}%
  \BibitemOpen
  \bibfield  {author} {\bibinfo {author} {\bibfnamefont {Y.}~\bibnamefont
  {Nakamura}}, \bibinfo {author} {\bibfnamefont {C.~D.}\ \bibnamefont {Chen}},
  \ and\ \bibinfo {author} {\bibfnamefont {J.~S.}\ \bibnamefont {Tsai}},\
  }\bibfield  {title} {\enquote {\bibinfo {title} {Spectroscopy of energy-level
  splitting between two macroscopic quantum states of charge coherently
  superposed by josephson coupling},}\ }\href {\doibase
  10.1103/PhysRevLett.79.2328} {\bibfield  {journal} {\bibinfo  {journal}
  {Phys. Rev. Lett.}\ }\textbf {\bibinfo {volume} {79}},\ \bibinfo {pages}
  {2328--2331} (\bibinfo {year} {1997})}\BibitemShut {NoStop}%
\bibitem [{\citenamefont {Krantz}\ \emph {et~al.}(2019)\citenamefont {Krantz},
  \citenamefont {Kjaergaard}, \citenamefont {Yan}, \citenamefont {Orlando},
  \citenamefont {Gustavsson},\ and\ \citenamefont {Oliver}}]{Krantz_2019}%
  \BibitemOpen
  \bibfield  {author} {\bibinfo {author} {\bibfnamefont {P.}~\bibnamefont
  {Krantz}}, \bibinfo {author} {\bibfnamefont {M.}~\bibnamefont {Kjaergaard}},
  \bibinfo {author} {\bibfnamefont {F.}~\bibnamefont {Yan}}, \bibinfo {author}
  {\bibfnamefont {T.~P.}\ \bibnamefont {Orlando}}, \bibinfo {author}
  {\bibfnamefont {S.}~\bibnamefont {Gustavsson}}, \ and\ \bibinfo {author}
  {\bibfnamefont {W.~D.}\ \bibnamefont {Oliver}},\ }\bibfield  {title}
  {\enquote {\bibinfo {title} {A quantum engineer{\textquotesingle}s guide to
  superconducting qubits},}\ }\href {\doibase 10.1063/1.5089550} {\bibfield
  {journal} {\bibinfo  {journal} {Applied Physics Reviews}\ }\textbf {\bibinfo
  {volume} {6}},\ \bibinfo {pages} {021318} (\bibinfo {year}
  {2019})}\BibitemShut {NoStop}%
\bibitem [{\citenamefont {Brooks}\ \emph {et~al.}(2013)\citenamefont {Brooks},
  \citenamefont {Kitaev},\ and\ \citenamefont {Preskill}}]{Brooks_2013}%
  \BibitemOpen
  \bibfield  {author} {\bibinfo {author} {\bibfnamefont {Peter}\ \bibnamefont
  {Brooks}}, \bibinfo {author} {\bibfnamefont {Alexei}\ \bibnamefont {Kitaev}},
  \ and\ \bibinfo {author} {\bibfnamefont {John}\ \bibnamefont {Preskill}},\
  }\bibfield  {title} {\enquote {\bibinfo {title} {Protected gates for
  superconducting qubits},}\ }\href {\doibase 10.1103/physreva.87.052306}
  {\bibfield  {journal} {\bibinfo  {journal} {Physical Review A}\ }\textbf
  {\bibinfo {volume} {87}} (\bibinfo {year} {2013}),\
  10.1103/physreva.87.052306}\BibitemShut {NoStop}%
\bibitem [{\citenamefont {Gu}\ \emph {et~al.}(2017)\citenamefont {Gu},
  \citenamefont {Kockum}, \citenamefont {Miranowicz}, \citenamefont {xi~Liu},\
  and\ \citenamefont {Nori}}]{Gu_2017}%
  \BibitemOpen
  \bibfield  {author} {\bibinfo {author} {\bibfnamefont {Xiu}\ \bibnamefont
  {Gu}}, \bibinfo {author} {\bibfnamefont {Anton~Frisk}\ \bibnamefont
  {Kockum}}, \bibinfo {author} {\bibfnamefont {Adam}\ \bibnamefont
  {Miranowicz}}, \bibinfo {author} {\bibfnamefont {Yu}~\bibnamefont {xi~Liu}},
  \ and\ \bibinfo {author} {\bibfnamefont {Franco}\ \bibnamefont {Nori}},\
  }\bibfield  {title} {\enquote {\bibinfo {title} {Microwave photonics with
  superconducting quantum circuits},}\ }\href {\doibase
  10.1016/j.physrep.2017.10.002} {\bibfield  {journal} {\bibinfo  {journal}
  {Physics Reports}\ }\textbf {\bibinfo {volume} {718-719}},\ \bibinfo {pages}
  {1--102} (\bibinfo {year} {2017})}\BibitemShut {NoStop}%
\bibitem [{\citenamefont {Gidney}\ and\ \citenamefont
  {Ekerå}(2021)}]{Gidney_2021}%
  \BibitemOpen
  \bibfield  {author} {\bibinfo {author} {\bibfnamefont {Craig}\ \bibnamefont
  {Gidney}}\ and\ \bibinfo {author} {\bibfnamefont {Martin}\ \bibnamefont
  {Ekerå}},\ }\bibfield  {title} {\enquote {\bibinfo {title} {How to factor
  2048 bit rsa integers in 8 hours using 20 million noisy qubits},}\ }\href
  {\doibase 10.22331/q-2021-04-15-433} {\bibfield  {journal} {\bibinfo
  {journal} {Quantum}\ }\textbf {\bibinfo {volume} {5}},\ \bibinfo {pages}
  {433} (\bibinfo {year} {2021})}\BibitemShut {NoStop}%
\bibitem [{\citenamefont {Tillich}\ and\ \citenamefont {Zemor}(2014)}]{HGP}%
  \BibitemOpen
  \bibfield  {author} {\bibinfo {author} {\bibfnamefont {Jean-Pierre}\
  \bibnamefont {Tillich}}\ and\ \bibinfo {author} {\bibfnamefont {Gilles}\
  \bibnamefont {Zemor}},\ }\bibfield  {title} {\enquote {\bibinfo {title}
  {Quantum {LDPC} codes with positive rate and minimum distance proportional to
  the square root of the blocklength},}\ }\href {\doibase
  10.1109/tit.2013.2292061} {\bibfield  {journal} {\bibinfo  {journal} {{IEEE}
  Transactions on Information Theory}\ }\textbf {\bibinfo {volume} {60}},\
  \bibinfo {pages} {1193--1202} (\bibinfo {year} {2014})}\BibitemShut {NoStop}%
\bibitem [{\citenamefont {Hastings}\ \emph {et~al.}(2021)\citenamefont
  {Hastings}, \citenamefont {Haah},\ and\ \citenamefont
  {O'Donnell}}]{TP_codes}%
  \BibitemOpen
  \bibfield  {author} {\bibinfo {author} {\bibfnamefont {Matthew~B.}\
  \bibnamefont {Hastings}}, \bibinfo {author} {\bibfnamefont {Jeongwan}\
  \bibnamefont {Haah}}, \ and\ \bibinfo {author} {\bibfnamefont {Ryan}\
  \bibnamefont {O'Donnell}},\ }\bibfield  {title} {\enquote {\bibinfo {title}
  {Fiber bundle codes: Breaking the $\sqrt{n}$ polylog($n$) barrier for quantum
  ldpc codes},}\ }in\ \href {\doibase 10.1145/3406325.3451005} {\emph {\bibinfo
  {booktitle} {Proceedings of the 53rd Annual ACM SIGACT Symposium on Theory of
  Computing}}},\ \bibinfo {series and number} {STOC 2021}\ (\bibinfo
  {publisher} {Association for Computing Machinery},\ \bibinfo {address} {New
  York, NY, USA},\ \bibinfo {year} {2021})\ p.\ \bibinfo {pages}
  {1276–1288}\BibitemShut {NoStop}%
\bibitem [{\citenamefont {Breuckmann}\ and\ \citenamefont
  {Eberhardt}(2021)}]{BP_codes}%
  \BibitemOpen
  \bibfield  {author} {\bibinfo {author} {\bibfnamefont {Nikolas~P.}\
  \bibnamefont {Breuckmann}}\ and\ \bibinfo {author} {\bibfnamefont {Jens~N.}\
  \bibnamefont {Eberhardt}},\ }\bibfield  {title} {\enquote {\bibinfo {title}
  {Balanced product quantum codes},}\ }\href {\doibase
  10.1109/tit.2021.3097347} {\bibfield  {journal} {\bibinfo  {journal} {{IEEE}
  Transactions on Information Theory}\ }\textbf {\bibinfo {volume} {67}},\
  \bibinfo {pages} {6653--6674} (\bibinfo {year} {2021})}\BibitemShut {NoStop}%
\bibitem [{\citenamefont {Panteleev}\ and\ \citenamefont
  {Kalachev}(2022{\natexlab{a}})}]{LP_codes}%
  \BibitemOpen
  \bibfield  {author} {\bibinfo {author} {\bibfnamefont {Pavel}\ \bibnamefont
  {Panteleev}}\ and\ \bibinfo {author} {\bibfnamefont {Gleb}\ \bibnamefont
  {Kalachev}},\ }\bibfield  {title} {\enquote {\bibinfo {title} {Quantum ldpc
  codes with almost linear minimum distance},}\ }\href {\doibase
  10.1109/TIT.2021.3119384} {\bibfield  {journal} {\bibinfo  {journal} {IEEE
  Transactions on Information Theory}\ }\textbf {\bibinfo {volume} {68}},\
  \bibinfo {pages} {213--229} (\bibinfo {year}
  {2022}{\natexlab{a}})}\BibitemShut {NoStop}%
\bibitem [{\citenamefont {Panteleev}\ and\ \citenamefont
  {Kalachev}(2022{\natexlab{b}})}]{Panteleev_2022}%
  \BibitemOpen
  \bibfield  {author} {\bibinfo {author} {\bibfnamefont {Pavel}\ \bibnamefont
  {Panteleev}}\ and\ \bibinfo {author} {\bibfnamefont {Gleb}\ \bibnamefont
  {Kalachev}},\ }\bibfield  {title} {\enquote {\bibinfo {title} {Asymptotically
  good quantum and locally testable classical ldpc codes},}\ }in\ \href
  {\doibase 10.1145/3519935.3520017} {\emph {\bibinfo {booktitle} {Proceedings
  of the 54th Annual ACM SIGACT Symposium on Theory of Computing}}},\ \bibinfo
  {series and number} {STOC 2022}\ (\bibinfo  {publisher} {Association for
  Computing Machinery},\ \bibinfo {address} {New York, NY, USA},\ \bibinfo
  {year} {2022})\ p.\ \bibinfo {pages} {375–388}\BibitemShut {NoStop}%
\bibitem [{\citenamefont {Leverrier}\ and\ \citenamefont
  {Zémor}(2022{\natexlab{a}})}]{qTanner_codes}%
  \BibitemOpen
  \bibfield  {author} {\bibinfo {author} {\bibfnamefont {Anthony}\ \bibnamefont
  {Leverrier}}\ and\ \bibinfo {author} {\bibfnamefont {Gilles}\ \bibnamefont
  {Zémor}},\ }\href@noop {} {\enquote {\bibinfo {title} {Quantum tanner
  codes},}\ } (\bibinfo {year} {2022}{\natexlab{a}}),\ \Eprint
  {http://arxiv.org/abs/2202.13641} {arXiv:2202.13641 [quant-ph]} \BibitemShut
  {NoStop}%
\bibitem [{\citenamefont {Dinur}\ \emph {et~al.}(2022)\citenamefont {Dinur},
  \citenamefont {Hsieh}, \citenamefont {Lin},\ and\ \citenamefont
  {Vidick}}]{dinur2022good}%
  \BibitemOpen
  \bibfield  {author} {\bibinfo {author} {\bibfnamefont {Irit}\ \bibnamefont
  {Dinur}}, \bibinfo {author} {\bibfnamefont {Min-Hsiu}\ \bibnamefont {Hsieh}},
  \bibinfo {author} {\bibfnamefont {Ting-Chun}\ \bibnamefont {Lin}}, \ and\
  \bibinfo {author} {\bibfnamefont {Thomas}\ \bibnamefont {Vidick}},\
  }\href@noop {} {\enquote {\bibinfo {title} {Good quantum ldpc codes with
  linear time decoders},}\ } (\bibinfo {year} {2022}),\ \Eprint
  {http://arxiv.org/abs/2206.07750} {arXiv:2206.07750 [quant-ph]} \BibitemShut
  {NoStop}%
\bibitem [{\citenamefont {Bravyi}\ \emph {et~al.}(2010)\citenamefont {Bravyi},
  \citenamefont {Poulin},\ and\ \citenamefont {Terhal}}]{BPT_2010}%
  \BibitemOpen
  \bibfield  {author} {\bibinfo {author} {\bibfnamefont {Sergey}\ \bibnamefont
  {Bravyi}}, \bibinfo {author} {\bibfnamefont {David}\ \bibnamefont {Poulin}},
  \ and\ \bibinfo {author} {\bibfnamefont {Barbara}\ \bibnamefont {Terhal}},\
  }\bibfield  {title} {\enquote {\bibinfo {title} {Tradeoffs for reliable
  quantum information storage in 2d systems},}\ }\href {\doibase
  10.1103/PhysRevLett.104.050503} {\bibfield  {journal} {\bibinfo  {journal}
  {Phys. Rev. Lett.}\ }\textbf {\bibinfo {volume} {104}},\ \bibinfo {pages}
  {050503} (\bibinfo {year} {2010})}\BibitemShut {NoStop}%
\bibitem [{\citenamefont {Baspin}\ and\ \citenamefont
  {Krishna}(2022)}]{Baspin_2022}%
  \BibitemOpen
  \bibfield  {author} {\bibinfo {author} {\bibfnamefont {Nou\'edyn}\
  \bibnamefont {Baspin}}\ and\ \bibinfo {author} {\bibfnamefont {Anirudh}\
  \bibnamefont {Krishna}},\ }\bibfield  {title} {\enquote {\bibinfo {title}
  {Quantifying nonlocality: How outperforming local quantum codes is
  expensive},}\ }\href {\doibase 10.1103/PhysRevLett.129.050505} {\bibfield
  {journal} {\bibinfo  {journal} {Phys. Rev. Lett.}\ }\textbf {\bibinfo
  {volume} {129}},\ \bibinfo {pages} {050505} (\bibinfo {year}
  {2022})}\BibitemShut {NoStop}%
\bibitem [{\citenamefont {Delfosse}\ \emph {et~al.}(2021)\citenamefont
  {Delfosse}, \citenamefont {Beverland},\ and\ \citenamefont
  {Tremblay}}]{delfosse2021}%
  \BibitemOpen
  \bibfield  {author} {\bibinfo {author} {\bibfnamefont {Nicolas}\ \bibnamefont
  {Delfosse}}, \bibinfo {author} {\bibfnamefont {Michael~E.}\ \bibnamefont
  {Beverland}}, \ and\ \bibinfo {author} {\bibfnamefont {Maxime~A.}\
  \bibnamefont {Tremblay}},\ }\href@noop {} {\enquote {\bibinfo {title} {Bounds
  on stabilizer measurement circuits and obstructions to local implementations
  of quantum ldpc codes},}\ } (\bibinfo {year} {2021}),\ \Eprint
  {http://arxiv.org/abs/2109.14599} {arXiv:2109.14599 [quant-ph]} \BibitemShut
  {NoStop}%
\bibitem [{\citenamefont {Baspin}\ \emph {et~al.}(2023)\citenamefont {Baspin},
  \citenamefont {Guruswami}, \citenamefont {Krishna},\ and\ \citenamefont
  {Li}}]{baspin2023improved}%
  \BibitemOpen
  \bibfield  {author} {\bibinfo {author} {\bibfnamefont {Nouédyn}\
  \bibnamefont {Baspin}}, \bibinfo {author} {\bibfnamefont {Venkatesan}\
  \bibnamefont {Guruswami}}, \bibinfo {author} {\bibfnamefont {Anirudh}\
  \bibnamefont {Krishna}}, \ and\ \bibinfo {author} {\bibfnamefont {Ray}\
  \bibnamefont {Li}},\ }\href@noop {} {\enquote {\bibinfo {title} {Improved
  rate-distance trade-offs for quantum codes with restricted connectivity},}\ }
  (\bibinfo {year} {2023}),\ \Eprint {http://arxiv.org/abs/2307.03283}
  {arXiv:2307.03283 [quant-ph]} \BibitemShut {NoStop}%
\bibitem [{\citenamefont {Pattison}\ \emph {et~al.}(2023)\citenamefont
  {Pattison}, \citenamefont {Krishna},\ and\ \citenamefont
  {Preskill}}]{pattison2023}%
  \BibitemOpen
  \bibfield  {author} {\bibinfo {author} {\bibfnamefont {Christopher~A.}\
  \bibnamefont {Pattison}}, \bibinfo {author} {\bibfnamefont {Anirudh}\
  \bibnamefont {Krishna}}, \ and\ \bibinfo {author} {\bibfnamefont {John}\
  \bibnamefont {Preskill}},\ }\href@noop {} {\enquote {\bibinfo {title}
  {Hierarchical memories: Simulating quantum ldpc codes with local gates},}\ }
  (\bibinfo {year} {2023}),\ \Eprint {http://arxiv.org/abs/2303.04798}
  {arXiv:2303.04798 [quant-ph]} \BibitemShut {NoStop}%
\bibitem [{\citenamefont {Gallager}(1962)}]{gallager_1962}%
  \BibitemOpen
  \bibfield  {author} {\bibinfo {author} {\bibfnamefont {R.}~\bibnamefont
  {Gallager}},\ }\bibfield  {title} {\enquote {\bibinfo {title} {Low-density
  parity-check codes},}\ }\href {\doibase 10.1109/TIT.1962.1057683} {\bibfield
  {journal} {\bibinfo  {journal} {IRE Transactions on Information Theory}\
  }\textbf {\bibinfo {volume} {8}},\ \bibinfo {pages} {21--28} (\bibinfo {year}
  {1962})}\BibitemShut {NoStop}%
\bibitem [{\citenamefont {{MacKay}}\ and\ \citenamefont
  {{Neal}}(1996)}]{random_ldpc}%
  \BibitemOpen
  \bibfield  {author} {\bibinfo {author} {\bibfnamefont {D.~J.~C.}\
  \bibnamefont {{MacKay}}}\ and\ \bibinfo {author} {\bibfnamefont {R.~M.}\
  \bibnamefont {{Neal}}},\ }\bibfield  {title} {\enquote {\bibinfo {title}
  {{Near Shannon limit performance of low density parity check codes}},}\
  }\href {\doibase 10.1049/el:19961141} {\bibfield  {journal} {\bibinfo
  {journal} {Electronics Letters}\ }\textbf {\bibinfo {volume} {32}},\ \bibinfo
  {pages} {1645} (\bibinfo {year} {1996})}\BibitemShut {NoStop}%
\bibitem [{\citenamefont {Bravyi}\ and\ \citenamefont
  {Terhal}(2009)}]{Bravyi_2009}%
  \BibitemOpen
  \bibfield  {author} {\bibinfo {author} {\bibfnamefont {Sergey}\ \bibnamefont
  {Bravyi}}\ and\ \bibinfo {author} {\bibfnamefont {Barbara}\ \bibnamefont
  {Terhal}},\ }\bibfield  {title} {\enquote {\bibinfo {title} {A no-go theorem
  for a two-dimensional self-correcting quantum memory based on stabilizer
  codes},}\ }\href {\doibase 10.1088/1367-2630/11/4/043029} {\bibfield
  {journal} {\bibinfo  {journal} {New Journal of Physics}\ }\textbf {\bibinfo
  {volume} {11}},\ \bibinfo {pages} {043029} (\bibinfo {year}
  {2009})}\BibitemShut {NoStop}%
\bibitem [{\citenamefont {Dennis}\ \emph {et~al.}(2002)\citenamefont {Dennis},
  \citenamefont {Kitaev}, \citenamefont {Landahl},\ and\ \citenamefont
  {Preskill}}]{Dennis_2002}%
  \BibitemOpen
  \bibfield  {author} {\bibinfo {author} {\bibfnamefont {Eric}\ \bibnamefont
  {Dennis}}, \bibinfo {author} {\bibfnamefont {Alexei}\ \bibnamefont {Kitaev}},
  \bibinfo {author} {\bibfnamefont {Andrew}\ \bibnamefont {Landahl}}, \ and\
  \bibinfo {author} {\bibfnamefont {John}\ \bibnamefont {Preskill}},\
  }\bibfield  {title} {\enquote {\bibinfo {title} {Topological quantum
  memory},}\ }\href {\doibase 10.1063/1.1499754} {\bibfield  {journal}
  {\bibinfo  {journal} {Journal of Mathematical Physics}\ }\textbf {\bibinfo
  {volume} {43}},\ \bibinfo {pages} {4452--4505} (\bibinfo {year}
  {2002})}\BibitemShut {NoStop}%
\bibitem [{\citenamefont {Fossorier}\ and\ \citenamefont
  {Lin}(1995)}]{Fossorier_1995}%
  \BibitemOpen
  \bibfield  {author} {\bibinfo {author} {\bibfnamefont {M.P.C.}\ \bibnamefont
  {Fossorier}}\ and\ \bibinfo {author} {\bibfnamefont {Shu}\ \bibnamefont
  {Lin}},\ }\bibfield  {title} {\enquote {\bibinfo {title} {Soft-decision
  decoding of linear block codes based on ordered statistics},}\ }\href
  {\doibase 10.1109/18.412683} {\bibfield  {journal} {\bibinfo  {journal} {IEEE
  Transactions on Information Theory}\ }\textbf {\bibinfo {volume} {41}},\
  \bibinfo {pages} {1379--1396} (\bibinfo {year} {1995})}\BibitemShut {NoStop}%
\bibitem [{\citenamefont {Panteleev}\ and\ \citenamefont
  {Kalachev}(2021)}]{Panteleev_2021}%
  \BibitemOpen
  \bibfield  {author} {\bibinfo {author} {\bibfnamefont {Pavel}\ \bibnamefont
  {Panteleev}}\ and\ \bibinfo {author} {\bibfnamefont {Gleb}\ \bibnamefont
  {Kalachev}},\ }\bibfield  {title} {\enquote {\bibinfo {title} {Degenerate
  quantum {LDPC} codes with good finite length performance},}\ }\href {\doibase
  10.22331/q-2021-11-22-585} {\bibfield  {journal} {\bibinfo  {journal}
  {Quantum}\ }\textbf {\bibinfo {volume} {5}},\ \bibinfo {pages} {585}
  (\bibinfo {year} {2021})}\BibitemShut {NoStop}%
\bibitem [{\citenamefont {Roffe}(2022)}]{Roffe_LDPC_Python_tools_2022}%
  \BibitemOpen
  \bibfield  {author} {\bibinfo {author} {\bibfnamefont {Joschka}\ \bibnamefont
  {Roffe}},\ }\href {https://pypi.org/project/ldpc/} {\enquote {\bibinfo
  {title} {{LDPC: Python tools for low density parity check codes}},}\ }
  (\bibinfo {year} {2022})\BibitemShut {NoStop}%
\bibitem [{\citenamefont {Kuo}\ \emph {et~al.}(2021)\citenamefont {Kuo},
  \citenamefont {Chern},\ and\ \citenamefont {Lai}}]{Kuo_2021}%
  \BibitemOpen
  \bibfield  {author} {\bibinfo {author} {\bibfnamefont {Kao-Yueh}\
  \bibnamefont {Kuo}}, \bibinfo {author} {\bibfnamefont {I-Chun}\ \bibnamefont
  {Chern}}, \ and\ \bibinfo {author} {\bibfnamefont {Ching-Yi}\ \bibnamefont
  {Lai}},\ }\bibfield  {title} {\enquote {\bibinfo {title} {Decoding of quantum
  data-syndrome codes via belief propagation},}\ }in\ \href {\doibase
  10.1109/isit45174.2021.9518018} {\emph {\bibinfo {booktitle} {2021 {IEEE}
  International Symposium on Information Theory ({ISIT})}}}\ (\bibinfo
  {publisher} {{IEEE}},\ \bibinfo {year} {2021})\BibitemShut {NoStop}%
\bibitem [{\citenamefont {Higgott}\ and\ \citenamefont
  {Breuckmann}(2023)}]{PRXQuantum.4.020332}%
  \BibitemOpen
  \bibfield  {author} {\bibinfo {author} {\bibfnamefont {Oscar}\ \bibnamefont
  {Higgott}}\ and\ \bibinfo {author} {\bibfnamefont {Nikolas~P.}\ \bibnamefont
  {Breuckmann}},\ }\bibfield  {title} {\enquote {\bibinfo {title} {Improved
  single-shot decoding of higher-dimensional hypergraph-product codes},}\
  }\href {\doibase 10.1103/PRXQuantum.4.020332} {\bibfield  {journal} {\bibinfo
   {journal} {PRX Quantum}\ }\textbf {\bibinfo {volume} {4}},\ \bibinfo {pages}
  {020332} (\bibinfo {year} {2023})}\BibitemShut {NoStop}%
\bibitem [{\citenamefont {Huang}\ and\ \citenamefont
  {Puri}(2023)}]{huang2023improved}%
  \BibitemOpen
  \bibfield  {author} {\bibinfo {author} {\bibfnamefont {Shilin}\ \bibnamefont
  {Huang}}\ and\ \bibinfo {author} {\bibfnamefont {Shruti}\ \bibnamefont
  {Puri}},\ }\href@noop {} {\enquote {\bibinfo {title} {Improved noisy syndrome
  decoding of quantum ldpc codes with sliding window},}\ } (\bibinfo {year}
  {2023}),\ \Eprint {http://arxiv.org/abs/2311.03307} {arXiv:2311.03307
  [quant-ph]} \BibitemShut {NoStop}%
\bibitem [{\citenamefont {Higgott}\ and\ \citenamefont
  {Gidney}(2023)}]{pymatching}%
  \BibitemOpen
  \bibfield  {author} {\bibinfo {author} {\bibfnamefont {Oscar}\ \bibnamefont
  {Higgott}}\ and\ \bibinfo {author} {\bibfnamefont {Craig}\ \bibnamefont
  {Gidney}},\ }\bibfield  {title} {\enquote {\bibinfo {title} {Sparse blossom:
  correcting a million errors per core second with minimum-weight matching},}\
  }\href@noop {} {\bibfield  {journal} {\bibinfo  {journal} {arXiv preprint
  arXiv:2303.15933}\ } (\bibinfo {year} {2023})}\BibitemShut {NoStop}%
\bibitem [{\citenamefont {Leu}\ \emph {et~al.}(2023)\citenamefont {Leu},
  \citenamefont {Gely}, \citenamefont {Weber}, \citenamefont {Smith},
  \citenamefont {Nadlinger},\ and\ \citenamefont {Lucas}}]{leu_fast_2023}%
  \BibitemOpen
  \bibfield  {author} {\bibinfo {author} {\bibfnamefont {AD}~\bibnamefont
  {Leu}}, \bibinfo {author} {\bibfnamefont {MF}~\bibnamefont {Gely}}, \bibinfo
  {author} {\bibfnamefont {MA}~\bibnamefont {Weber}}, \bibinfo {author}
  {\bibfnamefont {MC}~\bibnamefont {Smith}}, \bibinfo {author} {\bibfnamefont
  {DP}~\bibnamefont {Nadlinger}}, \ and\ \bibinfo {author} {\bibfnamefont
  {DM}~\bibnamefont {Lucas}},\ }\bibfield  {title} {\enquote {\bibinfo {title}
  {Fast, high-fidelity addressed single-qubit gates using efficient composite
  pulse sequences},}\ }\href@noop {} {\bibfield  {journal} {\bibinfo  {journal}
  {arXiv preprint arXiv:2305.06725}\ } (\bibinfo {year} {2023})}\BibitemShut
  {NoStop}%
\bibitem [{\citenamefont {Ding}\ \emph {et~al.}(2023)\citenamefont {Ding},
  \citenamefont {Hays}, \citenamefont {Sung}, \citenamefont {Kannan},
  \citenamefont {An}, \citenamefont {Di~Paolo}, \citenamefont {Karamlou},
  \citenamefont {Hazard}, \citenamefont {Azar}, \citenamefont {Kim} \emph
  {et~al.}}]{ding2023high}%
  \BibitemOpen
  \bibfield  {author} {\bibinfo {author} {\bibfnamefont {Leon}\ \bibnamefont
  {Ding}}, \bibinfo {author} {\bibfnamefont {Max}\ \bibnamefont {Hays}},
  \bibinfo {author} {\bibfnamefont {Youngkyu}\ \bibnamefont {Sung}}, \bibinfo
  {author} {\bibfnamefont {Bharath}\ \bibnamefont {Kannan}}, \bibinfo {author}
  {\bibfnamefont {Junyoung}\ \bibnamefont {An}}, \bibinfo {author}
  {\bibfnamefont {Agustin}\ \bibnamefont {Di~Paolo}}, \bibinfo {author}
  {\bibfnamefont {Amir~H}\ \bibnamefont {Karamlou}}, \bibinfo {author}
  {\bibfnamefont {Thomas~M}\ \bibnamefont {Hazard}}, \bibinfo {author}
  {\bibfnamefont {Kate}\ \bibnamefont {Azar}}, \bibinfo {author} {\bibfnamefont
  {David~K}\ \bibnamefont {Kim}},  \emph {et~al.},\ }\bibfield  {title}
  {\enquote {\bibinfo {title} {High-fidelity, frequency-flexible two-qubit
  fluxonium gates with a transmon coupler},}\ }\href@noop {} {\bibfield
  {journal} {\bibinfo  {journal} {arXiv preprint arXiv:2304.06087}\ } (\bibinfo
  {year} {2023})}\BibitemShut {NoStop}%
\bibitem [{\citenamefont {Zhang}\ \emph {et~al.}(2023)\citenamefont {Zhang},
  \citenamefont {Ding}, \citenamefont {Weiss}, \citenamefont {Huang},
  \citenamefont {Ma}, \citenamefont {Guinn}, \citenamefont {Sussman},
  \citenamefont {Chitta}, \citenamefont {Chen}, \citenamefont {Houck} \emph
  {et~al.}}]{zhang2023tunable}%
  \BibitemOpen
  \bibfield  {author} {\bibinfo {author} {\bibfnamefont {Helin}\ \bibnamefont
  {Zhang}}, \bibinfo {author} {\bibfnamefont {Chunyang}\ \bibnamefont {Ding}},
  \bibinfo {author} {\bibfnamefont {DK}~\bibnamefont {Weiss}}, \bibinfo
  {author} {\bibfnamefont {Ziwen}\ \bibnamefont {Huang}}, \bibinfo {author}
  {\bibfnamefont {Yuwei}\ \bibnamefont {Ma}}, \bibinfo {author} {\bibfnamefont
  {Charles}\ \bibnamefont {Guinn}}, \bibinfo {author} {\bibfnamefont {Sara}\
  \bibnamefont {Sussman}}, \bibinfo {author} {\bibfnamefont {Sai~Pavan}\
  \bibnamefont {Chitta}}, \bibinfo {author} {\bibfnamefont {Danyang}\
  \bibnamefont {Chen}}, \bibinfo {author} {\bibfnamefont {Andrew~A}\
  \bibnamefont {Houck}},  \emph {et~al.},\ }\bibfield  {title} {\enquote
  {\bibinfo {title} {Tunable inductive coupler for high fidelity gates between
  fluxonium qubits},}\ }\href@noop {} {\bibfield  {journal} {\bibinfo
  {journal} {arXiv preprint arXiv:2309.05720}\ } (\bibinfo {year}
  {2023})}\BibitemShut {NoStop}%
\bibitem [{\citenamefont {Cohen}\ \emph {et~al.}(2022)\citenamefont {Cohen},
  \citenamefont {Kim}, \citenamefont {Bartlett},\ and\ \citenamefont
  {Brown}}]{Cohen_2022}%
  \BibitemOpen
  \bibfield  {author} {\bibinfo {author} {\bibfnamefont {Lawrence~Z.}\
  \bibnamefont {Cohen}}, \bibinfo {author} {\bibfnamefont {Isaac~H.}\
  \bibnamefont {Kim}}, \bibinfo {author} {\bibfnamefont {Stephen~D.}\
  \bibnamefont {Bartlett}}, \ and\ \bibinfo {author} {\bibfnamefont
  {Benjamin~J.}\ \bibnamefont {Brown}},\ }\bibfield  {title} {\enquote
  {\bibinfo {title} {Low-overhead fault-tolerant quantum computing using
  long-range connectivity},}\ }\href {\doibase 10.1126/sciadv.abn1717}
  {\bibfield  {journal} {\bibinfo  {journal} {Science Advances}\ }\textbf
  {\bibinfo {volume} {8}} (\bibinfo {year} {2022}),\
  10.1126/sciadv.abn1717}\BibitemShut {NoStop}%
\bibitem [{\citenamefont {Xu}\ \emph {et~al.}(2023)\citenamefont {Xu},
  \citenamefont {Ataides}, \citenamefont {Pattison}, \citenamefont
  {Raveendran}, \citenamefont {Bluvstein}, \citenamefont {Wurtz}, \citenamefont
  {Vasic}, \citenamefont {Lukin}, \citenamefont {Jiang},\ and\ \citenamefont
  {Zhou}}]{xu2023constantoverhead}%
  \BibitemOpen
  \bibfield  {author} {\bibinfo {author} {\bibfnamefont {Qian}\ \bibnamefont
  {Xu}}, \bibinfo {author} {\bibfnamefont {J.~Pablo~Bonilla}\ \bibnamefont
  {Ataides}}, \bibinfo {author} {\bibfnamefont {Christopher~A.}\ \bibnamefont
  {Pattison}}, \bibinfo {author} {\bibfnamefont {Nithin}\ \bibnamefont
  {Raveendran}}, \bibinfo {author} {\bibfnamefont {Dolev}\ \bibnamefont
  {Bluvstein}}, \bibinfo {author} {\bibfnamefont {Jonathan}\ \bibnamefont
  {Wurtz}}, \bibinfo {author} {\bibfnamefont {Bane}\ \bibnamefont {Vasic}},
  \bibinfo {author} {\bibfnamefont {Mikhail~D.}\ \bibnamefont {Lukin}},
  \bibinfo {author} {\bibfnamefont {Liang}\ \bibnamefont {Jiang}}, \ and\
  \bibinfo {author} {\bibfnamefont {Hengyun}\ \bibnamefont {Zhou}},\
  }\href@noop {} {\enquote {\bibinfo {title} {Constant-overhead fault-tolerant
  quantum computation with reconfigurable atom arrays},}\ } (\bibinfo {year}
  {2023}),\ \Eprint {http://arxiv.org/abs/2308.08648} {arXiv:2308.08648
  [quant-ph]} \BibitemShut {NoStop}%
\bibitem [{\citenamefont {Horsman}\ \emph {et~al.}(2012)\citenamefont
  {Horsman}, \citenamefont {Fowler}, \citenamefont {Devitt},\ and\
  \citenamefont {Meter}}]{Horsman_2012}%
  \BibitemOpen
  \bibfield  {author} {\bibinfo {author} {\bibfnamefont {Dominic}\ \bibnamefont
  {Horsman}}, \bibinfo {author} {\bibfnamefont {Austin~G}\ \bibnamefont
  {Fowler}}, \bibinfo {author} {\bibfnamefont {Simon}\ \bibnamefont {Devitt}},
  \ and\ \bibinfo {author} {\bibfnamefont {Rodney~Van}\ \bibnamefont {Meter}},\
  }\bibfield  {title} {\enquote {\bibinfo {title} {Surface code quantum
  computing by lattice surgery},}\ }\href {\doibase
  10.1088/1367-2630/14/12/123011} {\bibfield  {journal} {\bibinfo  {journal}
  {New Journal of Physics}\ }\textbf {\bibinfo {volume} {14}},\ \bibinfo
  {pages} {123011} (\bibinfo {year} {2012})}\BibitemShut {NoStop}%
\bibitem [{\citenamefont {Brown}\ \emph {et~al.}(2017)\citenamefont {Brown},
  \citenamefont {Laubscher}, \citenamefont {Kesselring},\ and\ \citenamefont
  {Wootton}}]{Brown_2017}%
  \BibitemOpen
  \bibfield  {author} {\bibinfo {author} {\bibfnamefont {Benjamin~J.}\
  \bibnamefont {Brown}}, \bibinfo {author} {\bibfnamefont {Katharina}\
  \bibnamefont {Laubscher}}, \bibinfo {author} {\bibfnamefont {Markus~S.}\
  \bibnamefont {Kesselring}}, \ and\ \bibinfo {author} {\bibfnamefont
  {James~R.}\ \bibnamefont {Wootton}},\ }\bibfield  {title} {\enquote {\bibinfo
  {title} {Poking holes and cutting corners to achieve clifford gates with the
  surface code},}\ }\href {\doibase 10.1103/physrevx.7.021029} {\bibfield
  {journal} {\bibinfo  {journal} {Physical Review X}\ }\textbf {\bibinfo
  {volume} {7}} (\bibinfo {year} {2017}),\
  10.1103/physrevx.7.021029}\BibitemShut {NoStop}%
\bibitem [{\citenamefont {Krishna}\ and\ \citenamefont
  {Poulin}(2021)}]{HGP_FTgates}%
  \BibitemOpen
  \bibfield  {author} {\bibinfo {author} {\bibfnamefont {Anirudh}\ \bibnamefont
  {Krishna}}\ and\ \bibinfo {author} {\bibfnamefont {David}\ \bibnamefont
  {Poulin}},\ }\bibfield  {title} {\enquote {\bibinfo {title} {Fault-tolerant
  gates on hypergraph product codes},}\ }\href {\doibase
  10.1103/PhysRevX.11.011023} {\bibfield  {journal} {\bibinfo  {journal} {Phys.
  Rev. X}\ }\textbf {\bibinfo {volume} {11}},\ \bibinfo {pages} {011023}
  (\bibinfo {year} {2021})}\BibitemShut {NoStop}%
\bibitem [{\citenamefont {Quintavalle}\ \emph {et~al.}(2023)\citenamefont
  {Quintavalle}, \citenamefont {Webster},\ and\ \citenamefont
  {Vasmer}}]{quintavalle2022}%
  \BibitemOpen
  \bibfield  {author} {\bibinfo {author} {\bibfnamefont {Armanda~O.}\
  \bibnamefont {Quintavalle}}, \bibinfo {author} {\bibfnamefont {Paul}\
  \bibnamefont {Webster}}, \ and\ \bibinfo {author} {\bibfnamefont {Michael}\
  \bibnamefont {Vasmer}},\ }\bibfield  {title} {\enquote {\bibinfo {title}
  {Partitioning qubits in hypergraph product codes to implement logical
  gates},}\ }\href {\doibase 10.22331/q-2023-10-24-1153} {\bibfield  {journal}
  {\bibinfo  {journal} {Quantum}\ }\textbf {\bibinfo {volume} {7}},\ \bibinfo
  {pages} {1153} (\bibinfo {year} {2023})}\BibitemShut {NoStop}%
\bibitem [{\citenamefont {Bravyi}\ and\ \citenamefont
  {K\"onig}(2013)}]{Bravyi_2013}%
  \BibitemOpen
  \bibfield  {author} {\bibinfo {author} {\bibfnamefont {Sergey}\ \bibnamefont
  {Bravyi}}\ and\ \bibinfo {author} {\bibfnamefont {Robert}\ \bibnamefont
  {K\"onig}},\ }\bibfield  {title} {\enquote {\bibinfo {title} {Classification
  of topologically protected gates for local stabilizer codes},}\ }\href
  {\doibase 10.1103/PhysRevLett.110.170503} {\bibfield  {journal} {\bibinfo
  {journal} {Phys. Rev. Lett.}\ }\textbf {\bibinfo {volume} {110}},\ \bibinfo
  {pages} {170503} (\bibinfo {year} {2013})}\BibitemShut {NoStop}%
\bibitem [{\citenamefont {Knill}(2004)}]{knill2004}%
  \BibitemOpen
  \bibfield  {author} {\bibinfo {author} {\bibfnamefont {E.}~\bibnamefont
  {Knill}},\ }\href@noop {} {\enquote {\bibinfo {title} {Fault-tolerant
  postselected quantum computation: Schemes},}\ } (\bibinfo {year} {2004}),\
  \Eprint {http://arxiv.org/abs/quant-ph/0402171} {arXiv:quant-ph/0402171
  [quant-ph]} \BibitemShut {NoStop}%
\bibitem [{\citenamefont {Bravyi}\ and\ \citenamefont
  {Kitaev}(2005)}]{Bravyi_2005}%
  \BibitemOpen
  \bibfield  {author} {\bibinfo {author} {\bibfnamefont {Sergey}\ \bibnamefont
  {Bravyi}}\ and\ \bibinfo {author} {\bibfnamefont {Alexei}\ \bibnamefont
  {Kitaev}},\ }\bibfield  {title} {\enquote {\bibinfo {title} {Universal
  quantum computation with ideal clifford gates and noisy ancillas},}\ }\href
  {\doibase 10.1103/physreva.71.022316} {\bibfield  {journal} {\bibinfo
  {journal} {Physical Review A}\ }\textbf {\bibinfo {volume} {71}} (\bibinfo
  {year} {2005}),\ 10.1103/physreva.71.022316}\BibitemShut {NoStop}%
\bibitem [{\citenamefont {Bombin}(2015)}]{bombin2015gauge}%
  \BibitemOpen
  \bibfield  {author} {\bibinfo {author} {\bibfnamefont {H.}~\bibnamefont
  {Bombin}},\ }\href@noop {} {\enquote {\bibinfo {title} {Gauge color codes:
  Optimal transversal gates and gauge fixing in topological stabilizer
  codes},}\ } (\bibinfo {year} {2015}),\ \Eprint
  {http://arxiv.org/abs/1311.0879} {arXiv:1311.0879 [quant-ph]} \BibitemShut
  {NoStop}%
\bibitem [{\citenamefont {Kubica}\ and\ \citenamefont
  {Beverland}(2015)}]{kubica}%
  \BibitemOpen
  \bibfield  {author} {\bibinfo {author} {\bibfnamefont {Aleksander}\
  \bibnamefont {Kubica}}\ and\ \bibinfo {author} {\bibfnamefont {Michael~E.}\
  \bibnamefont {Beverland}},\ }\bibfield  {title} {\enquote {\bibinfo {title}
  {Universal transversal gates with color codes: A simplified approach},}\
  }\href {\doibase 10.1103/PhysRevA.91.032330} {\bibfield  {journal} {\bibinfo
  {journal} {Phys. Rev. A}\ }\textbf {\bibinfo {volume} {91}},\ \bibinfo
  {pages} {032330} (\bibinfo {year} {2015})}\BibitemShut {NoStop}%
\bibitem [{\citenamefont {Brown}(2020)}]{Brown_2020}%
  \BibitemOpen
  \bibfield  {author} {\bibinfo {author} {\bibfnamefont {Benjamin~J.}\
  \bibnamefont {Brown}},\ }\bibfield  {title} {\enquote {\bibinfo {title} {A
  fault-tolerant non-clifford gate for the surface code in two dimensions},}\
  }\href {\doibase 10.1126/sciadv.aay4929} {\bibfield  {journal} {\bibinfo
  {journal} {Science Advances}\ }\textbf {\bibinfo {volume} {6}} (\bibinfo
  {year} {2020}),\ 10.1126/sciadv.aay4929}\BibitemShut {NoStop}%
\bibitem [{\citenamefont {Karp}(1972)}]{SATto3SAT}%
  \BibitemOpen
  \bibfield  {author} {\bibinfo {author} {\bibfnamefont {Richard}\ \bibnamefont
  {Karp}},\ }\bibfield  {title} {\enquote {\bibinfo {title} {Reducibility among
  combinatorial problems},}\ \ }(\bibinfo {year} {1972})\ pp.\ \bibinfo {pages}
  {85--103}\BibitemShut {NoStop}%
\bibitem [{\citenamefont {Shor}(1997)}]{shor1997faulttolerant}%
  \BibitemOpen
  \bibfield  {author} {\bibinfo {author} {\bibfnamefont {Peter~W.}\
  \bibnamefont {Shor}},\ }\href@noop {} {\enquote {\bibinfo {title}
  {Fault-tolerant quantum computation},}\ } (\bibinfo {year} {1997}),\ \Eprint
  {http://arxiv.org/abs/quant-ph/9605011} {arXiv:quant-ph/9605011 [quant-ph]}
  \BibitemShut {NoStop}%
\bibitem [{\citenamefont {Kollar}\ \emph {et~al.}(2019)\citenamefont {Kollar},
  \citenamefont {Fitzpatrick},\ and\ \citenamefont {Houck}}]{Kollar_2019}%
  \BibitemOpen
  \bibfield  {author} {\bibinfo {author} {\bibfnamefont {Alicia~J.}\
  \bibnamefont {Kollar}}, \bibinfo {author} {\bibfnamefont {Mattias}\
  \bibnamefont {Fitzpatrick}}, \ and\ \bibinfo {author} {\bibfnamefont
  {Andrew~A.}\ \bibnamefont {Houck}},\ }\bibfield  {title} {\enquote {\bibinfo
  {title} {Hyperbolic lattices in circuit quantum electrodynamics},}\ }\href
  {https://doi.org/10.1038%2Fs41586-019-1348-3} {\bibfield  {journal} {\bibinfo
   {journal} {Nature}\ }\textbf {\bibinfo {volume} {571}},\ \bibinfo {pages}
  {45--50} (\bibinfo {year} {2019})}\BibitemShut {NoStop}%
\bibitem [{\citenamefont {Rosenberg}\ \emph {et~al.}(2017)\citenamefont
  {Rosenberg}, \citenamefont {Kim}, \citenamefont {Das}, \citenamefont {Yost},
  \citenamefont {Gustavsson}, \citenamefont {Hover}, \citenamefont {Krantz},
  \citenamefont {Melville}, \citenamefont {Racz}, \citenamefont {Samach},
  \citenamefont {Weber}, \citenamefont {Yan}, \citenamefont {Yoder},
  \citenamefont {Kerman},\ and\ \citenamefont {Oliver}}]{Rosenberg_2017}%
  \BibitemOpen
  \bibfield  {author} {\bibinfo {author} {\bibfnamefont {D.}~\bibnamefont
  {Rosenberg}}, \bibinfo {author} {\bibfnamefont {D.}~\bibnamefont {Kim}},
  \bibinfo {author} {\bibfnamefont {R.}~\bibnamefont {Das}}, \bibinfo {author}
  {\bibfnamefont {D.}~\bibnamefont {Yost}}, \bibinfo {author} {\bibfnamefont
  {S.}~\bibnamefont {Gustavsson}}, \bibinfo {author} {\bibfnamefont
  {D.}~\bibnamefont {Hover}}, \bibinfo {author} {\bibfnamefont
  {P.}~\bibnamefont {Krantz}}, \bibinfo {author} {\bibfnamefont
  {A.}~\bibnamefont {Melville}}, \bibinfo {author} {\bibfnamefont
  {L.}~\bibnamefont {Racz}}, \bibinfo {author} {\bibfnamefont {G.~O.}\
  \bibnamefont {Samach}}, \bibinfo {author} {\bibfnamefont {S.~J.}\
  \bibnamefont {Weber}}, \bibinfo {author} {\bibfnamefont {F.}~\bibnamefont
  {Yan}}, \bibinfo {author} {\bibfnamefont {J.~L.}\ \bibnamefont {Yoder}},
  \bibinfo {author} {\bibfnamefont {A.~J.}\ \bibnamefont {Kerman}}, \ and\
  \bibinfo {author} {\bibfnamefont {W.~D.}\ \bibnamefont {Oliver}},\ }\bibfield
   {title} {\enquote {\bibinfo {title} {3d integrated superconducting
  qubits},}\ }\href {https://doi.org/10.1038%2Fs41534-017-0044-0} {\bibfield
  {journal} {\bibinfo  {journal} {npj Quantum Information}\ }\textbf {\bibinfo
  {volume} {3}} (\bibinfo {year} {2017})}\BibitemShut {NoStop}%
\bibitem [{\citenamefont {Fowler}\ \emph {et~al.}(2010)\citenamefont {Fowler},
  \citenamefont {Wang}, \citenamefont {Hill}, \citenamefont {Ladd},
  \citenamefont {Meter},\ and\ \citenamefont {Hollenberg}}]{Fowler_2010}%
  \BibitemOpen
  \bibfield  {author} {\bibinfo {author} {\bibfnamefont {Austin~G.}\
  \bibnamefont {Fowler}}, \bibinfo {author} {\bibfnamefont {David~S.}\
  \bibnamefont {Wang}}, \bibinfo {author} {\bibfnamefont {Charles~D.}\
  \bibnamefont {Hill}}, \bibinfo {author} {\bibfnamefont {Thaddeus~D.}\
  \bibnamefont {Ladd}}, \bibinfo {author} {\bibfnamefont {Rodney~Van}\
  \bibnamefont {Meter}}, \ and\ \bibinfo {author} {\bibfnamefont {Lloyd C.~L.}\
  \bibnamefont {Hollenberg}},\ }\bibfield  {title} {\enquote {\bibinfo {title}
  {Surface code quantum communication},}\ }\href {\doibase
  10.1103/physrevlett.104.180503} {\bibfield  {journal} {\bibinfo  {journal}
  {Physical Review Letters}\ }\textbf {\bibinfo {volume} {104}} (\bibinfo
  {year} {2010}),\ 10.1103/physrevlett.104.180503}\BibitemShut {NoStop}%
\bibitem [{\citenamefont {Rozpedek}\ \emph {et~al.}(2021)\citenamefont
  {Rozpedek}, \citenamefont {Noh}, \citenamefont {Xu}, \citenamefont {Guha},\
  and\ \citenamefont {Jiang}}]{Rozpedek_2021}%
  \BibitemOpen
  \bibfield  {author} {\bibinfo {author} {\bibfnamefont {Filip}\ \bibnamefont
  {Rozpedek}}, \bibinfo {author} {\bibfnamefont {Kyungjoo}\ \bibnamefont
  {Noh}}, \bibinfo {author} {\bibfnamefont {Qian}\ \bibnamefont {Xu}}, \bibinfo
  {author} {\bibfnamefont {Saikat}\ \bibnamefont {Guha}}, \ and\ \bibinfo
  {author} {\bibfnamefont {Liang}\ \bibnamefont {Jiang}},\ }\bibfield  {title}
  {\enquote {\bibinfo {title} {Quantum repeaters based on concatenated bosonic
  and discrete-variable quantum codes},}\ }\href {\doibase
  10.1038/s41534-021-00438-7} {\bibfield  {journal} {\bibinfo  {journal} {npj
  Quantum Information}\ }\textbf {\bibinfo {volume} {7}} (\bibinfo {year}
  {2021}),\ 10.1038/s41534-021-00438-7}\BibitemShut {NoStop}%
\bibitem [{\citenamefont {et~al.}(2023)}]{moses_race_2023}%
  \BibitemOpen
  \bibfield  {author} {\bibinfo {author} {\bibfnamefont {S.~A.~Moses}\
  \bibnamefont {et~al.}},\ }\href@noop {} {\enquote {\bibinfo {title} {A race
  track trapped-ion quantum processor},}\ } (\bibinfo {year} {2023}),\ \Eprint
  {http://arxiv.org/abs/2305.03828} {arXiv:2305.03828 [quant-ph]} \BibitemShut
  {NoStop}%
\bibitem [{\citenamefont {Kielpinski}\ \emph {et~al.}(2002)\citenamefont
  {Kielpinski}, \citenamefont {Monroe},\ and\ \citenamefont
  {Wineland}}]{kielpinski_architecture_2002}%
  \BibitemOpen
  \bibfield  {author} {\bibinfo {author} {\bibfnamefont {D.}~\bibnamefont
  {Kielpinski}}, \bibinfo {author} {\bibfnamefont {C.}~\bibnamefont {Monroe}},
  \ and\ \bibinfo {author} {\bibfnamefont {D.~J.}\ \bibnamefont {Wineland}},\
  }\bibfield  {title} {\enquote {\bibinfo {title} {Architecture for a
  large-scale ion-trap quantum computer},}\ }\href {\doibase
  10.1038/nature00784} {\bibfield  {journal} {\bibinfo  {journal} {Nature}\
  }\textbf {\bibinfo {volume} {417}},\ \bibinfo {pages} {709--711} (\bibinfo
  {year} {2002})}\BibitemShut {NoStop}%
\bibitem [{\citenamefont {Harty}\ \emph {et~al.}(2014)\citenamefont {Harty},
  \citenamefont {Allcock}, \citenamefont {Ballance}, \citenamefont {Guidoni},
  \citenamefont {Janacek}, \citenamefont {Linke}, \citenamefont {Stacey},\ and\
  \citenamefont {Lucas}}]{harty_high-fidelity_2014}%
  \BibitemOpen
  \bibfield  {author} {\bibinfo {author} {\bibfnamefont {T.~P.}\ \bibnamefont
  {Harty}}, \bibinfo {author} {\bibfnamefont {D.~T.~C.}\ \bibnamefont
  {Allcock}}, \bibinfo {author} {\bibfnamefont {C.~J.}\ \bibnamefont
  {Ballance}}, \bibinfo {author} {\bibfnamefont {L.}~\bibnamefont {Guidoni}},
  \bibinfo {author} {\bibfnamefont {H.~A.}\ \bibnamefont {Janacek}}, \bibinfo
  {author} {\bibfnamefont {N.~M.}\ \bibnamefont {Linke}}, \bibinfo {author}
  {\bibfnamefont {D.~N.}\ \bibnamefont {Stacey}}, \ and\ \bibinfo {author}
  {\bibfnamefont {D.~M.}\ \bibnamefont {Lucas}},\ }\bibfield  {title} {\enquote
  {\bibinfo {title} {High-fidelity preparation, gates, memory, and readout of a
  trapped-ion quantum bit},}\ }\href {\doibase 10.1103/PhysRevLett.113.220501}
  {\bibfield  {journal} {\bibinfo  {journal} {Phys. Rev. Lett.}\ }\textbf
  {\bibinfo {volume} {113}},\ \bibinfo {pages} {220501} (\bibinfo {year}
  {2014})}\BibitemShut {NoStop}%
\bibitem [{\citenamefont {Clark}\ \emph {et~al.}(2021)\citenamefont {Clark},
  \citenamefont {Tinkey}, \citenamefont {Sawyer}, \citenamefont {Meier},
  \citenamefont {Burkhardt}, \citenamefont {Seck}, \citenamefont {Shappert},
  \citenamefont {Guise}, \citenamefont {Volin}, \citenamefont {Fallek},
  \citenamefont {Hayden}, \citenamefont {Rellergert},\ and\ \citenamefont
  {Brown}}]{clark_high-fidelity_2021}%
  \BibitemOpen
  \bibfield  {author} {\bibinfo {author} {\bibfnamefont {Craig~R.}\
  \bibnamefont {Clark}}, \bibinfo {author} {\bibfnamefont {Holly~N.}\
  \bibnamefont {Tinkey}}, \bibinfo {author} {\bibfnamefont {Brian~C.}\
  \bibnamefont {Sawyer}}, \bibinfo {author} {\bibfnamefont {Adam~M.}\
  \bibnamefont {Meier}}, \bibinfo {author} {\bibfnamefont {Karl~A.}\
  \bibnamefont {Burkhardt}}, \bibinfo {author} {\bibfnamefont {Christopher~M.}\
  \bibnamefont {Seck}}, \bibinfo {author} {\bibfnamefont {Christopher~M.}\
  \bibnamefont {Shappert}}, \bibinfo {author} {\bibfnamefont {Nicholas~D.}\
  \bibnamefont {Guise}}, \bibinfo {author} {\bibfnamefont {Curtis~E.}\
  \bibnamefont {Volin}}, \bibinfo {author} {\bibfnamefont {Spencer~D.}\
  \bibnamefont {Fallek}}, \bibinfo {author} {\bibfnamefont {Harley~T.}\
  \bibnamefont {Hayden}}, \bibinfo {author} {\bibfnamefont {Wade~G.}\
  \bibnamefont {Rellergert}}, \ and\ \bibinfo {author} {\bibfnamefont
  {Kenton~R.}\ \bibnamefont {Brown}},\ }\bibfield  {title} {\enquote {\bibinfo
  {title} {High-fidelity bell-state preparation with $^{40}{\mathrm{ca}}^{+}$
  optical qubits},}\ }\href {\doibase 10.1103/PhysRevLett.127.130505}
  {\bibfield  {journal} {\bibinfo  {journal} {Phys. Rev. Lett.}\ }\textbf
  {\bibinfo {volume} {127}},\ \bibinfo {pages} {130505} (\bibinfo {year}
  {2021})}\BibitemShut {NoStop}%
\bibitem [{\citenamefont {et~al.}(2022)}]{ryan-anderson_implementing_2022}%
  \BibitemOpen
  \bibfield  {author} {\bibinfo {author} {\bibfnamefont {C.~Ryan-Anderson}\
  \bibnamefont {et~al.}},\ }\href@noop {} {\enquote {\bibinfo {title}
  {Implementing fault-tolerant entangling gates on the five-qubit code and the
  color code},}\ } (\bibinfo {year} {2022}),\ \Eprint
  {http://arxiv.org/abs/2208.01863} {arXiv:2208.01863 [quant-ph]} \BibitemShut
  {NoStop}%
\bibitem [{\citenamefont {Postler}\ \emph {et~al.}(2022)\citenamefont
  {Postler}, \citenamefont {Heu$\beta$en}, \citenamefont {Pogorelov},
  \citenamefont {Rispler}, \citenamefont {Feldker}, \citenamefont {Meth},
  \citenamefont {Marciniak}, \citenamefont {Stricker}, \citenamefont
  {Ringbauer}, \citenamefont {Blatt}, \citenamefont {Schindler}, \citenamefont
  {Müller},\ and\ \citenamefont {Monz}}]{postler_demonstration_2022}%
  \BibitemOpen
  \bibfield  {author} {\bibinfo {author} {\bibfnamefont {Lukas}\ \bibnamefont
  {Postler}}, \bibinfo {author} {\bibfnamefont {Sascha}\ \bibnamefont
  {Heu$\beta$en}}, \bibinfo {author} {\bibfnamefont {Ivan}\ \bibnamefont
  {Pogorelov}}, \bibinfo {author} {\bibfnamefont {Manuel}\ \bibnamefont
  {Rispler}}, \bibinfo {author} {\bibfnamefont {Thomas}\ \bibnamefont
  {Feldker}}, \bibinfo {author} {\bibfnamefont {Michael}\ \bibnamefont {Meth}},
  \bibinfo {author} {\bibfnamefont {Christian~D.}\ \bibnamefont {Marciniak}},
  \bibinfo {author} {\bibfnamefont {Roman}\ \bibnamefont {Stricker}}, \bibinfo
  {author} {\bibfnamefont {Martin}\ \bibnamefont {Ringbauer}}, \bibinfo
  {author} {\bibfnamefont {Rainer}\ \bibnamefont {Blatt}}, \bibinfo {author}
  {\bibfnamefont {Philipp}\ \bibnamefont {Schindler}}, \bibinfo {author}
  {\bibfnamefont {Markus}\ \bibnamefont {Müller}}, \ and\ \bibinfo {author}
  {\bibfnamefont {Thomas}\ \bibnamefont {Monz}},\ }\bibfield  {title} {\enquote
  {\bibinfo {title} {Demonstration of fault-tolerant universal quantum gate
  operations},}\ }\href {\doibase 10.1038/s41586-022-04721-1} {\bibfield
  {journal} {\bibinfo  {journal} {Nature}\ }\textbf {\bibinfo {volume} {605}},\
  \bibinfo {pages} {675--680} (\bibinfo {year} {2022})}\BibitemShut {NoStop}%
\bibitem [{\citenamefont {Dalgoutte}\ and\ \citenamefont
  {Wilkinson}(1975)}]{dalgoutte_thin_1975}%
  \BibitemOpen
  \bibfield  {author} {\bibinfo {author} {\bibfnamefont {D.~G.}\ \bibnamefont
  {Dalgoutte}}\ and\ \bibinfo {author} {\bibfnamefont {C.~D.~W.}\ \bibnamefont
  {Wilkinson}},\ }\bibfield  {title} {\enquote {\bibinfo {title} {Thin grating
  couplers for integrated optics: an experimental and theoretical study},}\
  }\href {\doibase 10.1364/AO.14.002983} {\bibfield  {journal} {\bibinfo
  {journal} {Applied Optics}\ }\textbf {\bibinfo {volume} {14}},\ \bibinfo
  {pages} {2983--2998} (\bibinfo {year} {1975})},\ \bibinfo {note} {publisher:
  Optica Publishing Group}\BibitemShut {NoStop}%
\bibitem [{\citenamefont {Mehta}\ \emph {et~al.}(2016)\citenamefont {Mehta},
  \citenamefont {Bruzewicz}, \citenamefont {McConnell}, \citenamefont {Ram},
  \citenamefont {Sage},\ and\ \citenamefont
  {Chiaverini}}]{mehta_integrated_2016}%
  \BibitemOpen
  \bibfield  {author} {\bibinfo {author} {\bibfnamefont {Karan~K.}\
  \bibnamefont {Mehta}}, \bibinfo {author} {\bibfnamefont {Colin~D.}\
  \bibnamefont {Bruzewicz}}, \bibinfo {author} {\bibfnamefont {Robert}\
  \bibnamefont {McConnell}}, \bibinfo {author} {\bibfnamefont {Rajeev~J.}\
  \bibnamefont {Ram}}, \bibinfo {author} {\bibfnamefont {Jeremy~M.}\
  \bibnamefont {Sage}}, \ and\ \bibinfo {author} {\bibfnamefont {John}\
  \bibnamefont {Chiaverini}},\ }\bibfield  {title} {\enquote {\bibinfo {title}
  {Integrated optical addressing of an ion qubit},}\ }\href {\doibase
  10.1038/nnano.2016.139} {\bibfield  {journal} {\bibinfo  {journal} {Nature
  Nanotechnology}\ }\textbf {\bibinfo {volume} {11}},\ \bibinfo {pages}
  {1066--1070} (\bibinfo {year} {2016})},\ \bibinfo {note} {number: 12
  Publisher: Nature Publishing Group}\BibitemShut {NoStop}%
\bibitem [{\citenamefont {Niffenegger}\ \emph {et~al.}(2020)\citenamefont
  {Niffenegger}, \citenamefont {Stuart}, \citenamefont {Sorace-Agaskar},
  \citenamefont {Kharas}, \citenamefont {Bramhavar}, \citenamefont {Bruzewicz},
  \citenamefont {Loh}, \citenamefont {Maxson}, \citenamefont {McConnell},
  \citenamefont {Reens}, \citenamefont {West}, \citenamefont {Sage},\ and\
  \citenamefont {Chiaverini}}]{niffenegger_integrated_2020}%
  \BibitemOpen
  \bibfield  {author} {\bibinfo {author} {\bibfnamefont {R.~J.}\ \bibnamefont
  {Niffenegger}}, \bibinfo {author} {\bibfnamefont {J.}~\bibnamefont {Stuart}},
  \bibinfo {author} {\bibfnamefont {C.}~\bibnamefont {Sorace-Agaskar}},
  \bibinfo {author} {\bibfnamefont {D.}~\bibnamefont {Kharas}}, \bibinfo
  {author} {\bibfnamefont {S.}~\bibnamefont {Bramhavar}}, \bibinfo {author}
  {\bibfnamefont {C.~D.}\ \bibnamefont {Bruzewicz}}, \bibinfo {author}
  {\bibfnamefont {W.}~\bibnamefont {Loh}}, \bibinfo {author} {\bibfnamefont
  {R.~T.}\ \bibnamefont {Maxson}}, \bibinfo {author} {\bibfnamefont
  {R.}~\bibnamefont {McConnell}}, \bibinfo {author} {\bibfnamefont
  {D.}~\bibnamefont {Reens}}, \bibinfo {author} {\bibfnamefont {G.~N.}\
  \bibnamefont {West}}, \bibinfo {author} {\bibfnamefont {J.~M.}\ \bibnamefont
  {Sage}}, \ and\ \bibinfo {author} {\bibfnamefont {J.}~\bibnamefont
  {Chiaverini}},\ }\bibfield  {title} {\enquote {\bibinfo {title} {Integrated
  multi-wavelength control of an ion qubit},}\ }\href {\doibase
  10.1038/s41586-020-2811-x} {\bibfield  {journal} {\bibinfo  {journal}
  {Nature}\ }\textbf {\bibinfo {volume} {586}},\ \bibinfo {pages} {538--542}
  (\bibinfo {year} {2020})},\ \bibinfo {note} {number: 7830 Publisher: Nature
  Publishing Group}\BibitemShut {NoStop}%
\bibitem [{\citenamefont {Mehta}\ \emph {et~al.}(2020)\citenamefont {Mehta},
  \citenamefont {Zhang}, \citenamefont {Malinowski}, \citenamefont {Nguyen},
  \citenamefont {Stadler},\ and\ \citenamefont {Home}}]{mehta_integrated_2020}%
  \BibitemOpen
  \bibfield  {author} {\bibinfo {author} {\bibfnamefont {Karan~K.}\
  \bibnamefont {Mehta}}, \bibinfo {author} {\bibfnamefont {Chi}\ \bibnamefont
  {Zhang}}, \bibinfo {author} {\bibfnamefont {Maciej}\ \bibnamefont
  {Malinowski}}, \bibinfo {author} {\bibfnamefont {Thanh-Long}\ \bibnamefont
  {Nguyen}}, \bibinfo {author} {\bibfnamefont {Martin}\ \bibnamefont
  {Stadler}}, \ and\ \bibinfo {author} {\bibfnamefont {Jonathan~P.}\
  \bibnamefont {Home}},\ }\bibfield  {title} {\enquote {\bibinfo {title}
  {Integrated optical multi-ion quantum logic},}\ }\href {\doibase
  10.1038/s41586-020-2823-6} {\bibfield  {journal} {\bibinfo  {journal}
  {Nature}\ }\textbf {\bibinfo {volume} {586}},\ \bibinfo {pages} {533--537}
  (\bibinfo {year} {2020})},\ \bibinfo {note} {number: 7830 Publisher: Nature
  Publishing Group}\BibitemShut {NoStop}%
\bibitem [{\citenamefont {Kwon}\ \emph {et~al.}(2023)\citenamefont {Kwon},
  \citenamefont {Setzer}, \citenamefont {Gehl}, \citenamefont {Karl},
  \citenamefont {Van Der~Wall}, \citenamefont {Law}, \citenamefont {Stick},\
  and\ \citenamefont {McGuinness}}]{kwon_multi-site_2023}%
  \BibitemOpen
  \bibfield  {author} {\bibinfo {author} {\bibfnamefont {Joonhyuk}\
  \bibnamefont {Kwon}}, \bibinfo {author} {\bibfnamefont {William~J.}\
  \bibnamefont {Setzer}}, \bibinfo {author} {\bibfnamefont {Michael}\
  \bibnamefont {Gehl}}, \bibinfo {author} {\bibfnamefont {Nicholas}\
  \bibnamefont {Karl}}, \bibinfo {author} {\bibfnamefont {Jay}\ \bibnamefont
  {Van Der~Wall}}, \bibinfo {author} {\bibfnamefont {Ryan}\ \bibnamefont
  {Law}}, \bibinfo {author} {\bibfnamefont {Daniel}\ \bibnamefont {Stick}}, \
  and\ \bibinfo {author} {\bibfnamefont {Hayden~J.}\ \bibnamefont
  {McGuinness}},\ }\href {http://arxiv.org/abs/2308.14918} {\enquote {\bibinfo
  {title} {Multi-site {Integrated} {Optical} {Addressing} of {Trapped}
  {Ions}},}\ } (\bibinfo {year} {2023}),\ \bibinfo {note} {arXiv:2308.14918
  [quant-ph]}\BibitemShut {NoStop}%
\bibitem [{\citenamefont {Mordini}\ \emph {et~al.}(2024)\citenamefont
  {Mordini}, \citenamefont {Vasquez}, \citenamefont {Motohashi}, \citenamefont
  {Müller}, \citenamefont {Malinowski}, \citenamefont {Zhang}, \citenamefont
  {Mehta}, \citenamefont {Kienzler},\ and\ \citenamefont
  {Home}}]{mordini_multi-zone_2024}%
  \BibitemOpen
  \bibfield  {author} {\bibinfo {author} {\bibfnamefont {Carmelo}\ \bibnamefont
  {Mordini}}, \bibinfo {author} {\bibfnamefont {Alfredo~Ricci}\ \bibnamefont
  {Vasquez}}, \bibinfo {author} {\bibfnamefont {Yuto}\ \bibnamefont
  {Motohashi}}, \bibinfo {author} {\bibfnamefont {Mose}\ \bibnamefont
  {Müller}}, \bibinfo {author} {\bibfnamefont {Maciej}\ \bibnamefont
  {Malinowski}}, \bibinfo {author} {\bibfnamefont {Chi}\ \bibnamefont {Zhang}},
  \bibinfo {author} {\bibfnamefont {Karan~K.}\ \bibnamefont {Mehta}}, \bibinfo
  {author} {\bibfnamefont {Daniel}\ \bibnamefont {Kienzler}}, \ and\ \bibinfo
  {author} {\bibfnamefont {Jonathan~P.}\ \bibnamefont {Home}},\ }\href
  {http://arxiv.org/abs/2401.18056} {\enquote {\bibinfo {title} {Multi-zone
  trapped-ion qubit control in an integrated photonics {QCCD} device},}\ }
  (\bibinfo {year} {2024}),\ \bibinfo {note} {arXiv:2401.18056 [physics,
  physics:quant-ph]}\BibitemShut {NoStop}%
\bibitem [{\citenamefont {Lindenfelser}\ \emph {et~al.}(2017)\citenamefont
  {Lindenfelser}, \citenamefont {Marinelli}, \citenamefont {Negnevitsky},
  \citenamefont {Ragg},\ and\ \citenamefont
  {Home}}]{lindenfelser_cooling_2017}%
  \BibitemOpen
  \bibfield  {author} {\bibinfo {author} {\bibfnamefont {F.}~\bibnamefont
  {Lindenfelser}}, \bibinfo {author} {\bibfnamefont {M.}~\bibnamefont
  {Marinelli}}, \bibinfo {author} {\bibfnamefont {V.}~\bibnamefont
  {Negnevitsky}}, \bibinfo {author} {\bibfnamefont {S.}~\bibnamefont {Ragg}}, \
  and\ \bibinfo {author} {\bibfnamefont {J.~P.}\ \bibnamefont {Home}},\
  }\bibfield  {title} {\enquote {\bibinfo {title} {Cooling {Atomic} {Ions} with
  {Visible} and {Infra}-{Red} {Light}},}\ }\href {\doibase
  10.1088/1367-2630/aa7150} {\bibfield  {journal} {\bibinfo  {journal} {New
  Journal of Physics}\ }\textbf {\bibinfo {volume} {19}},\ \bibinfo {pages}
  {063041} (\bibinfo {year} {2017})}\BibitemShut {NoStop}%
\bibitem [{\citenamefont {Hendricks}\ \emph {et~al.}(2008)\citenamefont
  {Hendricks}, \citenamefont {Sørensen}, \citenamefont {Champenois},
  \citenamefont {Knoop},\ and\ \citenamefont
  {Drewsen}}]{hendricks_doppler_2008}%
  \BibitemOpen
  \bibfield  {author} {\bibinfo {author} {\bibfnamefont {Richard~J.}\
  \bibnamefont {Hendricks}}, \bibinfo {author} {\bibfnamefont {Jens~L.}\
  \bibnamefont {Sørensen}}, \bibinfo {author} {\bibfnamefont {Caroline}\
  \bibnamefont {Champenois}}, \bibinfo {author} {\bibfnamefont {Martina}\
  \bibnamefont {Knoop}}, \ and\ \bibinfo {author} {\bibfnamefont {Michael}\
  \bibnamefont {Drewsen}},\ }\bibfield  {title} {\enquote {\bibinfo {title}
  {Doppler cooling of calcium ions using a dipole-forbidden transition},}\
  }\href {\doibase 10.1103/PhysRevA.77.021401} {\bibfield  {journal} {\bibinfo
  {journal} {Physical Review A}\ }\textbf {\bibinfo {volume} {77}},\ \bibinfo
  {pages} {021401} (\bibinfo {year} {2008})},\ \bibinfo {note} {publisher:
  American Physical Society}\BibitemShut {NoStop}%
\bibitem [{\citenamefont {Pino}\ \emph {et~al.}(2021)\citenamefont {Pino},
  \citenamefont {Dreiling}, \citenamefont {Figgatt}, \citenamefont {Gaebler},
  \citenamefont {Moses}, \citenamefont {Allman}, \citenamefont {Baldwin},
  \citenamefont {Foss-Feig}, \citenamefont {Hayes}, \citenamefont {Mayer},
  \citenamefont {Ryan-Anderson},\ and\ \citenamefont
  {Neyenhuis}}]{pino_demonstration_2021}%
  \BibitemOpen
  \bibfield  {author} {\bibinfo {author} {\bibfnamefont {J.~M.}\ \bibnamefont
  {Pino}}, \bibinfo {author} {\bibfnamefont {J.~M.}\ \bibnamefont {Dreiling}},
  \bibinfo {author} {\bibfnamefont {C.}~\bibnamefont {Figgatt}}, \bibinfo
  {author} {\bibfnamefont {J.~P.}\ \bibnamefont {Gaebler}}, \bibinfo {author}
  {\bibfnamefont {S.~A.}\ \bibnamefont {Moses}}, \bibinfo {author}
  {\bibfnamefont {M.~S.}\ \bibnamefont {Allman}}, \bibinfo {author}
  {\bibfnamefont {C.~H.}\ \bibnamefont {Baldwin}}, \bibinfo {author}
  {\bibfnamefont {M.}~\bibnamefont {Foss-Feig}}, \bibinfo {author}
  {\bibfnamefont {D.}~\bibnamefont {Hayes}}, \bibinfo {author} {\bibfnamefont
  {K.}~\bibnamefont {Mayer}}, \bibinfo {author} {\bibfnamefont
  {C.}~\bibnamefont {Ryan-Anderson}}, \ and\ \bibinfo {author} {\bibfnamefont
  {B.}~\bibnamefont {Neyenhuis}},\ }\bibfield  {title} {\enquote {\bibinfo
  {title} {Demonstration of the trapped-ion quantum {CCD} computer
  architecture},}\ }\href {\doibase 10.1038/s41586-021-03318-4} {\bibfield
  {journal} {\bibinfo  {journal} {Nature}\ }\textbf {\bibinfo {volume} {592}},\
  \bibinfo {pages} {209--213} (\bibinfo {year} {2021})}\BibitemShut {NoStop}%
\bibitem [{\citenamefont {Landsman}\ \emph {et~al.}(2019)\citenamefont
  {Landsman}, \citenamefont {Wu}, \citenamefont {Leung}, \citenamefont {Zhu},
  \citenamefont {Linke}, \citenamefont {Brown}, \citenamefont {Duan},\ and\
  \citenamefont {Monroe}}]{landsman_two-qubit_2019}%
  \BibitemOpen
  \bibfield  {author} {\bibinfo {author} {\bibfnamefont {K.~A.}\ \bibnamefont
  {Landsman}}, \bibinfo {author} {\bibfnamefont {Y.}~\bibnamefont {Wu}},
  \bibinfo {author} {\bibfnamefont {P.~H.}\ \bibnamefont {Leung}}, \bibinfo
  {author} {\bibfnamefont {D.}~\bibnamefont {Zhu}}, \bibinfo {author}
  {\bibfnamefont {N.~M.}\ \bibnamefont {Linke}}, \bibinfo {author}
  {\bibfnamefont {K.~R.}\ \bibnamefont {Brown}}, \bibinfo {author}
  {\bibfnamefont {L.}~\bibnamefont {Duan}}, \ and\ \bibinfo {author}
  {\bibfnamefont {C.}~\bibnamefont {Monroe}},\ }\bibfield  {title} {\enquote
  {\bibinfo {title} {Two-qubit entangling gates within arbitrarily long chains
  of trapped ions},}\ }\href {\doibase 10.1103/PhysRevA.100.022332} {\bibfield
  {journal} {\bibinfo  {journal} {Phys. Rev. A}\ }\textbf {\bibinfo {volume}
  {100}},\ \bibinfo {pages} {022332} (\bibinfo {year} {2019})}\BibitemShut
  {NoStop}%
\bibitem [{\citenamefont {Wang}\ \emph {et~al.}(2020)\citenamefont {Wang},
  \citenamefont {Crain}, \citenamefont {Fang}, \citenamefont {Zhang},
  \citenamefont {Huang}, \citenamefont {Liang}, \citenamefont {Leung},
  \citenamefont {Brown},\ and\ \citenamefont {Kim}}]{wang_high-fidelity_2020}%
  \BibitemOpen
  \bibfield  {author} {\bibinfo {author} {\bibfnamefont {Ye}~\bibnamefont
  {Wang}}, \bibinfo {author} {\bibfnamefont {Stephen}\ \bibnamefont {Crain}},
  \bibinfo {author} {\bibfnamefont {Chao}\ \bibnamefont {Fang}}, \bibinfo
  {author} {\bibfnamefont {Bichen}\ \bibnamefont {Zhang}}, \bibinfo {author}
  {\bibfnamefont {Shilin}\ \bibnamefont {Huang}}, \bibinfo {author}
  {\bibfnamefont {Qiyao}\ \bibnamefont {Liang}}, \bibinfo {author}
  {\bibfnamefont {Pak~Hong}\ \bibnamefont {Leung}}, \bibinfo {author}
  {\bibfnamefont {Kenneth~R.}\ \bibnamefont {Brown}}, \ and\ \bibinfo {author}
  {\bibfnamefont {Jungsang}\ \bibnamefont {Kim}},\ }\bibfield  {title}
  {\enquote {\bibinfo {title} {High-fidelity two-qubit gates using a
  microelectromechanical-system-based beam steering system for individual qubit
  addressing},}\ }\href {\doibase 10.1103/PhysRevLett.125.150505} {\bibfield
  {journal} {\bibinfo  {journal} {Phys. Rev. Lett.}\ }\textbf {\bibinfo
  {volume} {125}},\ \bibinfo {pages} {150505} (\bibinfo {year}
  {2020})}\BibitemShut {NoStop}%
\bibitem [{\citenamefont {Monroe}\ \emph {et~al.}(2014)\citenamefont {Monroe},
  \citenamefont {Raussendorf}, \citenamefont {Ruthven}, \citenamefont {Brown},
  \citenamefont {Maunz}, \citenamefont {Duan},\ and\ \citenamefont
  {Kim}}]{monroe_large-scale_2014}%
  \BibitemOpen
  \bibfield  {author} {\bibinfo {author} {\bibfnamefont {C.}~\bibnamefont
  {Monroe}}, \bibinfo {author} {\bibfnamefont {R.}~\bibnamefont {Raussendorf}},
  \bibinfo {author} {\bibfnamefont {A.}~\bibnamefont {Ruthven}}, \bibinfo
  {author} {\bibfnamefont {K.~R.}\ \bibnamefont {Brown}}, \bibinfo {author}
  {\bibfnamefont {P.}~\bibnamefont {Maunz}}, \bibinfo {author} {\bibfnamefont
  {L.-M.}\ \bibnamefont {Duan}}, \ and\ \bibinfo {author} {\bibfnamefont
  {J.}~\bibnamefont {Kim}},\ }\bibfield  {title} {\enquote {\bibinfo {title}
  {Large-scale modular quantum-computer architecture with atomic memory and
  photonic interconnects},}\ }\href {\doibase 10.1103/PhysRevA.89.022317}
  {\bibfield  {journal} {\bibinfo  {journal} {Phys. Rev. A}\ }\textbf {\bibinfo
  {volume} {89}},\ \bibinfo {pages} {022317} (\bibinfo {year}
  {2014})}\BibitemShut {NoStop}%
\bibitem [{\citenamefont {Stephenson}\ \emph {et~al.}(2020)\citenamefont
  {Stephenson}, \citenamefont {Nadlinger}, \citenamefont {Nichol},
  \citenamefont {An}, \citenamefont {Drmota}, \citenamefont {Ballance},
  \citenamefont {Thirumalai}, \citenamefont {Goodwin}, \citenamefont {Lucas},\
  and\ \citenamefont {Ballance}}]{stephenson_2020}%
  \BibitemOpen
  \bibfield  {author} {\bibinfo {author} {\bibfnamefont {L.~J.}\ \bibnamefont
  {Stephenson}}, \bibinfo {author} {\bibfnamefont {D.~P.}\ \bibnamefont
  {Nadlinger}}, \bibinfo {author} {\bibfnamefont {B.~C.}\ \bibnamefont
  {Nichol}}, \bibinfo {author} {\bibfnamefont {S.}~\bibnamefont {An}}, \bibinfo
  {author} {\bibfnamefont {P.}~\bibnamefont {Drmota}}, \bibinfo {author}
  {\bibfnamefont {T.~G.}\ \bibnamefont {Ballance}}, \bibinfo {author}
  {\bibfnamefont {K.}~\bibnamefont {Thirumalai}}, \bibinfo {author}
  {\bibfnamefont {J.~F.}\ \bibnamefont {Goodwin}}, \bibinfo {author}
  {\bibfnamefont {D.~M.}\ \bibnamefont {Lucas}}, \ and\ \bibinfo {author}
  {\bibfnamefont {C.~J.}\ \bibnamefont {Ballance}},\ }\bibfield  {title}
  {\enquote {\bibinfo {title} {High-rate, high-fidelity entanglement of qubits
  across an elementary quantum network},}\ }\href {\doibase
  10.1103/PhysRevLett.124.110501} {\bibfield  {journal} {\bibinfo  {journal}
  {Phys. Rev. Lett.}\ }\textbf {\bibinfo {volume} {124}},\ \bibinfo {pages}
  {110501} (\bibinfo {year} {2020})}\BibitemShut {NoStop}%
\bibitem [{\citenamefont {Krutyanskiy}\ \emph {et~al.}(2019)\citenamefont
  {Krutyanskiy}, \citenamefont {Meraner}, \citenamefont {Schupp}, \citenamefont
  {Krcmarsky}, \citenamefont {Hainzer},\ and\ \citenamefont
  {Lanyon}}]{Krutyanskiy_2019}%
  \BibitemOpen
  \bibfield  {author} {\bibinfo {author} {\bibfnamefont {V.}~\bibnamefont
  {Krutyanskiy}}, \bibinfo {author} {\bibfnamefont {M.}~\bibnamefont
  {Meraner}}, \bibinfo {author} {\bibfnamefont {J.}~\bibnamefont {Schupp}},
  \bibinfo {author} {\bibfnamefont {V.}~\bibnamefont {Krcmarsky}}, \bibinfo
  {author} {\bibfnamefont {H.}~\bibnamefont {Hainzer}}, \ and\ \bibinfo
  {author} {\bibfnamefont {B.~P.}\ \bibnamefont {Lanyon}},\ }\bibfield  {title}
  {\enquote {\bibinfo {title} {Light-matter entanglement over
  50{\hspace{0.167em}}km of optical fibre},}\ }\href {\doibase
  10.1038/s41534-019-0186-3} {\bibfield  {journal} {\bibinfo  {journal} {npj
  Quantum Information}\ }\textbf {\bibinfo {volume} {5}} (\bibinfo {year}
  {2019}),\ 10.1038/s41534-019-0186-3}\BibitemShut {NoStop}%
\bibitem [{\citenamefont {Campbell}(2007)}]{campbell_distributed_2007}%
  \BibitemOpen
  \bibfield  {author} {\bibinfo {author} {\bibfnamefont {Earl~T.}\ \bibnamefont
  {Campbell}},\ }\bibfield  {title} {\enquote {\bibinfo {title} {Distributed
  quantum-information processing with minimal local resources},}\ }\href
  {\doibase 10.1103/PhysRevA.76.040302} {\bibfield  {journal} {\bibinfo
  {journal} {Phys. Rev. A}\ }\textbf {\bibinfo {volume} {76}},\ \bibinfo
  {pages} {040302} (\bibinfo {year} {2007})}\BibitemShut {NoStop}%
\bibitem [{\citenamefont {Browaeys}\ and\ \citenamefont
  {Lahaye}(2020)}]{Browaeys2020}%
  \BibitemOpen
  \bibfield  {author} {\bibinfo {author} {\bibfnamefont {Antoine}\ \bibnamefont
  {Browaeys}}\ and\ \bibinfo {author} {\bibfnamefont {Thierry}\ \bibnamefont
  {Lahaye}},\ }\bibfield  {title} {\enquote {\bibinfo {title} {{Many-body
  physics with individually controlled Rydberg atoms}},}\ }\href {\doibase
  10.1038/s41567-019-0733-z} {\bibfield  {journal} {\bibinfo  {journal} {Nature
  Physics}\ }\textbf {\bibinfo {volume} {16}},\ \bibinfo {pages} {132--142}
  (\bibinfo {year} {2020})}\BibitemShut {NoStop}%
\bibitem [{\citenamefont {Scholl}\ \emph {et~al.}(2021)\citenamefont {Scholl},
  \citenamefont {Schuler}, \citenamefont {Williams}, \citenamefont
  {Eberharter}, \citenamefont {Barredo}, \citenamefont {Schymik}, \citenamefont
  {Lienhard}, \citenamefont {Henry}, \citenamefont {Lang}, \citenamefont
  {Lahaye} \emph {et~al.}}]{scholl2021quantum}%
  \BibitemOpen
  \bibfield  {author} {\bibinfo {author} {\bibfnamefont {Pascal}\ \bibnamefont
  {Scholl}}, \bibinfo {author} {\bibfnamefont {Michael}\ \bibnamefont
  {Schuler}}, \bibinfo {author} {\bibfnamefont {Hannah~J}\ \bibnamefont
  {Williams}}, \bibinfo {author} {\bibfnamefont {Alexander~A}\ \bibnamefont
  {Eberharter}}, \bibinfo {author} {\bibfnamefont {Daniel}\ \bibnamefont
  {Barredo}}, \bibinfo {author} {\bibfnamefont {Kai-Niklas}\ \bibnamefont
  {Schymik}}, \bibinfo {author} {\bibfnamefont {Vincent}\ \bibnamefont
  {Lienhard}}, \bibinfo {author} {\bibfnamefont {Louis-Paul}\ \bibnamefont
  {Henry}}, \bibinfo {author} {\bibfnamefont {Thomas~C}\ \bibnamefont {Lang}},
  \bibinfo {author} {\bibfnamefont {Thierry}\ \bibnamefont {Lahaye}},  \emph
  {et~al.},\ }\bibfield  {title} {\enquote {\bibinfo {title} {Quantum
  simulation of 2d antiferromagnets with hundreds of rydberg atoms},}\
  }\href@noop {} {\bibfield  {journal} {\bibinfo  {journal} {Nature}\ }\textbf
  {\bibinfo {volume} {595}},\ \bibinfo {pages} {233--238} (\bibinfo {year}
  {2021})}\BibitemShut {NoStop}%
\bibitem [{\citenamefont {Ebadi}\ \emph {et~al.}(2021)\citenamefont {Ebadi},
  \citenamefont {Wang}, \citenamefont {Levine}, \citenamefont {Keesling},
  \citenamefont {Semeghini}, \citenamefont {Omran}, \citenamefont {Bluvstein},
  \citenamefont {Samajdar}, \citenamefont {Pichler}, \citenamefont {Ho} \emph
  {et~al.}}]{ebadi2021quantum}%
  \BibitemOpen
  \bibfield  {author} {\bibinfo {author} {\bibfnamefont {Sepehr}\ \bibnamefont
  {Ebadi}}, \bibinfo {author} {\bibfnamefont {Tout~T}\ \bibnamefont {Wang}},
  \bibinfo {author} {\bibfnamefont {Harry}\ \bibnamefont {Levine}}, \bibinfo
  {author} {\bibfnamefont {Alexander}\ \bibnamefont {Keesling}}, \bibinfo
  {author} {\bibfnamefont {Giulia}\ \bibnamefont {Semeghini}}, \bibinfo
  {author} {\bibfnamefont {Ahmed}\ \bibnamefont {Omran}}, \bibinfo {author}
  {\bibfnamefont {Dolev}\ \bibnamefont {Bluvstein}}, \bibinfo {author}
  {\bibfnamefont {Rhine}\ \bibnamefont {Samajdar}}, \bibinfo {author}
  {\bibfnamefont {Hannes}\ \bibnamefont {Pichler}}, \bibinfo {author}
  {\bibfnamefont {Wen~Wei}\ \bibnamefont {Ho}},  \emph {et~al.},\ }\bibfield
  {title} {\enquote {\bibinfo {title} {Quantum phases of matter on a 256-atom
  programmable quantum simulator},}\ }\href@noop {} {\bibfield  {journal}
  {\bibinfo  {journal} {Nature}\ }\textbf {\bibinfo {volume} {595}},\ \bibinfo
  {pages} {227--232} (\bibinfo {year} {2021})}\BibitemShut {NoStop}%
\bibitem [{\citenamefont {Young}\ \emph {et~al.}(2023)\citenamefont {Young},
  \citenamefont {Geller}, \citenamefont {Eckner}, \citenamefont {Schine},
  \citenamefont {Glancy}, \citenamefont {Knill},\ and\ \citenamefont
  {Kaufman}}]{young2023atomic}%
  \BibitemOpen
  \bibfield  {author} {\bibinfo {author} {\bibfnamefont {Aaron~W}\ \bibnamefont
  {Young}}, \bibinfo {author} {\bibfnamefont {Shawn}\ \bibnamefont {Geller}},
  \bibinfo {author} {\bibfnamefont {William~J}\ \bibnamefont {Eckner}},
  \bibinfo {author} {\bibfnamefont {Nathan}\ \bibnamefont {Schine}}, \bibinfo
  {author} {\bibfnamefont {Scott}\ \bibnamefont {Glancy}}, \bibinfo {author}
  {\bibfnamefont {Emanuel}\ \bibnamefont {Knill}}, \ and\ \bibinfo {author}
  {\bibfnamefont {Adam~M}\ \bibnamefont {Kaufman}},\ }\bibfield  {title}
  {\enquote {\bibinfo {title} {An atomic boson sampler},}\ }\href@noop {}
  {\bibfield  {journal} {\bibinfo  {journal} {arXiv preprint arXiv:2307.06936}\
  } (\bibinfo {year} {2023})}\BibitemShut {NoStop}%
\bibitem [{\citenamefont {Tao}\ \emph {et~al.}(2023)\citenamefont {Tao},
  \citenamefont {Ammenwerth}, \citenamefont {Gyger}, \citenamefont {Bloch},\
  and\ \citenamefont {Zeiher}}]{tao2023high}%
  \BibitemOpen
  \bibfield  {author} {\bibinfo {author} {\bibfnamefont {Renhao}\ \bibnamefont
  {Tao}}, \bibinfo {author} {\bibfnamefont {Maximilian}\ \bibnamefont
  {Ammenwerth}}, \bibinfo {author} {\bibfnamefont {Flavien}\ \bibnamefont
  {Gyger}}, \bibinfo {author} {\bibfnamefont {Immanuel}\ \bibnamefont {Bloch}},
  \ and\ \bibinfo {author} {\bibfnamefont {Johannes}\ \bibnamefont {Zeiher}},\
  }\bibfield  {title} {\enquote {\bibinfo {title} {High-fidelity detection of
  large-scale atom arrays in an optical lattice},}\ }\href@noop {} {\bibfield
  {journal} {\bibinfo  {journal} {arXiv preprint arXiv:2309.04717}\ } (\bibinfo
  {year} {2023})}\BibitemShut {NoStop}%
\bibitem [{\citenamefont {Eckner}\ \emph {et~al.}(2023)\citenamefont {Eckner},
  \citenamefont {Darkwah~Oppong}, \citenamefont {Cao}, \citenamefont {Young},
  \citenamefont {Milner}, \citenamefont {Robinson}, \citenamefont {Ye},\ and\
  \citenamefont {Kaufman}}]{eckner2023realizing}%
  \BibitemOpen
  \bibfield  {author} {\bibinfo {author} {\bibfnamefont {William~J}\
  \bibnamefont {Eckner}}, \bibinfo {author} {\bibfnamefont {Nelson}\
  \bibnamefont {Darkwah~Oppong}}, \bibinfo {author} {\bibfnamefont {Alec}\
  \bibnamefont {Cao}}, \bibinfo {author} {\bibfnamefont {Aaron~W}\ \bibnamefont
  {Young}}, \bibinfo {author} {\bibfnamefont {William~R}\ \bibnamefont
  {Milner}}, \bibinfo {author} {\bibfnamefont {John~M}\ \bibnamefont
  {Robinson}}, \bibinfo {author} {\bibfnamefont {Jun}\ \bibnamefont {Ye}}, \
  and\ \bibinfo {author} {\bibfnamefont {Adam~M}\ \bibnamefont {Kaufman}},\
  }\bibfield  {title} {\enquote {\bibinfo {title} {Realizing spin squeezing
  with rydberg interactions in an optical clock},}\ }\href@noop {} {\bibfield
  {journal} {\bibinfo  {journal} {Nature}\ ,\ \bibinfo {pages} {1--6}}
  (\bibinfo {year} {2023})}\BibitemShut {NoStop}%
\bibitem [{\citenamefont {Trotzky}\ \emph {et~al.}(2008)\citenamefont
  {Trotzky}, \citenamefont {Cheinet}, \citenamefont {Folling}, \citenamefont
  {Feld}, \citenamefont {Schnorrberger}, \citenamefont {Rey}, \citenamefont
  {Polkovnikov}, \citenamefont {Demler}, \citenamefont {Lukin},\ and\
  \citenamefont {Bloch}}]{trotzky2008time}%
  \BibitemOpen
  \bibfield  {author} {\bibinfo {author} {\bibfnamefont {Stefan}\ \bibnamefont
  {Trotzky}}, \bibinfo {author} {\bibfnamefont {Patrick}\ \bibnamefont
  {Cheinet}}, \bibinfo {author} {\bibfnamefont {S}~\bibnamefont {Folling}},
  \bibinfo {author} {\bibfnamefont {Michael}\ \bibnamefont {Feld}}, \bibinfo
  {author} {\bibfnamefont {Ute}\ \bibnamefont {Schnorrberger}}, \bibinfo
  {author} {\bibfnamefont {Ana~Maria}\ \bibnamefont {Rey}}, \bibinfo {author}
  {\bibfnamefont {Anatoli}\ \bibnamefont {Polkovnikov}}, \bibinfo {author}
  {\bibfnamefont {Eugene~A}\ \bibnamefont {Demler}}, \bibinfo {author}
  {\bibfnamefont {Mikhail~D}\ \bibnamefont {Lukin}}, \ and\ \bibinfo {author}
  {\bibfnamefont {Immanuel}\ \bibnamefont {Bloch}},\ }\bibfield  {title}
  {\enquote {\bibinfo {title} {Time-resolved observation and control of
  superexchange interactions with ultracold atoms in optical lattices},}\
  }\href@noop {} {\bibfield  {journal} {\bibinfo  {journal} {Science}\ }\textbf
  {\bibinfo {volume} {319}},\ \bibinfo {pages} {295--299} (\bibinfo {year}
  {2008})}\BibitemShut {NoStop}%
\bibitem [{\citenamefont {Kaufman}\ \emph {et~al.}(2015)\citenamefont
  {Kaufman}, \citenamefont {Lester}, \citenamefont {Foss-Feig}, \citenamefont
  {Wall}, \citenamefont {Rey},\ and\ \citenamefont
  {Regal}}]{kaufman2015entangling}%
  \BibitemOpen
  \bibfield  {author} {\bibinfo {author} {\bibfnamefont {AM}~\bibnamefont
  {Kaufman}}, \bibinfo {author} {\bibfnamefont {BJ}~\bibnamefont {Lester}},
  \bibinfo {author} {\bibfnamefont {M}~\bibnamefont {Foss-Feig}}, \bibinfo
  {author} {\bibfnamefont {ML}~\bibnamefont {Wall}}, \bibinfo {author}
  {\bibfnamefont {AM}~\bibnamefont {Rey}}, \ and\ \bibinfo {author}
  {\bibfnamefont {CA}~\bibnamefont {Regal}},\ }\bibfield  {title} {\enquote
  {\bibinfo {title} {Entangling two transportable neutral atoms via local spin
  exchange},}\ }\href@noop {} {\bibfield  {journal} {\bibinfo  {journal}
  {Nature}\ }\textbf {\bibinfo {volume} {527}},\ \bibinfo {pages} {208--211}
  (\bibinfo {year} {2015})}\BibitemShut {NoStop}%
\bibitem [{\citenamefont {Barredo}\ \emph {et~al.}(2016)\citenamefont
  {Barredo}, \citenamefont {De~L{\'e}s{\'e}leuc}, \citenamefont {Lienhard},
  \citenamefont {Lahaye},\ and\ \citenamefont {Browaeys}}]{barredo2016atom}%
  \BibitemOpen
  \bibfield  {author} {\bibinfo {author} {\bibfnamefont {Daniel}\ \bibnamefont
  {Barredo}}, \bibinfo {author} {\bibfnamefont {Sylvain}\ \bibnamefont
  {De~L{\'e}s{\'e}leuc}}, \bibinfo {author} {\bibfnamefont {Vincent}\
  \bibnamefont {Lienhard}}, \bibinfo {author} {\bibfnamefont {Thierry}\
  \bibnamefont {Lahaye}}, \ and\ \bibinfo {author} {\bibfnamefont {Antoine}\
  \bibnamefont {Browaeys}},\ }\bibfield  {title} {\enquote {\bibinfo {title}
  {An atom-by-atom assembler of defect-free arbitrary two-dimensional atomic
  arrays},}\ }\href@noop {} {\bibfield  {journal} {\bibinfo  {journal}
  {Science}\ }\textbf {\bibinfo {volume} {354}},\ \bibinfo {pages} {1021--1023}
  (\bibinfo {year} {2016})}\BibitemShut {NoStop}%
\bibitem [{\citenamefont {Endres}\ \emph {et~al.}(2016)\citenamefont {Endres},
  \citenamefont {Bernien}, \citenamefont {Keesling}, \citenamefont {Levine},
  \citenamefont {Anschuetz}, \citenamefont {Krajenbrink}, \citenamefont
  {Senko}, \citenamefont {Vuletic}, \citenamefont {Greiner},\ and\
  \citenamefont {Lukin}}]{endres2016atom}%
  \BibitemOpen
  \bibfield  {author} {\bibinfo {author} {\bibfnamefont {Manuel}\ \bibnamefont
  {Endres}}, \bibinfo {author} {\bibfnamefont {Hannes}\ \bibnamefont
  {Bernien}}, \bibinfo {author} {\bibfnamefont {Alexander}\ \bibnamefont
  {Keesling}}, \bibinfo {author} {\bibfnamefont {Harry}\ \bibnamefont
  {Levine}}, \bibinfo {author} {\bibfnamefont {Eric~R}\ \bibnamefont
  {Anschuetz}}, \bibinfo {author} {\bibfnamefont {Alexandre}\ \bibnamefont
  {Krajenbrink}}, \bibinfo {author} {\bibfnamefont {Crystal}\ \bibnamefont
  {Senko}}, \bibinfo {author} {\bibfnamefont {Vladan}\ \bibnamefont {Vuletic}},
  \bibinfo {author} {\bibfnamefont {Markus}\ \bibnamefont {Greiner}}, \ and\
  \bibinfo {author} {\bibfnamefont {Mikhail~D}\ \bibnamefont {Lukin}},\
  }\bibfield  {title} {\enquote {\bibinfo {title} {Atom-by-atom assembly of
  defect-free one-dimensional cold atom arrays},}\ }\href@noop {} {\bibfield
  {journal} {\bibinfo  {journal} {Science}\ }\textbf {\bibinfo {volume}
  {354}},\ \bibinfo {pages} {1024--1027} (\bibinfo {year} {2016})}\BibitemShut
  {NoStop}%
\bibitem [{\citenamefont {Singh}\ \emph {et~al.}(2023)\citenamefont {Singh},
  \citenamefont {Bradley}, \citenamefont {Anand}, \citenamefont {Ramesh},
  \citenamefont {White},\ and\ \citenamefont {Bernien}}]{singh2023mid}%
  \BibitemOpen
  \bibfield  {author} {\bibinfo {author} {\bibfnamefont {K}~\bibnamefont
  {Singh}}, \bibinfo {author} {\bibfnamefont {CE}~\bibnamefont {Bradley}},
  \bibinfo {author} {\bibfnamefont {S}~\bibnamefont {Anand}}, \bibinfo {author}
  {\bibfnamefont {V}~\bibnamefont {Ramesh}}, \bibinfo {author} {\bibfnamefont
  {R}~\bibnamefont {White}}, \ and\ \bibinfo {author} {\bibfnamefont
  {H}~\bibnamefont {Bernien}},\ }\bibfield  {title} {\enquote {\bibinfo {title}
  {Mid-circuit correction of correlated phase errors using an array of
  spectator qubits},}\ }\href@noop {} {\bibfield  {journal} {\bibinfo
  {journal} {Science}\ }\textbf {\bibinfo {volume} {380}},\ \bibinfo {pages}
  {1265--1269} (\bibinfo {year} {2023})}\BibitemShut {NoStop}%
\bibitem [{\citenamefont {Lis}\ \emph {et~al.}(2023)\citenamefont {Lis},
  \citenamefont {Senoo}, \citenamefont {McGrew}, \citenamefont {R{\"o}nchen},
  \citenamefont {Jenkins},\ and\ \citenamefont {Kaufman}}]{lis2023mid}%
  \BibitemOpen
  \bibfield  {author} {\bibinfo {author} {\bibfnamefont {Joanna~W}\
  \bibnamefont {Lis}}, \bibinfo {author} {\bibfnamefont {Aruku}\ \bibnamefont
  {Senoo}}, \bibinfo {author} {\bibfnamefont {William~F}\ \bibnamefont
  {McGrew}}, \bibinfo {author} {\bibfnamefont {Felix}\ \bibnamefont
  {R{\"o}nchen}}, \bibinfo {author} {\bibfnamefont {Alec}\ \bibnamefont
  {Jenkins}}, \ and\ \bibinfo {author} {\bibfnamefont {Adam~M}\ \bibnamefont
  {Kaufman}},\ }\bibfield  {title} {\enquote {\bibinfo {title} {Mid-circuit
  operations using the omg-architecture in neutral atom arrays},}\ }\href@noop
  {} {\bibfield  {journal} {\bibinfo  {journal} {arXiv preprint
  arXiv:2305.19266}\ } (\bibinfo {year} {2023})}\BibitemShut {NoStop}%
\bibitem [{\citenamefont {Norcia}\ \emph {et~al.}(2023)\citenamefont {Norcia},
  \citenamefont {Cairncross}, \citenamefont {Barnes}, \citenamefont
  {Battaglino}, \citenamefont {Brown}, \citenamefont {Brown}, \citenamefont
  {Cassella}, \citenamefont {Chen}, \citenamefont {Coxe}, \citenamefont {Crow}
  \emph {et~al.}}]{norcia2023mid}%
  \BibitemOpen
  \bibfield  {author} {\bibinfo {author} {\bibfnamefont {MA}~\bibnamefont
  {Norcia}}, \bibinfo {author} {\bibfnamefont {WB}~\bibnamefont {Cairncross}},
  \bibinfo {author} {\bibfnamefont {K}~\bibnamefont {Barnes}}, \bibinfo
  {author} {\bibfnamefont {P}~\bibnamefont {Battaglino}}, \bibinfo {author}
  {\bibfnamefont {A}~\bibnamefont {Brown}}, \bibinfo {author} {\bibfnamefont
  {MO}~\bibnamefont {Brown}}, \bibinfo {author} {\bibfnamefont {K}~\bibnamefont
  {Cassella}}, \bibinfo {author} {\bibfnamefont {C-A}\ \bibnamefont {Chen}},
  \bibinfo {author} {\bibfnamefont {R}~\bibnamefont {Coxe}}, \bibinfo {author}
  {\bibfnamefont {D}~\bibnamefont {Crow}},  \emph {et~al.},\ }\bibfield
  {title} {\enquote {\bibinfo {title} {Mid-circuit qubit measurement and
  rearrangement in a $^{137}$ yb atomic array},}\ }\href@noop {} {\bibfield
  {journal} {\bibinfo  {journal} {arXiv preprint arXiv:2305.19119}\ } (\bibinfo
  {year} {2023})}\BibitemShut {NoStop}%
\bibitem [{\citenamefont {Scholl}\ \emph {et~al.}(2023)\citenamefont {Scholl},
  \citenamefont {Shaw}, \citenamefont {Tsai}, \citenamefont {Finkelstein},
  \citenamefont {Choi},\ and\ \citenamefont {Endres}}]{scholl2023erasure}%
  \BibitemOpen
  \bibfield  {author} {\bibinfo {author} {\bibfnamefont {Pascal}\ \bibnamefont
  {Scholl}}, \bibinfo {author} {\bibfnamefont {Adam~L}\ \bibnamefont {Shaw}},
  \bibinfo {author} {\bibfnamefont {Richard Bing-Shiun}\ \bibnamefont {Tsai}},
  \bibinfo {author} {\bibfnamefont {Ran}\ \bibnamefont {Finkelstein}}, \bibinfo
  {author} {\bibfnamefont {Joonhee}\ \bibnamefont {Choi}}, \ and\ \bibinfo
  {author} {\bibfnamefont {Manuel}\ \bibnamefont {Endres}},\ }\bibfield
  {title} {\enquote {\bibinfo {title} {Erasure conversion in a high-fidelity
  rydberg quantum simulator},}\ }\href@noop {} {\bibfield  {journal} {\bibinfo
  {journal} {arXiv preprint arXiv:2305.03406}\ } (\bibinfo {year}
  {2023})}\BibitemShut {NoStop}%
\bibitem [{\citenamefont {Pause}\ \emph {et~al.}(2023)\citenamefont {Pause},
  \citenamefont {Preuschoff}, \citenamefont {Sch{\"a}ffner}, \citenamefont
  {Schlosser},\ and\ \citenamefont {Birkl}}]{pause2023reservoir}%
  \BibitemOpen
  \bibfield  {author} {\bibinfo {author} {\bibfnamefont {Lars}\ \bibnamefont
  {Pause}}, \bibinfo {author} {\bibfnamefont {Tilman}\ \bibnamefont
  {Preuschoff}}, \bibinfo {author} {\bibfnamefont {Dominik}\ \bibnamefont
  {Sch{\"a}ffner}}, \bibinfo {author} {\bibfnamefont {Malte}\ \bibnamefont
  {Schlosser}}, \ and\ \bibinfo {author} {\bibfnamefont {Gerhard}\ \bibnamefont
  {Birkl}},\ }\bibfield  {title} {\enquote {\bibinfo {title} {Reservoir-based
  deterministic loading of single-atom tweezer arrays},}\ }\href@noop {}
  {\bibfield  {journal} {\bibinfo  {journal} {Physical Review Research}\
  }\textbf {\bibinfo {volume} {5}},\ \bibinfo {pages} {L032009} (\bibinfo
  {year} {2023})}\BibitemShut {NoStop}%
\bibitem [{\citenamefont {Xia}\ \emph {et~al.}(2015)\citenamefont {Xia},
  \citenamefont {Lichtman}, \citenamefont {Maller}, \citenamefont {Carr},
  \citenamefont {Piotrowicz}, \citenamefont {Isenhower},\ and\ \citenamefont
  {Saffman}}]{xia2015randomized}%
  \BibitemOpen
  \bibfield  {author} {\bibinfo {author} {\bibfnamefont {T}~\bibnamefont
  {Xia}}, \bibinfo {author} {\bibfnamefont {M}~\bibnamefont {Lichtman}},
  \bibinfo {author} {\bibfnamefont {K}~\bibnamefont {Maller}}, \bibinfo
  {author} {\bibfnamefont {AW}~\bibnamefont {Carr}}, \bibinfo {author}
  {\bibfnamefont {MJ}~\bibnamefont {Piotrowicz}}, \bibinfo {author}
  {\bibfnamefont {L}~\bibnamefont {Isenhower}}, \ and\ \bibinfo {author}
  {\bibfnamefont {M}~\bibnamefont {Saffman}},\ }\bibfield  {title} {\enquote
  {\bibinfo {title} {Randomized benchmarking of single-qubit gates in a 2d
  array of neutral-atom qubits},}\ }\href@noop {} {\bibfield  {journal}
  {\bibinfo  {journal} {Physical review letters}\ }\textbf {\bibinfo {volume}
  {114}},\ \bibinfo {pages} {100503} (\bibinfo {year} {2015})}\BibitemShut
  {NoStop}%
\bibitem [{\citenamefont {Wang}\ \emph {et~al.}(2016)\citenamefont {Wang},
  \citenamefont {Kumar}, \citenamefont {Wu},\ and\ \citenamefont
  {Weiss}}]{wang2016single}%
  \BibitemOpen
  \bibfield  {author} {\bibinfo {author} {\bibfnamefont {Yang}\ \bibnamefont
  {Wang}}, \bibinfo {author} {\bibfnamefont {Aishwarya}\ \bibnamefont {Kumar}},
  \bibinfo {author} {\bibfnamefont {Tsung-Yao}\ \bibnamefont {Wu}}, \ and\
  \bibinfo {author} {\bibfnamefont {David~S}\ \bibnamefont {Weiss}},\
  }\bibfield  {title} {\enquote {\bibinfo {title} {Single-qubit gates based on
  targeted phase shifts in a 3d neutral atom array},}\ }\href@noop {}
  {\bibfield  {journal} {\bibinfo  {journal} {Science}\ }\textbf {\bibinfo
  {volume} {352}},\ \bibinfo {pages} {1562--1565} (\bibinfo {year}
  {2016})}\BibitemShut {NoStop}%
\bibitem [{\citenamefont {Sheng}\ \emph {et~al.}(2018)\citenamefont {Sheng},
  \citenamefont {He}, \citenamefont {Xu}, \citenamefont {Guo}, \citenamefont
  {Wang}, \citenamefont {Xiong}, \citenamefont {Liu}, \citenamefont {Wang},\
  and\ \citenamefont {Zhan}}]{sheng2018high}%
  \BibitemOpen
  \bibfield  {author} {\bibinfo {author} {\bibfnamefont {Cheng}\ \bibnamefont
  {Sheng}}, \bibinfo {author} {\bibfnamefont {Xiaodong}\ \bibnamefont {He}},
  \bibinfo {author} {\bibfnamefont {Peng}\ \bibnamefont {Xu}}, \bibinfo
  {author} {\bibfnamefont {Ruijun}\ \bibnamefont {Guo}}, \bibinfo {author}
  {\bibfnamefont {Kunpeng}\ \bibnamefont {Wang}}, \bibinfo {author}
  {\bibfnamefont {Zongyuan}\ \bibnamefont {Xiong}}, \bibinfo {author}
  {\bibfnamefont {Min}\ \bibnamefont {Liu}}, \bibinfo {author} {\bibfnamefont
  {Jin}\ \bibnamefont {Wang}}, \ and\ \bibinfo {author} {\bibfnamefont
  {Mingsheng}\ \bibnamefont {Zhan}},\ }\bibfield  {title} {\enquote {\bibinfo
  {title} {High-fidelity single-qubit gates on neutral atoms in a
  two-dimensional magic-intensity optical dipole trap array},}\ }\href@noop {}
  {\bibfield  {journal} {\bibinfo  {journal} {Physical review letters}\
  }\textbf {\bibinfo {volume} {121}},\ \bibinfo {pages} {240501} (\bibinfo
  {year} {2018})}\BibitemShut {NoStop}%
\bibitem [{\citenamefont {Jenkins}\ \emph {et~al.}(2022)\citenamefont
  {Jenkins}, \citenamefont {Lis}, \citenamefont {Senoo}, \citenamefont
  {McGrew},\ and\ \citenamefont {Kaufman}}]{jenkins2022ytterbium}%
  \BibitemOpen
  \bibfield  {author} {\bibinfo {author} {\bibfnamefont {Alec}\ \bibnamefont
  {Jenkins}}, \bibinfo {author} {\bibfnamefont {Joanna~W}\ \bibnamefont {Lis}},
  \bibinfo {author} {\bibfnamefont {Aruku}\ \bibnamefont {Senoo}}, \bibinfo
  {author} {\bibfnamefont {William~F}\ \bibnamefont {McGrew}}, \ and\ \bibinfo
  {author} {\bibfnamefont {Adam~M}\ \bibnamefont {Kaufman}},\ }\bibfield
  {title} {\enquote {\bibinfo {title} {Ytterbium nuclear-spin qubits in an
  optical tweezer array},}\ }\href@noop {} {\bibfield  {journal} {\bibinfo
  {journal} {Physical Review X}\ }\textbf {\bibinfo {volume} {12}},\ \bibinfo
  {pages} {021027} (\bibinfo {year} {2022})}\BibitemShut {NoStop}%
\bibitem [{\citenamefont {Ma}\ \emph {et~al.}(2022)\citenamefont {Ma},
  \citenamefont {Burgers}, \citenamefont {Liu}, \citenamefont {Wilson},
  \citenamefont {Zhang},\ and\ \citenamefont {Thompson}}]{ma2022universal}%
  \BibitemOpen
  \bibfield  {author} {\bibinfo {author} {\bibfnamefont {Shuo}\ \bibnamefont
  {Ma}}, \bibinfo {author} {\bibfnamefont {Alex~P}\ \bibnamefont {Burgers}},
  \bibinfo {author} {\bibfnamefont {Genyue}\ \bibnamefont {Liu}}, \bibinfo
  {author} {\bibfnamefont {Jack}\ \bibnamefont {Wilson}}, \bibinfo {author}
  {\bibfnamefont {Bichen}\ \bibnamefont {Zhang}}, \ and\ \bibinfo {author}
  {\bibfnamefont {Jeff~D}\ \bibnamefont {Thompson}},\ }\bibfield  {title}
  {\enquote {\bibinfo {title} {Universal gate operations on nuclear spin qubits
  in an optical tweezer array of yb 171 atoms},}\ }\href@noop {} {\bibfield
  {journal} {\bibinfo  {journal} {Physical Review X}\ }\textbf {\bibinfo
  {volume} {12}},\ \bibinfo {pages} {021028} (\bibinfo {year}
  {2022})}\BibitemShut {NoStop}%
\bibitem [{\citenamefont {Suchara}\ \emph {et~al.}(2015)\citenamefont
  {Suchara}, \citenamefont {Cross},\ and\ \citenamefont
  {Gambetta}}]{suchara2015leakage}%
  \BibitemOpen
  \bibfield  {author} {\bibinfo {author} {\bibfnamefont {Martin}\ \bibnamefont
  {Suchara}}, \bibinfo {author} {\bibfnamefont {Andrew~W}\ \bibnamefont
  {Cross}}, \ and\ \bibinfo {author} {\bibfnamefont {Jay~M}\ \bibnamefont
  {Gambetta}},\ }\bibfield  {title} {\enquote {\bibinfo {title} {Leakage
  suppression in the toric code},}\ }in\ \href@noop {} {\emph {\bibinfo
  {booktitle} {2015 IEEE International Symposium on Information Theory
  (ISIT)}}}\ (\bibinfo {organization} {IEEE},\ \bibinfo {year} {2015})\ pp.\
  \bibinfo {pages} {1119--1123}\BibitemShut {NoStop}%
\bibitem [{\citenamefont {Berthusen}\ and\ \citenamefont
  {Gottesman}(2023)}]{Berthusen:2023lpm}%
  \BibitemOpen
  \bibfield  {author} {\bibinfo {author} {\bibfnamefont {Noah}\ \bibnamefont
  {Berthusen}}\ and\ \bibinfo {author} {\bibfnamefont {Daniel}\ \bibnamefont
  {Gottesman}},\ }\bibfield  {title} {\enquote {\bibinfo {title} {{Partial
  Syndrome Measurement for Hypergraph Product Codes}},}\ }\href@noop {} {\
  (\bibinfo {year} {2023})},\ \Eprint {http://arxiv.org/abs/2306.17122}
  {arXiv:2306.17122 [quant-ph]} \BibitemShut {NoStop}%
\bibitem [{\citenamefont {Delfosse}\ and\ \citenamefont
  {Nickerson}(2021)}]{Delfosse_2021}%
  \BibitemOpen
  \bibfield  {author} {\bibinfo {author} {\bibfnamefont {Nicolas}\ \bibnamefont
  {Delfosse}}\ and\ \bibinfo {author} {\bibfnamefont {Naomi~H.}\ \bibnamefont
  {Nickerson}},\ }\bibfield  {title} {\enquote {\bibinfo {title} {Almost-linear
  time decoding algorithm for topological codes},}\ }\href {\doibase
  10.22331/q-2021-12-02-595} {\bibfield  {journal} {\bibinfo  {journal}
  {Quantum}\ }\textbf {\bibinfo {volume} {5}},\ \bibinfo {pages} {595}
  (\bibinfo {year} {2021})}\BibitemShut {NoStop}%
\bibitem [{\citenamefont {H{\'{e}ctor Bomb{\'{\i}}n}}(2015)}]{Bombin_2015}%
  \BibitemOpen
  \bibfield  {author} {\bibinfo {author} {\bibnamefont {H{\'{e}ctor
  Bomb{\'{\i}}n}}},\ }\bibfield  {title} {\enquote {\bibinfo {title}
  {Single-shot fault-tolerant quantum error correction},}\ }\href {\doibase
  10.1103/physrevx.5.031043} {\bibfield  {journal} {\bibinfo  {journal}
  {Physical Review X}\ }\textbf {\bibinfo {volume} {5}} (\bibinfo {year}
  {2015}),\ 10.1103/physrevx.5.031043}\BibitemShut {NoStop}%
\bibitem [{\citenamefont {Kubica}\ and\ \citenamefont
  {Vasmer}(2022)}]{Kubica_2022}%
  \BibitemOpen
  \bibfield  {author} {\bibinfo {author} {\bibfnamefont {Aleksander}\
  \bibnamefont {Kubica}}\ and\ \bibinfo {author} {\bibfnamefont {Michael}\
  \bibnamefont {Vasmer}},\ }\bibfield  {title} {\enquote {\bibinfo {title}
  {Single-shot quantum error correction with the three-dimensional subsystem
  toric code},}\ }\href {\doibase 10.1038/s41467-022-33923-4} {\bibfield
  {journal} {\bibinfo  {journal} {Nature Communications}\ }\textbf {\bibinfo
  {volume} {13}} (\bibinfo {year} {2022}),\
  10.1038/s41467-022-33923-4}\BibitemShut {NoStop}%
\bibitem [{\citenamefont {Stahl}(2023)}]{stahl2023}%
  \BibitemOpen
  \bibfield  {author} {\bibinfo {author} {\bibfnamefont {Charles}\ \bibnamefont
  {Stahl}},\ }\href@noop {} {\enquote {\bibinfo {title} {Single-shot quantum
  error correction in intertwined toric codes},}\ } (\bibinfo {year} {2023}),\
  \Eprint {http://arxiv.org/abs/2307.08118} {arXiv:2307.08118 [quant-ph]}
  \BibitemShut {NoStop}%
\bibitem [{\citenamefont {Richardson}\ and\ \citenamefont
  {Urbanke}(2008)}]{Richardson2008}%
  \BibitemOpen
  \bibfield  {author} {\bibinfo {author} {\bibfnamefont {Tom}\ \bibnamefont
  {Richardson}}\ and\ \bibinfo {author} {\bibfnamefont {Rüdiger}\ \bibnamefont
  {Urbanke}},\ }\href@noop {} {\emph {\bibinfo {title} {Modern Coding
  Theory}}}\ (\bibinfo  {publisher} {Cambridge University Press},\ \bibinfo
  {year} {2008})\BibitemShut {NoStop}%
\bibitem [{\citenamefont {Roffe}\ \emph {et~al.}(2020)\citenamefont {Roffe},
  \citenamefont {White}, \citenamefont {Burton},\ and\ \citenamefont
  {Campbell}}]{Roffe_2020}%
  \BibitemOpen
  \bibfield  {author} {\bibinfo {author} {\bibfnamefont {Joschka}\ \bibnamefont
  {Roffe}}, \bibinfo {author} {\bibfnamefont {David~R.}\ \bibnamefont {White}},
  \bibinfo {author} {\bibfnamefont {Simon}\ \bibnamefont {Burton}}, \ and\
  \bibinfo {author} {\bibfnamefont {Earl}\ \bibnamefont {Campbell}},\
  }\bibfield  {title} {\enquote {\bibinfo {title} {Decoding across the quantum
  low-density parity-check code landscape},}\ }\href {\doibase
  10.1103/physrevresearch.2.043423} {\bibfield  {journal} {\bibinfo  {journal}
  {Physical Review Research}\ }\textbf {\bibinfo {volume} {2}} (\bibinfo {year}
  {2020}),\ 10.1103/physrevresearch.2.043423}\BibitemShut {NoStop}%
\bibitem [{\citenamefont {Leverrier}\ and\ \citenamefont
  {Zémor}(2022{\natexlab{b}})}]{leverrier2022efficient}%
  \BibitemOpen
  \bibfield  {author} {\bibinfo {author} {\bibfnamefont {Anthony}\ \bibnamefont
  {Leverrier}}\ and\ \bibinfo {author} {\bibfnamefont {Gilles}\ \bibnamefont
  {Zémor}},\ }\href@noop {} {\enquote {\bibinfo {title} {Efficient decoding up
  to a constant fraction of the code length for asymptotically good quantum
  codes},}\ } (\bibinfo {year} {2022}{\natexlab{b}}),\ \Eprint
  {http://arxiv.org/abs/2206.07571} {arXiv:2206.07571 [quant-ph]} \BibitemShut
  {NoStop}%
\bibitem [{\citenamefont {Leverrier}(2022)}]{qTanner_rotated_surface}%
  \BibitemOpen
  \bibfield  {author} {\bibinfo {author} {\bibfnamefont {Anthony}\ \bibnamefont
  {Leverrier}},\ }\href@noop {} {\enquote {\bibinfo {title} {Mapping the toric
  code to the rotated toric code},}\ }\bibinfo {howpublished}
  {\url{https://github.com/errorcorrectionzoo/eczoo_data/files/9210173/rotated.pdf}}
  (\bibinfo {year} {2022})\BibitemShut {NoStop}%
\bibitem [{\citenamefont {Manes}\ and\ \citenamefont
  {Claes}(2023)}]{manes2023}%
  \BibitemOpen
  \bibfield  {author} {\bibinfo {author} {\bibfnamefont {Argyris~Giannisis}\
  \bibnamefont {Manes}}\ and\ \bibinfo {author} {\bibfnamefont {Jahan}\
  \bibnamefont {Claes}},\ }\href@noop {} {\enquote {\bibinfo {title}
  {Distance-preserving stabilizer measurements in hypergraph product codes},}\
  } (\bibinfo {year} {2023}),\ \Eprint {http://arxiv.org/abs/2308.15520}
  {arXiv:2308.15520 [quant-ph]} \BibitemShut {NoStop}%
\bibitem [{\citenamefont {Quintavalle}\ \emph {et~al.}(2021)\citenamefont
  {Quintavalle}, \citenamefont {Vasmer}, \citenamefont {Roffe},\ and\
  \citenamefont {Campbell}}]{Quintavalle_2021}%
  \BibitemOpen
  \bibfield  {author} {\bibinfo {author} {\bibfnamefont {Armanda~O.}\
  \bibnamefont {Quintavalle}}, \bibinfo {author} {\bibfnamefont {Michael}\
  \bibnamefont {Vasmer}}, \bibinfo {author} {\bibfnamefont {Joschka}\
  \bibnamefont {Roffe}}, \ and\ \bibinfo {author} {\bibfnamefont {Earl~T.}\
  \bibnamefont {Campbell}},\ }\bibfield  {title} {\enquote {\bibinfo {title}
  {Single-shot error correction of three-dimensional homological product
  codes},}\ }\href {\doibase 10.1103/prxquantum.2.020340} {\bibfield  {journal}
  {\bibinfo  {journal} {{PRX} Quantum}\ }\textbf {\bibinfo {volume} {2}}
  (\bibinfo {year} {2021}),\ 10.1103/prxquantum.2.020340}\BibitemShut {NoStop}%
\bibitem [{\citenamefont {Sipser}\ and\ \citenamefont
  {Spielman}(1996)}]{sipser1996}%
  \BibitemOpen
  \bibfield  {author} {\bibinfo {author} {\bibfnamefont {M.}~\bibnamefont
  {Sipser}}\ and\ \bibinfo {author} {\bibfnamefont {D.A.}\ \bibnamefont
  {Spielman}},\ }\bibfield  {title} {\enquote {\bibinfo {title} {Expander
  codes},}\ }\href {\doibase 10.1109/18.556667} {\bibfield  {journal} {\bibinfo
   {journal} {IEEE Transactions on Information Theory}\ }\textbf {\bibinfo
  {volume} {42}},\ \bibinfo {pages} {1710--1722} (\bibinfo {year}
  {1996})}\BibitemShut {NoStop}%
\bibitem [{\citenamefont {Leverrier}\ \emph {et~al.}(2015)\citenamefont
  {Leverrier}, \citenamefont {Tillich},\ and\ \citenamefont
  {Zemor}}]{Leverrier_2015}%
  \BibitemOpen
  \bibfield  {author} {\bibinfo {author} {\bibfnamefont {Anthony}\ \bibnamefont
  {Leverrier}}, \bibinfo {author} {\bibfnamefont {Jean-Pierre}\ \bibnamefont
  {Tillich}}, \ and\ \bibinfo {author} {\bibfnamefont {Gilles}\ \bibnamefont
  {Zemor}},\ }\bibfield  {title} {\enquote {\bibinfo {title} {Quantum expander
  codes},}\ }in\ \href {\doibase 10.1109/focs.2015.55} {\emph {\bibinfo
  {booktitle} {2015 IEEE 56th Annual Symposium on Foundations of Computer
  Science}}}\ (\bibinfo  {publisher} {IEEE},\ \bibinfo {year}
  {2015})\BibitemShut {NoStop}%
\bibitem [{\citenamefont {Fawzi}\ \emph {et~al.}(2020)\citenamefont {Fawzi},
  \citenamefont {Grospellier},\ and\ \citenamefont {Leverrier}}]{Fawzi_2020}%
  \BibitemOpen
  \bibfield  {author} {\bibinfo {author} {\bibfnamefont {Omar}\ \bibnamefont
  {Fawzi}}, \bibinfo {author} {\bibfnamefont {Antoine}\ \bibnamefont
  {Grospellier}}, \ and\ \bibinfo {author} {\bibfnamefont {Anthony}\
  \bibnamefont {Leverrier}},\ }\bibfield  {title} {\enquote {\bibinfo {title}
  {Constant overhead quantum fault tolerance with quantum expander codes},}\
  }\href {\doibase 10.1145/3434163} {\bibfield  {journal} {\bibinfo  {journal}
  {Commun. ACM}\ }\textbf {\bibinfo {volume} {64}},\ \bibinfo {pages}
  {106–114} (\bibinfo {year} {2020})}\BibitemShut {NoStop}%
\bibitem [{\citenamefont {Grospellier}\ and\ \citenamefont
  {Krishna}(2019)}]{grospellier2019}%
  \BibitemOpen
  \bibfield  {author} {\bibinfo {author} {\bibfnamefont {Antoine}\ \bibnamefont
  {Grospellier}}\ and\ \bibinfo {author} {\bibfnamefont {Anirudh}\ \bibnamefont
  {Krishna}},\ }\href@noop {} {\enquote {\bibinfo {title} {Numerical study of
  hypergraph product codes},}\ } (\bibinfo {year} {2019}),\ \Eprint
  {http://arxiv.org/abs/1810.03681} {arXiv:1810.03681 [quant-ph]} \BibitemShut
  {NoStop}%
\end{thebibliography}%

\end{document}